\documentclass[article]{JHEP}
\usepackage{amsmath,epsfig}
\usepackage{amssymb,amsfonts}
\usepackage{latexsym}
\usepackage{graphicx}
\usepackage{epsfig}
\relax
\renewcommand{\theequation}{\arabic{section}.\arabic{equation}}
\renewcommand{\theequation}{\arabic{section}.\arabic{subsection}.\arabic{equation}}
\renewcommand{\theequation}{\arabic{section}.\arabic{equation}}
\def\be{\begin{equation}}
\def\ee{\end{equation}}
\def\bea{\begin{eqnarray}}
\def\eea{\end{eqnarray}}
\def\be{\begin{equation}}
\def\ee{\end{equation}}

\newcommand\fverb{\setbox\pippobox=\hbox\bgroup\verb}
\newcommand\fverbdo{\egroup\medskip\noindent%
                        \fbox{\unhbox\pippobox}\ }
\newcommand\fverbit{\egroup\item[\fbox{\unhbox\pippobox}]}
\newbox\pippobox
\def\F{\Phi}

\def\e{\epsilon}
\def\h{\eta}
\def\ka{\kappa}
\def\z{\zeta}
\def\m{\mu}
\def\n{\nu}
\def\r{\rho}
\def\s{\sigma}
\def\th{\theta}
\def\t{\tau}
\def\sp{\;\;\;,\;\;\;}

\def\p{\partial}

\def\f{\varphi}
\def\a{\alpha}
\def\b{\beta}
\def\d{\delta}
\def\l{\lambda}
\def\g{\gamma}
\def\G{\Gamma}
\def\w{\omega}
\def\W{\Omega}
\def\la{\langle}
\def\ra{\rangle}
\def\ib{\langle 1 \rangle}
\def\ba{\begin{eqnarray}}
\def\ea{\end{eqnarray}}
\def\nb{\nonumber}
\def\jh{{\hat{\jmath}}}

\title{D-branes and BCFT in Hpp-wave backgrounds}
\author{Giuseppe D'Appollonio\\
Department of Mathematics, King's College\\
The Strand, London WC2R 2LS, U.K.\\
and\\
LPTHE, Universit{\'e} Paris VI, 4 pl Jussieu, \\
75252 Paris cedex 05, FRANCE\\
{\tt E-mail: giuseppe@mth.kcl.ac.uk}}
\author{
Elias Kiritsis\\
CPHT, Ecole Polytechnique,\\
 91128, Palaiseau, FRANCE\\
  UMR du CNRS 7644.\\
and\\
Department of Physics, University of Crete\\
71003 Heraklion, GREECE\\
{\tt E-mail: kiritsis@cpht.polytechnique.fr} }

\preprint{\hepth{yymmxxx} \\ CPHT RR 012.0304 \\ KCL-MTH-04-15
 \\ LPTHE-04-28}

\abstract{ In this paper we study two classes of symmetric
D-branes in the Nappi-Witten gravitational wave, namely D2 and $S
1$ branes. We solve the sewing constraints and determine the
bulk-boundary couplings and the boundary three-point couplings.
For the D2 brane our solution gives the first explicit results for
the structure constants of the twisted symmetric branes in a WZW
model. We also compute the boundary four-point functions,
providing examples of open string four-point amplitudes in a
curved background. We finally discuss the annulus amplitudes, the
relation with branes in $AdS_3$ and in $S^3$ and the analogy
between the open string couplings in the $H_4$ model and the
couplings for magnetized and intersecting branes. }

\begin{document}

\maketitle 

\section{Introduction}

The study of gravitational waves as string theory backgrounds began more
than fifteen years ago \cite{orig}-\cite{nw}. They were proposed as the
most convenient starting point for extending the analysis of the
properties of string theory from the familiar vacua given by the product
of flat space and a compact manifold to the less explored curved,
non-compact  space-times. The main reason was that already from the point
of view of general relativity  the gravitational waves are some of the
simplest time-dependent backgrounds. They admit a covariantly constant
null Killing vector, most of their curvature invariants vanish and there
is no particle creation. Another distinctive feature, which is
particularly relevant for string theory, is that it is always possible to
fix the light-cone gauge for the quantization of the world-sheet action.
Recently it was also realized that the gravitational waves play an
important role in the study of  gauge/string duality.

Motivated by the
notion of Penrose limits \cite{pen1}-\cite{pen2}, it was argued
\cite{mald} that such backgrounds are dual to modified large-N limits of
gauge theories. This observation has opened new avenues in understanding
stringy aspects of the gauge theory/string theory duality. Indeed, the
Green-Schwarz action for the pp-wave that is obtained by the Penrose limit
of $AdS_5 \times S^5$, can be quantized in the light-cone gauge, even
though there is a non trivial R-R flux \cite{metsaev,mt}. A lot of
progress has been made since, and this is reviewed in
\cite{review}-\cite{review3}. It should be noted however that at the
moment there is no generally accepted theory of the full holographic
correspondence, although several proposals have been put forward
\cite{hol1}-\cite{hol5}.

Given the relevance for both the study of string theory in non-compact
curved space-times and the gauge theory/string theory  duality, we believe
that it is important to obtain a clear and detailed understanding of the
string dynamics at least in some particular gravitational wave
backgrounds. With this aim in mind, it is natural that our choice falls
upon a class of gravitational waves that are supported by a NS-NS flux and
have an exact CFT description as WZW models. This is the class of the WZW
models based on the Heisenberg groups $H_{2n+2}$, $n \ge 1$. The first
example in this family was discovered  by Nappi and Witten \cite{nw} and
the others were introduced in \cite{ks1},\cite{kehagias}. These models,
unlike those with the same metric but supported by a R-R flux, can be
quantized in a covariant way using standard perturbative string theory
techniques. The presence of the affine symmetry algebra then imposes
additional constraints that can lead to the complete solution of the
model\footnote{See \cite{papasolv,papasolv1} for a study of other
interesting pp-waves without affine symmetry algebras.}. Since the study
of string and brane dynamics in non-compact curved backgrounds is a
difficult arena, all models that can be solved exactly are a source of
useful information. Unfortunately, only very few examples are available
and they essentially amount to the Liouville model \cite{dol,zzl,tliou},
to $AdS_3$ \cite{tads,moog1} and its cosets \cite{2dbh}-\cite{2dbh2}.

In \cite{dk} we added another entry to this list, solving the $H_4$ model,
which describes the propagation of a string in a four-dimensional
gravitational wave. The structure of the closed string spectrum turned out
to be very similar to the one established for $AdS_3$ \cite{moog1}. It can
be organized in highest-weight and spectral-flowed representations of the
affine $\hat {\cal H}_4$ algebra and there are two distinct classes of
states. For generic values of the light-cone momentum $p$, the states
belong to the discrete representations of the $\hat {\cal H}_4$ algebra.
They correspond to short strings that are confined by the background
fields in closed orbits in the plane transverse to the two light-cone
coordinates. Whenever $\m \a\,' p \in \mathbb{Z}$, with $\m$ a parameter
of the pp-wave metric, the states belong to the continuous representations
of the $\hat {\cal H}_4$ algebra and correspond to long strings that move
freely in the transverse plane.

In \cite{dk}, we computed all the three
and four-point correlation functions of primary vertex operators, thus
providing all the structure constants for this non-compact WZW model. We
also showed explicitly that the spectral-flowed representations are
necessary for the consistency of the model since they appear in the
intermediate channel of four-point amplitudes with external highest-weight
states. In \cite{bdkz} we performed a similar analysis for the $H_{6}$
model. This model already displays all the new features of the
higher-dimensional cases, namely the existence of enhanced symmetry points
and the necessity of introducing representations that satisfy a modified
highest-weight condition, which generalize the concept of spectral-flowed
representations. The $H_6$ model is also relevant for the $AdS_3/CFT_2$
correspondence, being the Penrose limit of $AdS_3 \times S^3$.

As conformal field theories, the $H_4$  model and its higher dimensional
versions deserve attention not only because of the rich and interesting
structure we have just outlined but also because there are several
relations between them and other important models. Most of these relations
follow directly from the original idea of Penrose of considering the
gravitational waves as limits of other space-times. From the point of view
of the world-sheet $\s$-model, the Penrose limit that connects two
backgrounds both having an exact description as WZW models can be
interpreted as a contraction of the underlying current algebra \cite{ors}.
As such, the $H_4$ model captures the limiting behavior of backgrounds of
the form $\mathbb{R} \times S^3$ and $AdS_3 \times S^1$.  In \cite{dk} we
analyzed  the contraction of $\mathbb{R} \times SU(2)_k$ to $H_4$ and in
\cite{bdkz} the contraction of $SL(2,\mathbb{R})_k \times SU(2)_k$ to
$H_6$.

 The structure of the algebra changes drastically in the contraction
process. The structure of the space-time is also drastically changing
during the Penrose limit. Despite this,  we have shown that it is possible
to take the limit of the CFT operators and of the dynamical quantities
such as the correlation functions in a controlled way. Another interesting
relation stems from the free-field realization of the $H_4$ model
introduced in \cite{kk,kk2}. The $H_4$ primary vertex operators can be
represented using the twist fields of the orbifold obtained as the
quotient of the plane by a rotation and a dictionary can be established
connecting the amplitudes computed in the $H_{4}$ model and the amplitudes
computed in the orbifold CFT \cite{dk}.

In this paper we complete our analysis of the $H_4$ model by studying
D-branes in the Nappi-Witten gravitational wave and the dynamics of their
open string excitations. The dynamics of open strings in curved space-time
is even less understood than its closed counterpart and again we have at
our disposal a very limited number of exactly solved examples. What is
typically accessible, are the boundary states that have been studied for
the Liouville branes \cite{zzbl1,zzbl2}, for branes in $AdS_3$
\cite{kutads,oog2,schads} and for the 2d black hole \cite{schsyl,fnp}. For
all the other quantities such as the bulk-boundary and the three-point
boundary couplings only partial results exist \cite{schads} and their
computation proved to be an extremely difficult task. The only exception
is the Liouville model for which the complete solution is available
\cite{zzbl1,zzbl2,hoso,pons,tbl}. In this paper we will provide the
complete solution for the BCFT pertaining to the two classes of symmetric
branes of the $H_4$ model.

D-branes in pp-wave backgrounds have already been the object of several
studies and we summarize here only the main results. D-branes in R-R
supported pp-waves have been discussed in the light-cone gauge and various
aspects of their physics have been analyzed \cite{gr1}-\cite{sk2}.
Interesting world-volume theories have been argued to exist on such branes
\cite{ganor}. D-branes in NS-NS supported pp-wave have also been studied,
since they are relevant for the Penrose limits of little string theory
and, unlike the RR supported pp-waves, they are amenable to study using
boundary CFT methods \cite{dbr,dbr2},\cite{dpp}-\cite{dpp3}. Our aim in
this paper is not to describe the most general brane configuration in the
Nappi-Witten gravitational wave. We focus instead on two particular
classes of branes which preserve half of the background isometries and we
clarify the closed and open string dynamics in full detail.

The $H_4$ model has two families of symmetric D-branes \cite{dbr,dbr2},
namely $D2$ and $S 1$ branes. We solve in both cases the consistency BCFT
conditions \cite{lew,lew2} and obtain the BCFT data, that is,  the
bulk-boundary and the three-point boundary couplings. The bulk-boundary
couplings for the $D2$ branes can be found in Eq. $(\ref{d2bb1})$ and
$(\ref{d2bb2})$ while the three-point boundary couplings are in Eq.
$(\ref{sd214})$, $(\ref{b28})$, $(\ref{d2o1})$ and $(\ref{d2o2})$. The
bulk-boundary couplings for the $S 1$ branes are in Eq. $(\ref{bbc1m})$
while the boundary three-point couplings can be found in Eq.
$(\ref{b-3-+0})$ $(\ref{b-3000})$ and $(\ref{bex})$. To our knowledge,
with the notable exception of the Liouville model
\cite{zzbl1,zzbl2,hoso,pons,tbl}, this is the first {\em complete}
tree-level solution of D-brane dynamics in a curved non-compact
background.

The D2 branes are the twisted symmetric branes of the $H_4$ model. Their
world-volume covers the two light-cone directions and one direction in the
transverse plane. The induced metric is that of a pp-wave in one dimension
less and they also carry a null electromagnetic flux. As such, they provide an
interesting example of curved branes in a curved space-time. The spectrum
of open strings starting and ending on the same brane or stretched between
different branes contains all the representations of the $\hat{\cal H}_4$
algebra.

There are many similarities between the D2 branes in $H_4$ and
the $AdS_2$ branes in $AdS_3$. This is not surprising since they can be
considered as Penrose limits of specific branes in $AdS_3\times S^1$ or in
$\mathbb{R}\times S^3$. More precisely, if we start from $\mathbb{R}\times
S^3$ they arise from $S^2$ branes with Neumann boundary condition in time
while if we start from $AdS_3\times S^1$ they arise from the $AdS_2$
branes. The relation between the $H_4$ vertex operators and the orbifold
twist fields leads in this case to an analogy between the D2 branes in the
Nappi-Witten wave and configurations of intersecting branes in flat space
\cite{angles}.

The $S 1$ branes are the untwisted symmetric branes of the $H_4$
model. They have Dirichlet boundary conditions on the two
light-cone coordinates and their world volume covers the
transverse plane with an induced flat Euclidean metric. They are
also supported by a world-volume electric field whose magnitude
determines their localization in one of the light-cone
coordinates. Having a non-trivial boundary condition along the
real time direction they are examples of S-branes \cite{S-branes}.
In fact, the Penrose limit relates the $S 1$ branes to either $S^2$
branes with a Dirichlet boundary condition in time in
$\mathbb{R}\times S^3$ or to the $H_2$ branes in $AdS_3\times
S^1$.

The world-volume of the $S 1$ branes shrinks to a point
whenever their light-cone position is given by $\m u = 2 \pi n$.
For these values of the coordinate $u$, there is another class of
symmetric branes with a cylindrical world-volume. These branes
extend along the light-cone direction $v$ and have a fixed radius
in the transverse plane. We will not discuss them in detail in
this paper. We also mention that the $S 1$ branes are a special
case of a more general class of non-symmetric branes that we
discuss from the DBI point of view. Finally, the behavior of the
open strings attached to the $S 1$ branes is very similar to the
behavior of open strings ending on magnetized branes in flat
space \cite{acny}.

Our solution of the $H_4$ model with boundary should be useful not
only to improve our understanding of the closed and open string
dynamics in curved space-times but also to clarify some properties
of both compact and non-compact WZW models. Indeed, as it is
widely appreciated by now, only when studied in the presence of a
boundary a conformal field theory reveals its full richness. Among
other results we provide the first example of structure constants
for twisted symmetric branes in a WZW model (the D2 branes) and of
open four-point functions in a curved background.

While the
physical interpretation of the amplitudes in the presence of the
$D2$ branes is straightforward, the interpretation of the
amplitudes for the $S 1$ branes is less evident, in particular
when they involve open string states stretched between different
S-branes which can be put on shell. They might play the role of
boundary conditions at spatial infinity but at fixed time
(specified by the S-brane in question). In fact thinking of an
S-brane as a standard soliton supported by a scalar
\cite{S-branes}, we can imagine that they appear because of the
special initial state of the scalars. In this context, the open
string insertions can be interpreted as small variations on the
initial data that ``creates" the branes. It still remains to be
seen whether such a setup may be realized in a problem with
physical interest and whether the $S$-branes can be relevant for
cosmology.

In this paper, we also discuss the annulus amplitudes. Our
results, even though suggestive, are not conclusive and the
relation between the open and the closed string channel of the
annulus certainly deserves further study.

The structure of this paper is the following: In section \ref{a} we review
the geometry of the symmetric branes of the $H_4$ WZW model, first
considered in \cite{dbr,dbr2}. We also discuss the relationship between
branes in the Nappi-Witten gravitational wave and branes in $AdS_3 \times
S^1$ and $\mathbb{R}\times S^3$. In section \ref{semic} we evaluate the
bulk-boundary couplings using the semi-classical wave functions. In
section \ref{spectrum} we discuss the spectrum of the boundary operators.
In section \ref{sc} we solve the Cardy-Lewellen constraints
\cite{lew,lew2} and display the structure constants of the boundary
theory. In section \ref{fp} we discuss the four-point amplitudes. In
section \ref{annulus} we analyze the annulus amplitudes and discuss the
contraction of the $\mathbb{R} \times SU(2)_k$ WZW model with boundary. In
section \ref{dbi} we analyze the physics of the branes in the Nappi-Witten
gravitational wave using the Dirac-Born-Infeld action. Finally in section
\ref{f} we suggest some interesting lines of further research. Several
technical details are collected in the appendices.

\section{Branes in $H_4$ \label{a}}
\renewcommand{\theequation}{\arabic{section}.\arabic{equation}}

The Nappi-Witten gravitational wave \cite{nw} is a  curved
homogeneous lorentzian space. The metric \be ds^2 = - 2 du dv -
\frac{\m^2 r^2}{4} du^2 + dr^2 + r^2 d \f^2 \ ,  \label{a01} \ee
solves the Einstein equations with a constant null stress-energy
tensor, provided in our case by the 2-form field-strength \be H =
\m r dr \wedge d \f \wedge du \ . \label{a01a} \ee The light-cone
and the radial coordinates are related to the cartesian ones by $u
= \frac{t+x}{\sqrt{2}}$, $v = \frac{t-x}{\sqrt{2}}$ and $r e^{i
\f} = \chi + i \xi$. As given before, the metric is in the
so-called Brinkman form. The change of coordinates \be u = x^+ \ ,
\hspace{0.6cm} v = x^- + \frac{\m}{8} (y_1^2+y_2^2) \sin \m x^+ \
, \hspace{0.6cm} \chi = y_1 \sin \frac{\m x^+}{2}  \ ,
\hspace{0.6cm} \xi = y_2 \sin \frac{\m x^+}{2}  \ , \label{btr}
\ee gives the metric in Rosen form \be ds^2 = -2 dx^+dx^- +\sin^2
\frac{\m x^+}{2} (dy_1^2 + dy^2_2) \ , \hspace{1cm} H = \m \sin^2
\frac{\m x^+}{2} dy_1 \wedge dy_2 \wedge dx^+ \  . \label{a001}
\ee In the following, both Brinkman and Rosen coordinates will be
useful. The two-dimensional $\s$-model that describes the
propagation of a string in this background is a WZW model based on
the Heisenberg group $H_4$ \cite{nw}. The commutation relations
are \be [P^+,P^-] = - 2i \m K \ , \hspace{1cm} [J,P^{\pm}] = \mp i
\m P^{\pm} \ , \label{a02} \ee where the generators $J$ and $K$
are anti-hermitian and $(P^+)^{\dagger} = P^-$. Even though the
group is not semi-simple, there is a non-degenerate invariant
symmetric form given by \be 2 \langle J,K \rangle = \langle
P^+,P^- \rangle \ , \label{a03} \ee which can be used to express
the stress-energy tensor as a bilinear in the currents. For a
detailed discussion of this model we refer the reader to
\cite{dk}.

Since this gravitational wave is a WZW model, we can study in considerable
detail the symmetric branes, that is the branes that preserve a linear
combination of the left and right affine algebras. The symmetric branes
fall in families which are in one-to-one correspondence with the
automorphisms of the current algebra. Given such an automorphism
$\Lambda$, the relevant boundary CFT is defined by the following boundary
conditions on the affine currents \be \left [ J^a(z) - \Lambda\left (\bar
J^a(\bar z)\right) \right ] |_{z = \bar z} = 0 \ . \label{a04} \ee
Equivalently we can introduce for each symmetric brane a boundary state
$|B \ra \ra$ that satisfies \be \left [ J^a_{m} + \Lambda( \bar J^a_{-m} )
\right ] |B \ra \ra = 0 \ . \label{a05} \ee The geometry of the symmetric
branes in a group manifold has a simple and elegant description. Their
world-volume coincides with the (twisted) conjugacy classes \cite{as,ffs}
\be {\it C}_{g}^{\Lambda} = \{ \Lambda_G(h^{-1})gh \ , \forall h \in G \}
\ , \label{a06} \ee where $\Lambda_G$ is the group automorphism induced by
$\Lambda$. A generic automorphism can be written as the composition of the
adjoint action of a group element $g_0$ and of an outer automorphism $\W$
\be \Lambda = \W \circ Ad_{g_0} \ . \label{a07} \ee Since two families of
branes that differ only in the choice of the inner automorphism $Ad_{g_0}$
are related by the left action of the group, we can set without loss of
generality $g_0=1$ and restrict our attention to families of branes
associated with distinct outer automorphisms $\W$. The $H_4$ algebra
admits a non-trivial outer automorphism $\Omega$ which acts on the
currents as charge conjugation \be \Omega(P^\pm) = P^\mp \ , \hspace{1cm}
\Omega(J) = - J \ , \hspace{1cm} \Omega(K) = -K \ . \label{a08} \ee
As
such, we have two families of symmetric branes, wrapped on the conjugacy
classes ${\it C}_g$ and on the twisted conjugacy classes ${\it C}^\W_g$
respectively. In the following, we will briefly review their geometry which
was first discussed in \cite{dbr,dbr2}.

The $H_4$ conjugacy classes ${\it C}_g(u,\h)$ are characterized by
two parameters, the constant value of the coordinate $u$ and the
constant value of the invariant \be \h = v - \frac{\m r^2}{4}
\cot{\frac{\m u}{2}} \ . \label{radius} \ee Their geometric
description is particularly simple in Rosen coordinates where the
brane world-volume coincides with the wave-fronts since $x^- =
\h$. These branes are thus Euclidean two-planes with an
$x^+$-dependent scale factor and a two-form field-strength \be
{\cal F}_{12} \equiv B_{12} + 2 \pi \a^{'} F_{12} = - \frac{\sin
\m x^+}{2} \ . \label{sfluxr} \ee As usual, ${\cal F}$ is the
gauge invariant combination that appears in the Dirac-Born-Infeld
action. These branes have a non-trivial boundary condition on the
real time coordinate and can be called, following the modern
terminology, $S 1$-branes \cite{S-branes}. As we will show
at the end of this section, they are related by the Penrose limit
to the $H_2$ branes in $AdS_3 \times S^1$ or to the $S^2$ branes
in $\mathbb{R}\times S^3$ with a Dirichlet boundary condition
along the time direction.

The brane world-volume degenerates to a
point whenever $\m x^+ = 2 \pi n$, $n \in \mathbb{Z}$. Indeed if
we start from $\m x^+ = 2 \pi n$ and change the value of $x^+$
until we reach $\m x^+ = \pi + 2 \pi n$, we interpolate between a
point-like world-volume and a flat, infinite two-dimensional world-volume.
This is very similar to what happens in flat space when we turn on
a magnetic field on a brane and send its field-strength to
infinity. As we will show in more detail later, there are several
analogies between the untwisted symmetric branes of the $H_4$
model and branes in flat space with a magnetic field on the
world-volume. In particular, the open strings stretched between two
$S 1$ branes behave very similarly to the open strings in a
magnetic field \cite{acny}.

In Brinkman coordinates, the metric induced on the two-dimensional
world-volume is trivial and the flux is \be {\cal F}_{r\f} = B_{r
\f} + 2 \pi \a^{'} F_{r \f} = - r \cot \frac{\m u}{2} \ .
\label{sfluxb} \ee When $2 \pi n < \m u < \pi + 2 \pi n$ we can
parameterize the world-volume with $(r,\f)$ and consider the brane
as a point-like object which appears at the point $(u,v=\h,r=0)$
and then moves away along the $x$ axis at the velocity of the
light while simultaneously expanding  in a circle in the
transverse plane according to Eq. $(\ref{radius})$. When $\pi + 2
\pi n < \m u < 2 \pi n$ we have the time-reversed process, where
an infinite circle coming from spatial infinity in the transverse
plane shrinks to a point. Finally when $\m u = \pi + 2 \pi n$ the
brane is a two-plane with a fixed light-cone position also in
Brinkman coordinates.

When $\m u = 2 \pi n$, the geometry of the conjugacy classes
changes. In Rosen coordinates we  notice that the
two-dimensional planes degenerate to points. However, the
transformation $(\ref{btr})$ is singular when $\m u = 2 \pi n$.
Indeed, the analysis of the conjugacy classes shows that there are
other possibilities for the symmetric branes at $\m u = 2 \pi n$:
when $r=0$ we have points with a fixed value of $v$ and when $r
\ne 0$ we have a cylinder of radius $r$ extended along the null
direction $v$. The geometry of the conjugacy classes is summarized
in the following table

\begin{center}

\begin{tabular}{|p{2cm}|p{2cm}|p{2.4cm}|}
\hline $u \ne 2 \pi n$ & $\h$ & {\rm $S 1$ branes } \\ \hline $u =
2 \pi n$ & v, \, r = 0 & {\rm S(-1) branes} \\ \hline $u= 2 \pi n$
& r $\ne$ 0 & {\rm null branes} \\ \hline
\end{tabular}

\end{center}

In this paper we will often denote the two parameters that
identify an $S 1$ branes with a single letter $a \equiv
(u_a,\h_a)$. It would be interesting to identify all these branes
as bound states of some set of elementary branes. For instance, the
$D2$ branes in $S^3$ were shown to arise as bound states of $D0$
branes \cite{volker}. In a similar spirit and with similar
techniques we could choose as fundamental branes the point-like
branes, that is the degenerate conjugacy classes for $\m u = 2 \pi
n$ and try to identify the $S 1$ branes and the cylindrical branes
as bound states of them.

The second class of symmetric branes we are interested in, wrap the
twisted conjugacy classes ${\cal C}^\W_g(\chi)$ which are
parameterized by a single invariant \be \chi = r\cos{\f} \ . \ee In
Brinkman coordinates, they have a simple description as $D2$ branes
whose world-volume extends along the $(u,v,\xi)$ directions and is
localized in the $\chi$ direction. They have a non-trivial induced
metric, which describes a gravitational wave, and a null
world-volume flux $F_{u \xi} = \frac{\m \chi}{2}$. The $D2$ branes
of the $H_4$ model thus provide an interesting example of curved
branes in a curved space-time.

The Nappi-Witten gravitational wave is the Penrose limit of two
simple and interesting space-times, $\mathbb{R}\times S^3$ and
$AdS_3 \times S^1$. For both space-times there is an exact CFT
description in terms of WZW models based respectively on
$\mathbb{R}\times SU(2)_k$ and $SL(2,\mathbb{R})_k \times U(1)$,
where the level $k$ is related to the radius of curvature. From
the CFT point of view, the existence of the Penrose limit relating
the Nappi-Witten wave and $\mathbb{R}\times S^3$ or $AdS_3 \times
S^1$ corresponds to the fact that the $H_4$ current algebra is a
contraction of the current algebras underlying the two original
space-times \cite{ors}.

We close this section by discussing
the relation implied by the Penrose limit between the symmetric
branes of the $H_4$ WZW model and the symmetric branes in
$\mathbb{R}\times SU(2)_k$ and in $SL(2,\mathbb{R})_k \times U(1)$.
The latter branes have been studied in \cite{as, ffs, bd, bacpet}. In the
first case, we will see that the $S 1$ branes arise from the $S^2$
branes in $S^3$ with a Dirichlet boundary condition in the time
direction and that the $D2$ branes arise from rotated $S^2$ branes
in $S^3$ with a Neumann boundary condition in the time direction.
In the second case, we will identify the $D2$ branes as the limit
of the $AdS_2$ branes in $AdS_3$ with a Neumann boundary condition
in $S^1$ and the $S 1$ branes as the limit of the $H_2$ branes in
$AdS_3$ with a Dirichlet boundary condition in $S^1$. A similar
discussion of the Penrose limit applied simultaneously to the
space-time and to a brane contained in it can be found in
\cite{sz}.

We start with $\mathbb{R}\times S^3$. Using the following
standard parametrization for the $SU(2)$ group manifold  \be
g(\a,\b,\g) =
\begin{pmatrix} \cos \g e^{i \a} & i \sin \g e^{-i \b} \\
 i \sin \g e^{i \b} & \cos \g e^{-i \a}
\end{pmatrix} \ , \ee
the metric and the two-form field-strength read \be ds^2 = k \left
[ - dt^2 + \cos^2 \g d\a^2+d \g^2 +\sin^2\g d\b^2 \right ]
 \ , \hspace{1cm} H_{\a\b\g} = - k \sin 2 \g . \label{d6} \ee
Along the time direction we can impose either Neumann or Dirichlet
boundary conditions. As for the symmetric branes in $S^3$, they
wrap the $SU(2)$ conjugacy classes which are two-spheres
characterized by a constant value of ${\rm tr}(g) = 2 \cos \g \cos
\a$. Since $SU(2)$ does not have any external automorphism, the
other possible symmetric branes are two-spheres shifted by the
action of a group element $R(\tilde \a,\tilde \b,\tilde \g) \in
SU(2)$ and characterized by a constant value of ${\rm tr} (R g)$.
In order to take the Penrose limit we first perform the change of
variables \be \a = \frac{\m u}{2} - \frac{2v}{\m k} \ ,
\hspace{1cm} t = \frac{\m u}{2} \ , \hspace{1cm} \g =
\frac{r}{\sqrt{k}} \ , \hspace{1cm} \b = \f  \ , \label{d7} \ee
and then send $k \rightarrow \infty$. In the limit \be \cos{\g}
\cos{\a} \sim \cos{\frac{\m u}{2}} + \frac{2 \h}{k} \sin{\frac{\m
u}{2}}  \ . \ee
We observe that an $S^2$ brane with a Dirichlet
boundary condition along the time direction becomes an $S 1$ brane
labeled by the parameters $u$ and $\h$. Note that we have
obtained the untwisted $H_4$ branes starting from branes whose
world-volume does not contain the null geodesic used to define the
Penrose limit. It is then natural to consider the limit of branes
whose world-volume contains the null geodesic. For this purpose, we
consider a family of $S^2$ branes rotated by $R(0,0,\pi/2)$ and
with a Neumann boundary condition along the time direction. In the
limit, the parameter that characterizes the new family of branes
behaves as follows \be \sin{\g}\cos{\b} \sim \frac{r}{\sqrt{k}}
\cos \f \ , \ee and they become the $D2$ branes of the $H_4$ model. We
can proceed in a similar way for the Penrose limit of $AdS_3
\times S^1$ and of its symmetric branes. We write the background
in global coordinates \be ds^2 = k \left [ - \cosh^2 \r d\tau^2+d
\r^2 +\sinh^2\r d\f^2 \right ] + k dx^2 \ , \hspace{1cm} H_{\r \f
\tau} = k \sinh 2 \r \ , \label{d1} \ee and define the Penrose
limit introducing \be \tau = \frac{\m u}{2} + \frac{2v}{\m k} \ ,
\hspace{1cm} x = \frac{\m u}{2} \ , \hspace{1cm} \r =
\frac{r}{\sqrt{k}} \ , \label{d2} \ee and then sending $k$ to
infinity. The conjugacy classes are characterized by a constant
value of $\cosh \r \cos \tau$. In the limit \be \cosh \r \cos \t
\sim \cos \frac{\m u}{2} - \frac{2 \h}{k} \sin \frac{\m u}{2} \ .
\label{d3} \ee We observe that an untwisted symmetric brane of
$AdS_3$ with a Dirichlet boundary condition in the $S^1$ factor
gives rise to an $S 1$ brane in $H_4$. Note that for large $k$, the
classes we are considering are precisely those with $|\cosh \r
\cos \tau| \le 1$. This is precisely the $H_2$ branes and the degenerate
branes in $AdS_3$ \cite{bacpet}. The null
geodesic used in the limit is not contained in the brane
world-volume. On the other hand, the twisted conjugacy class with
a Neumann boundary condition in the $S^1$ factor, are characterized
by a constant value of $\sinh \r \cos \f$ and contains the null
geodesic. Since in the limit \be \sinh \r \cos \f \sim
\frac{r}{\sqrt{k}} \cos \f \ , \label{d4} \ee we observe that the $D2$
branes of the $H_4$ model are the Penrose limit of the $AdS_2$ branes.

A more detailed description of the Penrose limit for the different
types of branes, using coordinate systems adapted to their
world-volumes, can be found in Appendix \ref{plb}. In section
\ref{annulus} we will extend the analysis performed in \cite{dk}
and discuss the contraction of the boundary $\mathbb{R} \times
SU(2)_k$ WZW model which is the world-sheet analogue of the Penrose
limit applied simultaneously to the space-time and to the brane
contained in it.

\section{Review of the bulk spectrum and semiclassical analysis
\label{semic}}

The structure of the Hilbert space of the $H_4$ WZW model
\cite{dk} is very similar to the structure of the Hilbert space of
the $SL(2,\mathbb{R})_k$ WZW model, clarified by Maldacena and
Ooguri \cite{moog1}. Together with the standard highest-weight
representations of the affine algebra, restricted by a unitarity
constraint, there are other representations that satisfy a
modified highest-weight condition. These new representations are
related to the standard ones by the operation of spectral flow,
which is an automorphism of the current algebra. In our case there
are three classes of highest-weight affine representations,
reviewed in appendix \ref{rep}. To each affine representations we
associate a primary chiral vertex operator \ba && \F^\pm_{\pm
p,\jh}(z,x) \ , \hspace{1cm} 0 < \m p < 1 \ , \hspace{1cm} \jh \in
\mathbb{R}
\ , \nb \\
&& \F^0_{s,\jh}(z,x) \ , \hspace{1.3cm} s > 0 \ ,  \hspace{1.9cm}
\jh \in [-\m/2,\m/2) \ . \ea For the $\F_{\pm p, \jh}^\pm$ vertex
operators, $p$ is the eigenvalue of $K$ and $\jh$ the highest
(lowest) eigenvalue of $J$. For the $\F_{s,\jh}^0$ vertex
operators, $s$ is related to the Casimir of the representation and
$\jh$ is the fractional part of the eigenvalues of $J$. Here $z$
is a coordinate on the world-sheet and $x$ a charge variable we
introduced to keep track of the infinite number of components of
the $H_4$ representations. On the charge variables the $H_4$
algebra is realized in terms of the operators \be P^{\pm} =
\sqrt{2} \m p x \ , \hspace{0.6cm} P^{\mp} = \sqrt{2} \p_x \ ,
\hspace{0.6cm} J = i(\jh \pm \m x \p_x) \ , \hspace{0.6cm} K = \pm
ip \ , \label{opm} \ee when considering its action on $\F^\pm_{\pm
p,\jh}$ and by the operators \be P^+ = s x \ , \hspace{0.6cm} P^-
= \frac{s}{x} \ , \hspace{0.6cm} J = i(\jh + \m x \p_x) \ ,
\hspace{0.6cm} K = 0 \ , \label{oo} \ee when considering its
action on  $\F^0_{s,\jh}$. States with $\m p = \m \hat p + w$ with
$0 < \m \hat p <1$ and $w \in \mathbb{N}$ fall into spectral-flowed
discrete representations $\Sigma_{\pm w}(\F^\pm_{\pm \m
\hat p, \jh})$ while states with $\m p = w$ with $w \in
\mathbb{Z}$ fall into spectral-flowed continuous representations
$\Sigma_{w}(\F^0_{s,\jh})$. We also recall the relation \be
\Sigma_{-1}[\F^+_{p,\jh}] = \F^-_{1-p,\jh} \ . \label{relpm} \ee
We will denote the image under spectral flow of a representation
$\a$ and the corresponding vertex operators either by
$\Sigma_{w}[\F_\a]$, as we did before, or by the introduction of a
further index $\F_{\a;w}$. Finally the operator content of the
bulk theory is given by the charge conjugation modular invariant.
We then have the operators \ba \F^\pm_{\pm p,\jh}(z,\bar z|x, \bar
x) &=&
\F^\pm_{\pm p,\jh}(z,x) \F^\mp_{\mp p,-\jh}(\bar z, \bar x) \ , \nb \\
\F^0_{s,\jh}(z, \bar z|x, \bar x) &=& \F^0_{s,\jh}(z,x)
\F^0_{s,-\jh} ( \bar z, -\bar x^{-1} ) \ , \ea as well as their
images under an equal amount of spectral flow in the left and
right sectors. The currents that generate the affine algebra
preserved by the boundary conditions are given by the combination
$J+\Omega(\bar J)$. A more detailed description of the
representation theory of the affine $\hat{\cal H}_4$ algebra can
be found for instance in \cite{dk}.

Before performing the exact CFT analysis we will try to clarify
the physics of the model in a semiclassical approximation. In
doing so we will gain some intuition about the spectrum of the
boundary operators and on the form of the bulk-boundary and the
three-point boundary couplings. We use the following
parametrization for the group elements \be g = e^{\frac{u}{2}J}
e^{\frac{i\tilde{w}}{2}P^-+\frac{iw}{2}P^+} e^{\frac{u}{2}J+vK} \
, \ee where $w = r e^{i \f} = \chi + i \xi$. The isometry
generators are $K = - \bar K = \p_v$ and \ba \tiny J &=& \p_u
-\frac{i \m}{2} (w \p -\tilde{w} \tilde{\p}) \ , \hspace{0.4cm}
\small P^+ = -\frac{e^{-i \m \frac{u}{2}}}{2} [ 4 i \tilde{\p} +
\mu w  \p_v  ] \ , \hspace{0.4cm} \small P^- = -\frac{e^{i \m
\frac{u}{2}}}{2} \left[ 4 i  \p - \mu \tilde{w} \p_v \right ] \ ,
\nb \\
\small \bar{J} &=& -\p_u -\frac{i \m}{2} (w \p -\tilde{w}
\tilde{\p}) \ , \hspace{0.4cm} \small \bar{P}^+ = \frac{e^{i \m
\frac{u}{2}}}{2} [    4 i  \tilde{\p} - \m w \p_v ] \ ,
\hspace{0.4cm} \small \bar{P}^- =  \frac{e^{-i \m \frac{u}{2}}}{2}
\left[ 4 i  \p + \mu \tilde{w} \p_v \right ] \ . \nb \ea To each
vertex operator we can associate a semiclassical wave function.
For the discrete representations they are \ba \F^{+}_{p,\jh} &=&
e^{ipv+i \jh u - \frac{\m p}{4} w \bar w - i \frac{\m p}{\sqrt{2}}
w \bar x e^{\frac{i \m u}{2}}
+ i \frac{\m p}{\sqrt{2}} \bar w  x e^{\frac{i \m u}{2}} - \m p x \bar x e^{i \m u}} \ , \nb \\
\F^{-}_{-p,\jh} &=& e^{-ipv+i \jh u - \frac{\m p}{4} w \bar w + i
\frac{\m p}{ \sqrt{2}} w  x e^{-\frac{i \m u}{2}} - i \frac{\m
p}{\sqrt{2}} \bar w  \bar x e^{-\frac{i \m u}{2}} - \m p x \bar x
e^{-i \m u}} \ , \label{wavefd} \ea where $p > 0$, $\jh \in
\mathbb{R}$ and $x$, $\bar x$ are two independent charge
variables. The states that belong to these representations are
confined in periodic orbits in the transverse plane, familiar from
the quantum mechanical problem of a charged particle in a magnetic
field. For the continuous representations we have \be \F^0_{s,\jh}
= e^{ iju + \frac{i s}{2} \left [ \frac{\bar w}{\bar x} e^{-i
\frac{\m u}{2}} + w \bar{x} e^{i \frac{\m u}{2}} \right ]} \sum_{n
\in \mathbb{Z}}~ (x \bar{x})^n ~e^{i n \m u}   \ , \label{wavefc}
\ee where $s \ge 0$, $\jh \in [-\m/2,\m/2)$ and $x = e^{i \a}$,
$\bar{x} = e^{i \bar{\a}}$ are two independent phases. The states
that belong to the continuous representations can move freely in
the transverse plane: the parameter $s$ is the modulus of the
momentum and we can identify $\g = \frac{\a-\bar{\a}}{2}$ with its
phase. These wave functions can be expanded in modes which
represent the different components of the $H_4 \times H_4$
representations. It is also easy to compute semiclassical
expressions for the bulk two and three-point functions.

We can proceed in a similar way for the states confined on the
brane world-volume. In the case of the D2 branes, according to the
boundary conditions (\ref{a04}), the generators of the unbroken
background isometries are \be K - \bar K = 2 \p_v \ , \hspace{1cm}
J - \bar J = 2 \p_u \ , \hspace{1cm} P^{\pm} + \bar P^{\mp} =
e^{\mp i \m \frac{u}{2}}\left[ \pm 2 \p_\xi -  i \mu \xi \p_v
\right] \ . \ee They satisfy the commutation relation of the
Heisenberg algebra and the brane spectrum, exactly as the bulk
spectrum, can be organized in terms of $H_4$ representations. We
then introduce three types of vertex operators for the open
strings that live on a $D2$ brane localized in the $\chi$
direction \ba \Psi^{\chi\chi}_{p,\jh} &=& e^{i \frac{p}{2}v + i
\frac{\jh}{2} u - \frac{\m p}{8} \xi^2 + \frac{\m p \xi
x}{\sqrt{2}}e^{\frac{i \m u}{2}} -\frac{\m p}{2}x^2e^{i \m u}  } \
, \hspace{1.6cm} p > 0 \ ,
\hspace{0.4cm} \jh \in \mathbb{R}  \nb \\
\Psi^{\chi\chi}_{-p,\jh} &=& e^{-i \frac{p}{2}v + i \frac{\jh}{2}
u - \frac{\m p}{8} \xi^2 -\frac{\m p \xi x}{\sqrt{2}}e^{-\frac{i
\m u}{2}} -\frac{\m p}{2}x^2 e^{-i \m u} } \ ,  \hspace{1cm} p > 0
\ ,
\hspace{0.4cm} \jh \in \mathbb{R}  \\
\Psi^{\chi\chi}_{s,\jh}  &=& e^{ i \frac{\jh}{2}u +\frac{i}{2} \xi
s } \sum_{n \in \mathbb{Z}} x^{n} e^{i n \frac{\m u-\pi}{2}} \ ,
\hspace{3.2cm} s \in \mathbb{R} \ , \hspace{0.4cm} \jh \in \left
[-\frac{\m}{2},\frac{\m}{2} \right ) \ . \nb \label{sm31} \ea The
operators that act on the charge variable $x$ are given by
$(\ref{opm}, \ref{oo})$. The wave functions for the discrete
representations are easily recognized as the generating functions
for the Hermite polynomials. The open string states, as it was
already the case for the closed string states, are trapped in
periodic orbits in the $\xi$ direction of the brane world-volume
unless their light-cone momentum is an integer. Note that for the
boundary operators in the continuous representations $s$ can be an
arbitrary real number.

Semiclassical expressions for the couplings between the bulk and
the boundary operators can now be computed as overlap integrals of
the corresponding wave functions on the brane world-volume. As a
first example consider the bulk-boundary coupling $\la
\F^\pm_{p,\jh_1} \Psi^{\chi \chi}_{\mp q,\jh_2} \ra$. It is easily
evaluated as \be \int_\chi \F^\pm_{p,\jh_1} \Psi^{\chi \chi}_{\mp
q,\jh_2} =  e^{  - \m p x_2(x_1 + \bar x_1)} (x_1-\bar x_1)^{|\n|}
\sqrt{\frac{2\pi}{\m p}} \frac{i^{|\n|}}{2^{|\n|}|\n|!} \left [ \m
p  \ \right ]^{\frac{|\n|}{2}} e^{-\frac{\m p \chi^2}{4}}
H_{|\n|}\left ( \pm \sqrt{\frac{\m p}{2}} \chi \right ) \ ,
\label{sc28} \ee where \be \int_\chi \equiv \int du dv d \xi d
\chi^{'} \d(\chi^{'} -\chi)  \ , \ee denotes a space-time integral
restricted to the brane world-volume. The integral is non-zero
only when $q=2p$ and $|\n| \in \mathbb{N}$ where $\n =
-2\jh_1-\jh_2$. The other possible bulk-boundary coupling is given
by \ba && \int_{\chi} \F^0_{s,\jh_1} \Psi^{\chi \chi}_{t,\jh_2} =
8 \pi^2  (x_1 \bar x_1)^{\n/2} \sum_{m \in \mathbb{Z}} \left [
\frac{x_2^2}{x_1\bar x_1} \right ]^{\frac{m}{2}} \nb \\ &&\left [
\d(t+2s \sin \g) e^{i \chi s \cos \g} +(-1)^\n \d(t-2s \sin
\g)e^{-i \chi s \cos \g}\right ] \ , \ea where $x_i = e^{i \a_i}$.
Since the D2 branes are invariant under translations along the two
light-cone directions, only operators with $p=0$ and $\jh=0, 1/2$
couple to their world-volume. Their one-point functions can be
derived from the previous bulk-boundary coupling upon setting
$t=0$ and integrating over $\a_2$ \ba \la \F^0_{s,0} \ra &=& 8
\pi^2 \, \frac{\cos \chi s}{s} \,
[ \d(\g)+\d(\g-\pi) ]  \ , \nb \\
\la \F^0_{s,1/2} \ra &=& 8 \pi^2 i \, \frac{\sin \chi s}{s x} \, [
\d(\g)+\d(\g-\pi) ] \ .  \ea

The discussion for the $S 1$ branes is very similar. The relevant
isometry generators are $K + \bar K = 0$ and \be J + \bar J = - i
\m(w \p -\tilde{w} \tilde{\p}) \ , \hspace{0.4cm} P^+ + \bar P^+ =
- 4 \sin \frac{\m u}{2} \ \tilde{\p} \ , \hspace{0.4cm} P^- + \bar
P^- = 4 \sin \frac{\m u}{2} \ \p \ . \ee They realize the algebra
of the rigid motions of the plane and as a consequence the open
string states that live on an $S 1$ brane labeled by $a \equiv
(u_a,\h_a)$ can only belong to the continuous representations of
the Heisenberg algebra. The semiclassical wave functions are \be
\Psi^{aa}_{s} = e^{-\frac{s}{4 \sin \frac{\m u_a}{2}} \left ( \bar
w x - \frac{w}{x} \right )} \ . \ee The action of the zero-modes
of the currents on the charge variable $x$ is always given by
$(\ref{oo})$. As we did for the $D2$ branes we can now extract the
bulk-boundary couplings from the overlap of the wave functions \ba
&& \int_{u_a,\h_a} \F^\pm_{p,\jh}(x_1) \Psi^{aa}_{s}(x_2) =  \nb \\
&\pm& \frac{8 \pi i}{\m p} \sin \frac{\m u_a }{2} \ e^{ \pm i p
\h_a + i \left ( \jh \mp \frac{\m}{2} \right ) u_a - \m p x_1 \bar
x_1 - \frac{s}{\sqrt{2}} \left ( \frac{x_1}{x_2^{\pm 1}} + \bar
x_1 x_2^{\pm 1} \right ) - \frac{s^2}{4 \m p}\left ( 1 \pm i \cot
\frac{\m u_a}{2} \right ) } \ . \ea  We see that all the discrete
representations as well as the identity field have a non-vanishing
one-point function, as expected since the $S 1$ branes brake the
translational invariance in the light-cone directions. When $ \m u
= 2 \pi n$ the geometry of the conjugacy classes changes. The
$\F^\pm$ vertex operators have a non-vanishing one-point function
only in the presence of the $S(-1)$ branes, localized at the
origin of the transverse plane and at arbitrary positions in the
light-cone directions \be \la \F^\pm_{p,\jh} \ra_{u,v} = \frac{8
\pi}{\mu p} e^{\pm i p v + i \jh u -\m p x \bar x }  \ . \ee
Translational invariance in the transverse plane being now broken,
also all the continuous representations have a non-vanishing
coupling \be \la \F^0_{s,\jh} \ra = 2 \pi e^{i \jh u} \d(\a+\bar
\a) \ . \ee If we consider instead the cylindrical branes,
extended along the $v$ direction and with a fixed radius $r$ in
the transverse plane, only the $\F^0_{s,\jh}$ vertex operators
have a non-vanishing coupling \be \la \F^0_{s,\jh} \ra_{u,r} =
e^{i \jh u } J_0(s r )\d (\a+\bar \a) \ \ . \ee In the following
we will not discuss in detail the cylindrical branes even though
it would be interesting to extend our exact CFT analysis to them
as well.

\section{Spectrum of the boundary operators \label{spectrum}}

In the previous section we discussed the semiclassical open string
spectrum for the two families of symmetric branes we are studying in this
paper. According to the semiclassical analysis, the states of open strings
that live on a D2 brane, form all possible representations of the
$\hat{\cal H}_4$ algebra. The states of open strings that live on an $S 1$
brane, can only belong to the continuous representations. In this section,
we provide a detailed description of the spectrum of the open strings and
extend the analysis also to the open strings that end on different branes.
In close analogy with the bulk primary vertex operators, we introduce for
each representation $\a$ of the affine algebra, a boundary primary vertex
operator $\psi_\a^{ab}(t,x)$, which depends on the insertion point on the
real axis $t$ and on a charge variable $x$. The two upper indices label
the two branes on which the open string ends. This is the same as  the two
boundary conditions, the vertex operator interpolates between.

Let us start with a D2 brane localized at $\chi=0$. The space of
states ${\cal H}_{00}$ contains in this case all possible
representations of the $\hat{\cal H}_4$ algebra \ba && \Sigma_{\pm
w} \left [ \psi^{\chi\chi}_{ \pm p,\jh} \right ] \ ,
\hspace{0.8cm} 0 < \m p < 1 \ , \hspace{0.8cm} \jh \in
\mathbb{R} \ , \hspace{0.8cm} w \in \mathbb{N} \ , \nb \\
&& \Sigma_w \left [ \psi^{\chi\chi}_{s,\jh} \right ] \ ,
\hspace{0.8cm}
 s \in \mathbb{R} \ , \hspace{0.8cm} \jh \in [-\m /2, \m/2  ) \ ,
 \hspace{0.8cm} w \in \mathbb{Z} \ . \label{bs1} \ea
This is very similar to what happens for the $AdS_2$ brane localized at
$\psi=0$ in $AdS_3$, where the spectrum is also given by the holomorphic
square root of the bulk spectrum \cite{oog1}. Similarly to that case, an
observation about the spectral flow is in order \cite{oog1}. Given a
solution $g(u_0,v_0,r_0e^{i\f_0}) \in H_4$ of the classical equation of
motion, a new solution can be generated by the action of the spectral flow
\be \Sigma_w[g](u,v,r e^{i \f}) \equiv e^{\frac{w
\tau^+}{\mu}J}g(u_0,v_0,r_0e^{i \f_0}) e^{\frac{w \tau^-}{\mu}J} \ , \ee
where $\tau^{\pm} = \t \pm \s$. Therefore \be u = u_0 + \frac{2 w}{\mu} \,
\t \ , \hspace{0.8cm} v = v_0 \ , \hspace{0.8cm} re^{i \f} = r_0e^{i
(\f_0+w \s)} \ . \label{sfd2} \ee As we can see from $(\ref{sfd2})$, when
$\chi \ne 0$ only the spectral flow by an even integer is a symmetry of
the D2 brane spectrum. Spectral flow by an odd integer maps a string
living on a brane sitting at $\chi$ to a string stretched between a brane
at $\chi$ and a brane at $-\chi$. As a consequence, we cannot use in the
general case the spectral flow by an odd integer, to generate the complete
brane spectrum, as we did in (\ref{bs1}) for a brane sitting at the
origin. However, using the relation $(\ref{relpm})$, it is easy to verify
that for the discrete representations, it is enough to consider the
spectral flow by an even integer in order to obtain all possible values of
the light-cone momentum. For instance, a state carrying a light-cone
momentum $\m p +2w-1$, belongs to the representation
$\Sigma_{2w}\left[\psi^{\chi\chi}_{-(1-p),\jh}\right]$.

For the continuous representations, the spectral flow by an even integer
is not enough. We have to proceed in a different way \cite{oog1}. We start
with the vertex operator $\psi^{-\chi \chi}_{s,\jh}$ and take its image
under the spectral flow by an odd integer $2w+1$. In this way, as
explained before, we obtain the vertex operator pertaining to a string
ending on the brane at $\chi$ and carrying an odd light-cone momentum.
This asymmetry between the even and the odd spectral-flowed continuous
representations will be also manifest in the annulus amplitude, as we will
see in section \ref{annulus}. Summarizing, the spectrum of a brane
localized at $\chi \ne 0$, is given by \ba && \Sigma_{2w} \left [
\psi^{\chi\chi}_{ \pm p,\jh} \right ] \ , \hspace{0.8cm} 0 < \m p < 1 \ ,
\hspace{0.8cm} \jh \in
\mathbb{R} \ , \hspace{0.8cm} w \in \mathbb{Z} \ , \nb \\
&& \Sigma_{2w} \left [ \psi^{\chi\chi}_{s,\jh} \right ] \ ,
\hspace{0.8cm}
 s \in \mathbb{R} \ , \hspace{0.8cm} \jh \in [-\m /2, \m/2  ) \ ,
 \hspace{0.8cm} w \in \mathbb{Z} \ , \nb \\
&& \Sigma_{2w+1} \left [ \psi^{-\chi\chi}_{s,\jh} \right ] \ ,
\hspace{0.8cm}
 s \in \mathbb{R} \ , \hspace{0.8cm} \jh \in [-\m /2, \m/2  ) \ ,
 \hspace{0.8cm} w \in \mathbb{Z} \ . \ea
The same structure of the spectrum holds for the strings ending on two
different branes localized at $\chi_1$ and $\chi_2$ respectively. The only
difference is that the possible values of the parameter $s$ that label the
continuous representations are now constrained by $2 \pi^2 s^2 \ge
[\chi_1-e^{i\pi w}\chi_2]^2$. The minimal value of $s$ simply reflects the
tension of the string stretched between the two branes. The fact that the
bound depends on whether the amount of spectral flow is even or odd nicely
reflects the different behavior of the continuous representations under
the action of the spectral flow.

We now turn to the $S 1$ branes. As in the previous section, we use the
short-hand notation $a \equiv (u_a,\h_a)$ for the boundary labels. The $S
1$ branes are Cardy branes. Therefore, a relation can be established
between the parameters labeling the conjugacy classes and the quantum
numbers of the $\hat {\cal H}_4$ representations \be \m u = \pm 2 \pi (\m
p + w) \ , \hspace{1cm} 2\h = \pi (2 \jh \pm 2 p \mp 1) \ . \label{rel}
\ee As usual, $0 < \m p <1 $ and $w \in \mathbb{N}$. We will derive this
relation in section \ref{annulus}, both by studying the annulus amplitudes
of the $H_4$ model and by taking the Penrose limit of the $\mathbb{R}
\times SU(2)_k$ WZW model. In close analogy with the D-branes in flat
space, the quantum numbers $p$ and $\jh$ together with the spectral flow
parameter $w$, are related to the distance along the $u$ and the $v$
direction respectively. Using the relation (\ref{rel}), we can associate
to a brane with labels $(u_a,\h_a)$ the parameters $p_a$, $\jh_a$ and
$w_a$. This is useful because as it is the case for the Cardy branes in a
RCFT,  the spectrum of open strings $\psi^{ab}$ stretched from the brane
$b$ to the brane $a$ is encoded in the fusion product $\Sigma_{\pm
w_b}[\F_{\pm p_b,\jh_b}] \otimes \Sigma_{\mp w_a}[\F_{\mp p_a,-\jh_a}]$ of
the two corresponding chiral vertex operators.

As mentioned earlier, the open strings ending on the same brane belong to
the continuous representations of the $\hat{\cal H}_4$ algebra. The
Hilbert space decomposes as \be {\cal H}_{aa} = \int_0^\infty ds s \ \hat
V_{s,0;0} \ , \ee and the corresponding vertex operators are
$\psi^{aa}_{s,0;0}$. The open strings that end on two different $S 1$
branes with labels $a$ and $b$ can belong to any of the highest-weight
representations of the $\hat {\cal H}_4$ algebra as well as to their
images under spectral flow. The precise representation  depends on the
distance between the two branes along the $u$ direction. We introduce the
following notation: when $p_b-p_a>0$ we set $p_b-p_a = p^{ab} + w^{ab}$,
with $0 < p^{ab} <1$ and $w^{ab} \in \mathbb{N}$. We also define $\jh^{ab}
= \jh_b-\jh_a$. The brane spectrum is then given by \be {\cal H}_{ab} =
\sum_{n=0}^\infty \hat V_{p^{ab},\jh^{ab}-n;w^{ab}} \ , \ee and the vertex
operators are $\psi^{ab}_{p^{ab},\jh^{ab}-n;w^{ab}}$. Similarly when $p_b
- p_a < 0$ we set $p_b - p_a = p^{ab} + w^{ab}$ with $-1 < p^{ab} < 0$,
$-w_{ab} \in \mathbb{N}$. The brane spectrum is then given by \be {\cal
H}_{ab} = \sum_{n=0}^\infty \hat V_{p^{ab},\jh^{ab}+n;w^{ab}} \ , \ee and
the vertex operators are $\psi^{ab}_{p^{ab},\jh^{ab}+n;w^{ab}}$, $n \in
\mathbb{N}$. Finally when $p_b-p_a$ is an integer  we set $p_b-p_a =
w^{ab}$ and we have \be {\cal H}_{ab} = \int_{0}^\infty ds \hat
V_{s,\jh^{ab};w^{ab}} \ . \ee The vertex operators read
$\psi^{ab}_{s,\jh^{ab};w^{ab}}$. We assume that for our non-compact WZW
model, the space ${\cal H}_{ab}$ decomposes as expected from the fusion
rules. This assumption will be confirmed by the complete solution of the
model that is presented in the next section.

For the $S 1$ branes, the spectral-flowed representations appear whenever
the distance along the $u$ direction between two branes exceeds
$\frac{2\pi}{\m}$. In fact, the action of the spectral flow amounts to \be
\Sigma_w[g](u,v,r e^{i \f}) \equiv e^{\frac{w
\tau^+}{\mu}J}g(u_0,v_0,r_0e^{i \f_0}) e^{-\frac{w \tau^-}{\mu}J} \ , \ee
that is \be u = u_0 + \frac{2 w}{\mu} \, \s \ , \hspace{0.8cm} v = v_0 \ ,
\hspace{0.8cm} re^{i \f} = r_0e^{i (\f_0 - w \t)} \ . \label{sfs1} \ee
Note that in this case, the spectral flow is a symmetry for every integer
$w$ and maps a string stretched between two branes localized at $u_a$ and
$u_b$ to a string stretched from $u_a$ to $u_b+2 \pi w$. In the following,
we will derive most of our results assuming that all the representations
are highest-weight representations of the current algebra. It is however
not difficult to extend our results to amplitudes involving spectral-flowed
states, using the free-field realization \cite{kk,kk2}, as we did
for the closed string amplitudes in \cite{dk}.

\section{Structure constants \label{sc}}

A boundary conformal field theory is completely specified by three sets of
structure constants: the couplings between three bulk or three boundary
fields and the couplings between one bulk and one boundary field. These
structure constants satisfy a set of factorization constraints first
derived by Cardy and Lewellen \cite{lew,lew2} (see also
\cite{pss,ingo,zuber}). For the sake of clarity, we will briefly review
the sewing constraints for a generic CFT and in the next section we will
write them explicitly for the Nappi-Witten model.

 For a CFT defined on the
upper-half plane, there are two sets of fields. The first set contains the
bulk fields $\f_{i,\bar \i}(z,\bar{z})$, inserted in the interior of the
upper-half plane and characterized by the quantum numbers $(i,\bar \i)$.
These quantum numbers
 specify the representations of the left and right chiral algebras. The
 second set contains the
 boundary  fields $\psi^{ab}_i(t)$, inserted on the boundary, (here the real
axis). They are characterized by two boundary conditions $a$ and $b$. They
are also characterized by the quantum number $i$,  which labels the
representations of the linear combination of the left and right affine
algebras left unbroken by the boundary conditions. There are three sets of
OPEs (we adopt with minor modifications the notation used in \cite{ingo})
\be \f_{i,\bar \i}(z_1,\bar z_1)\f_{j,\bar \j}(z_2,\bar z_2) \sim \sum_{k}
(z_1-z_2)^{h_k-h_i-h_j} (\bar z_1-\bar z_2)^{h_{\bar k}-h_{\bar
\i}-h_{\bar \j}} C_{(i,\bar \i),(j,\bar \j)}{}{}{}^{(k,\bar k)} \f_{k,
\bar k}(z_2,\bar z_2) \ , \label{Obb} \ee \be \f_{i,\bar \i}(t+iy) \sim
\sum_{j} (2y)^{h_j-h_i-h_{\bar \i}}  \ {}^a B_{(i,\bar \i)}^j
\psi^{aa}_j(t) \ , \label{Occ} \ee \be \psi_i^{ab}(t_1) \psi_j^{bc}(t_2)
\sim \sum_{k} (t_1-t_2)^{h_k-h_i-h_j} C^{abc,k}_{ij} \psi^{ac}_k(t_2) \ ,
\label{Obc} \ee where $t_1 < t_2$. Here and in the following, the symbol
$1$ will always denote the identity field and $\ib_a$ the one-point
function of the identity with boundary condition $a$ on the real axis.
Indexes are raised and lowered using
\be
d^{ab}_{ij} = C^{aba,1}_{ij}\ib_a
= C^{bab,1}_{ji} \ib_b
\ee
 and for consistency we require that
$C_{11}^{aaa,1}=1$. Any correlation function on a Riemann surface with
boundaries and with arbitrary insertions of bulk and boundary operators
can be constructed by sewing together the basic amplitudes that correspond
to the structure constants displayed in $(\ref{Obb})-(\ref{Obc})$. The
sewing constraints follow from the requirement that all possible ways of
decomposing a given amplitude into the basic amplitudes lead to the same
answer. The resulting constraints involve, besides the structure
constants, the fusing matrices ${\bf F}_{p q}  \tiny
\begin{bmatrix} j & k
\\ i & l \end{bmatrix}$ and the
modular $S$ matrix.
The fusing matrices ${\bf F}_{p q}  \tiny
\begin{bmatrix} j & k
\\ i & l \end{bmatrix}$, by definition, implement the duality transformations of
the conformal blocks pertaining to the four-point amplitudes.
Our conventions for the fusing matrices can be found
in appendix \ref{ff}.

Sonoda \cite{sonoda} has analyzed the sewing
constraints for  Riemann surfaces without boundaries, and proved that the
CFT correlation functions are unambiguously defined provided that the
four-point functions on the sphere are crossing symmetric and that the
one-point functions on the torus are modular covariant. Cardy and Lewellen
extended these results to Riemann surfaces with boundaries
\cite{lew,lew2}.
They proved that all the amplitudes are unambiguously defined provided that
the structure constants satisfy  four additional constraints. The first
constraint is a quadratic constraint that follows from two different
factorization limits of the bulk two-point function $\la
\f_i(z_1)\f_j(z_2) \ra_a$ and reads \be {}^aB^l_i \, {}^aB^{\check{l}}_j
\, C_{l \check{l}}^{aaa,1} = \sum_k C_{ij}{}{}^k \, {}^aB^1_k \,
 {\bf F}_{k l}  \small  \begin{bmatrix} \w(\bar \i)  & i \\ \w (\bar \j) &
 j
\end{bmatrix} \ .
\label{2closed} \ee Here$\check{l}$ is the representation
conjugate to $l$, and $\w$ represents the action of an external
automorphism $\W$ on the representations of the chiral algebra. A
non-trivial $\W$ has to be taken into account when considering for
instance the symmetric branes of a WZW model. This constraint is a
particular case of a more general one that follows from the
factorization of a three-point function with two bulk and one
boundary field.

The second constraint follows from two distinct
factorization limits ($t_2 \sim t_3$ and $t_1 \sim t_2$
respectively) of the four-point boundary correlator \be \la
\psi^{ab}_i(t_1) \psi^{bc}_j(t_2) \psi^{cd}_k(t_3)
\psi^{da}_l(t_4) \ra \ , \ee and it reads \be C^{bcd,n}_{jk}
C^{abd,\check{l}}_{in} C^{ada,1}_{\check{l}l} = \sum_r
C^{abc,\check r}_{ij}C^{cda,r}_{kl}C^{aca,1}_{\check{r}r}
 {\bf F}_{r n}  \small  \begin{bmatrix} j & k \\ i & l
\end{bmatrix} \ .
\label{4open} \ee

The third constraint, relates the
bulk-boundary couplings ${}^a B_i^j$ and  the boundary three-point
couplings $C_{ij}^{abc,k}$.
It follows from the requirement of locality
for a bulk field in a three-point function with two boundary
fields \be \la \f_i(z) \psi^{ab}_j(t_1) \psi^{ba}_k(t_2) \ra \ .
\ee It can be written as follows \be {}^b B_i^l
C^{abb,\check{k}}_{jl}C^{aba,1}_{\check{k}k} = \sum_{r,n} {}^a
B_i^r C^{aba,\check{r}}_{jk} C^{aaa,1}_{r\check{r}} e^{i \pi \left
(2 h_n-h_j-h_k-2h_i+\frac{h_r+h_l}{2} \right )} {\bf F}_{r n}
\small  \begin{bmatrix} k & i \\ j & \w(\bar \i)
\end{bmatrix} {\bf F}_{n l}  \small  \begin{bmatrix} \w(\bar \i) & i \\ j & k
\end{bmatrix}\ .
\label{1cl-2op} \ee

The fourth and last constraint involves the
boundary one-point functions on the cylinder. When the boundary
field is the identity, this constraint reduces to the Cardy
constraint which relates the open and the closed string channel of
the vacuum annulus amplitude. We postpone the analysis of this
constraint to section \ref{annulus}.

In the following,  we will determine the bulk-boundary couplings
${}^a B^j_i$ and the boundary three-point couplings
$C^{abc}_{ijk}$ for the two classes of symmetric branes of the
$H_4$ WZW model. We will use as an input the structure constants
and the fusing matrices of the model,  as computed in
\cite{dk}. Here, we only recall the two and the three-point
couplings, while the fusing matrices are collected for convenience
of the reader in appendix $\ref{ff}$. In order to write our
formulae in a simple way, we shall leave in the following the
dependence on the world-sheet variables $z_i$ and $t_i$
understood. We will write each bulk amplitude as the product of
three terms, two kinematical parts ${\cal K}$ and $\bar {\cal K}$,
which contain the dependence on the charge variables of the left
and right chiral algebras, and a dynamical part ${\cal C}$. The
form of ${\cal K}$ and $\bar {\cal K}$ for the two and the
three-point functions is completely fixed by the current algebra
Ward identities\footnote{The factor ${\cal K}$ for the
three-point function,
is the group-theoretic Clebsch-Gordan coefficient of the classical algebra
 $H_4$.}. Another convention we will commonly use is  \be
\n = - \sum_{i=1}^n \jh_i \ , \ee for an $n$-point amplitude with
primary vertex operators carrying the labels $\jh_i$. The
non-trivial two-point functions are \ba \la \F^+_{p_1,\jh_1}
\F^-_{p_2,\jh_2} \ra &=&
\d(p_1-p_2)\d(\jh_1+\jh_2)e^{-p(x_1x_2+\bar x_1+\bar x_2)} \  ,
 \\
\la \F^0_{s_1,\jh_1}\F^0_{s_2,\jh_2} \ra &=&
\frac{\d(s_1-s_2)}{s_1}(2\pi)^2\d(\a_1-\a_2) \d(\bar \a_1-\bar
\a_2) \ , \hspace{0.68cm} \n = 0, 1 \ . \nb \ea The bulk
three-point couplings between three discrete representations are
\ba {\cal
C}_{(p_1,\jh_1),(p_2,\jh_2)}{}{}^{(p_1+p_2,\jh_1+\jh_2+n)} &=&
\frac{1}{n!} \left [ \frac{\g (p_1+p_2)}{\g(p_1)\g(p_2)} \right
]^{\frac{1}{2}+n} \ , \label{qpppm}  \\ {\cal
C}_{(p_1,\jh_1),(-p_2,\jh_2)}{}{}^{(p_1-p_2,\jh_1+\jh_2-n)}  &=&
\frac{1}{n!} \left [ \frac{\g(p_1)}{\g(p_2)\g(p_1-p_2)} \right
]^{\frac{1}{2}+n} \ , \hspace{0.6cm}    p_1 > p_2 \ ,
\label{qppmp} \ \ea where $\gamma(x) = \Gamma(x)/ \Gamma(1-x)$.
The corresponding kinematical factors are \ba {\cal
K}_{(p_1,\jh_1),(p_2,\jh_2),(-p_3,\jh_3)} &=&
e^{-x_3(p_1x_1+p_2x_2)}(x_1-x_2)^{\n}
\ , \nb \\
{\cal K}_{(p_1,\jh_1),(-p_2,\jh_2),(p_3,\jh_3)} &=&
e^{-x_1(p_2x_2+p_3x_3)}(x_2-x_3)^{-\n} \ , \label{kin1}\ea with
similar expressions for $\bar {\cal K}$. The coupling between one
continuous and two discrete representations is \be \ {\cal
C}_{(p,\jh_1),(-p,\jh_2)}{}{}^{(s,\{\jh_1+\jh_2\})} =
e^{\frac{s^2}{2} \s(p)} \ , \label{cpmo2} \ee  where $\psi(x) =
\frac{d \ln{\G(x)}}{dx}$ and we defined \be \s(p) \equiv
\psi(p)+\psi(1-p)-2\psi(1) \ . \ee We also have \be {\cal
K}_{(p,\jh_1),(-p,\jh_2),(s,\jh_3)} =
e^{-px_1x_2-\frac{s}{\sqrt{2}}\left(x_2x_3+\frac{x_1}{x_3}\right)}x_3^{\n}
\ . \label{kin2}\ee Finally, the coupling between three $\F^0$
vertex operators simply reflects the conservation of the momentum
in the transverse plane. Therefore it is non-zero only when \be
s_{3}^2 = s_{1}^2+s_{2}^2-2s_{1}s_{2} \cos{\th} \ , \hspace{1cm}
s_{3}e^{i\f} = s_{1}-s_{2}e^{i \th} \ . \hspace{0.6cm}
\label{conserv} \ee  It can be written as \be  {\cal K} \bar {\cal
K} {\cal C}_{(s_1,\jh_1),(s_2,\jh_2),(s_3,\jh_3)} = (2\pi)^4 (x_1
\bar x_1)^{\n}\frac{ \d(\a_{21}+\bar{\a}_{21})\d(\a_{31} +
\bar{\a}_{31})\d(\a_{21}-\th)\d(\a_{31}-\f)}{\pi
\sqrt{4s_{1}^2s_{2}^2-(s_{3}^2-s_{1}^2-s_{2}^2)^2}}
 \ , \label{cooo} \ee where
$\a_{ij}=\a_i-\a_j$  and $\n \in \mathbb{Z}$.

We have now at our disposal all the information required to solve
the $H_4$ WZW model with boundary. We will consider first the $D2$
branes and use the sewing constraints to fix their structure
constants. We will then study the $S 1$ branes that, as we already
explained, are the Cardy branes of the model. In this case we
compute the bulk-boundary couplings and verify that, for our
non-compact WZW model, the boundary three-point couplings are
also proportional to the fusing matrices. More details concerning
the steps leading to the solution, can be found in appendix
\ref{dsc}.

\subsection{The D2 branes}
\setcounter{equation}{0}
\renewcommand{\theequation}{\arabic{section}.\arabic{subsection}.\arabic{equation}}

In this section we compute the structure constants for the
maximally symmetric D2 branes of the Nappi-Witten model. From the
physical point of view, the couplings we derive are important since
they provide examples of open and closed string interactions in a
non-compact, curved space-time. They are also interesting from a more formal
point of view, since they represent the first example of the
structure constants for the twisted symmetric branes of a WZW
model. They could be a useful guide, leading  to  a general answer,
extending the one that already exists for the Cardy branes. We
will first derive the bulk-boundary couplings and then the
boundary three-point couplings for strings ending on the same D2
brane at $\chi=0$. Once these couplings are obtained, it is easy to
generalize the solution to strings ending on different branes at
arbitrary position in the $\chi$ direction.

The boundary two-point functions are \ba \la
\psi^{\chi_1\chi_2}_{p_1,\jh_1}(x_1)\psi_{p_2,\jh_2}^{\chi_2\chi_1}(x_2)\ra
&=& C^{\chi_1\chi_2\chi_1,1}_{(p_1,\jh_1),(p_2,\jh_2)}\ib_{\chi_1}
e^{-px_1x_2} \ , \nb \\
\la
\psi^{\chi_1\chi_2}_{s_1,\jh_1}(x_1)\psi_{s_2,\jh_2}^{\chi_2\chi_1}(x_2)\ra
&=& C^{\chi_1\chi_2\chi_1,1}_{(s_1,\jh_1),(s_2,\jh_2)}\ib_{\chi_1}
2 \pi \d(s_1+s_2) \d(\a_2-\a_1) x_1^{-\jh_1-\jh_2} \ .  \ea The
second two-point function is non-zero only when $\jh_1+\jh_2=0,
-1$. The bulk-boundary couplings have the following form \be \la
\F^\pm_{p,\jh_1}(x_1)\psi^{\chi \chi}_{\mp 2p,\jh_2}(x_2) \ra =
e^{-px_2(x_1+\bar x_1)}(x_1 - \bar x_1)^{\pm \n} \, {}^\chi
B_{(\pm p, \jh)}^{(\pm 2p,-\jh_2)} \ib_\chi \ . \ee Here $\n =
-2\jh_1-\jh_2$ and the coupling is non-zero only when $\jh_2 = -2
\jh_1 \mp n$, $n \in \mathbb{N}$. In order to write the couplings
for the $\F^0_{s,\jh}$ vertex operators, it is convenient to
introduce two new angles defined by \be \a = \b+\g \ ,
\hspace{1cm} \bar \a = \b - \g \ , \hspace{1cm} 0 \le \b \le 2 \pi
\ , \hspace{1cm} -\pi \le \g \le \pi \ . \ee We obtain  \be \la
\F^0_{s,\jh_1}(x_1)\psi^{\chi \chi}_{t,\jh_2}(x_2) \ra = \pi
(x_1\bar x_1)^{\frac{\n}{2}}\sum_{m \in \mathbb{Z}} \left [ -
\frac{x_2^2}{x_1 \bar x_1} \right ]^{m} \left [ \d(\g-\th) +
\d(\g-\th-\pi) \right ] \, {}^\chi B^t_s  \, \ib_\chi \ , \ee
where $t = 2s\sin \th $ and $\n = -2\jh_1-\jh_2 \in \mathbb{Z}$.
The coupling with the identity can be obtained by setting $t=0$
and integrating over $\a_2$ \be \la \F^0_{s,\jh_1}(x) \ra = \pi
(x\bar x)^{\frac{\n}{2}} [\d( \g) + \d(\g-\pi)] {}^\chi B^1_s \,
\ib_\chi \ . \ee There are two non-trivial bulk two-point
functions that may be used to derive the constraint for the
bulk-boundary couplings, namely $\la \F^0_{s,\jh_1} \F^0_{t,\jh_2}
\ra$ and $\la \F^+_{p,\jh_1} \F^-_{-p,\jh_2} \ra$. The first one
is very similar to the corresponding amplitude in flat space.
Indeed, the form of ${}^\chi B_{s,\jh_1}^{t,\jh_2}$ turns out to
be very simple \ba {}^\chi B_{s,\jh_1}^{t,\jh_2} &=&
\frac{1}{s}\cos\left [\sqrt{2} \chi s \cos
\th \right] \ , \hspace{1cm} \n \in 2 \mathbb{Z} \ , \nb \\
{}^\chi B_{s,\jh_1}^{t,\jh_2} &=&  \frac{i}{s}  \sin\left
[\sqrt{2} \chi s \cos \th \right] \ , \hspace{1cm} \n \in 2
\mathbb{Z}+1 \ , \label{d2bb1} \ea with  $t^2 = 4 s^2 \sin^2 \th$
and $\th \in [0,\pi)$. The second correlator leads to the
following constraint \ba && {}^\chi B_{(p,\jh_1)}^{(2p,2\jh_1 +
n)} {}^\chi B_{(-p,\jh_2)}^{(-2p,2\jh_2 - n + \n)}
C^{\chi \chi \chi,1}_{(2p,2\jh_1+n),(-2p,2\jh_2-n+\n)} \nb \\
&=& \int_0^\infty ds s \, {\cal
C}_{(p,\jh_1),(-p,\jh_2)}{}^{(s,\jh)} \ {}^\chi B_s^1 \ {\bf
F}_{(s,\jh),(2p,2\jh_1+n)} \small  \begin{bmatrix} (p,\jh_1) &
(p,\jh_1)
\\ (-p,\jh_2) & (-p,\jh_2) \end{bmatrix} \ ,
\ea which  is solved by \be {}^\chi B_{(\pm p,\jh)}^{(\pm
2p,2\jh_1 \pm n)} =  \frac{1}{\sqrt{2}} \frac{{\rm
i}^n}{2^{\frac{n}{2}} n!} \left ( \pi^2 \cot \pi \m p \right
)^{\frac{1}{4}} \left [ \frac{\g(2 \m p)}{\g^2(\m p)} \right
]^{\frac{n}{2}+\frac{1}{4}} e^{-\frac{\chi^2}{2 \pi \cot \pi \m
p}} H_n\left ( \pm \frac{\chi}{\sqrt{\pi \cot \pi \m p}} \right )
\ . \label{d2bb2} \ee This exact result and the semiclassical
computation (\ref{sc28}) agree in the limit  $\m p << 1$. Actually
the two results differ only in that $\m p $ has to be replaced by
$\tan (\pi \m p)$ in the argument of both the exponential and the
Hermite polynomials and that the overall powers of $\m p$ have to
become powers of $\g(\m p)$. Note also the similarity of this
coupling with the square-root of a bulk coupling of the form
${\cal C}_{++-}$ $(\ref{qpppm})$, as expected on general grounds.
Finally, it is interesting to remark that the coupling vanishes
when $\m p = 1/2$ and it has to be replaced by the coupling with a
spectral-flowed boundary operator. These couplings can be computed
either using the free-field realization of \cite{kk,kk2,dk} or by
studying the factorization of a four-point amplitude with suitable
external momenta.

We now proceed to the computation of the boundary three-point
couplings. The D2 branes of the $H_4$ WZW model are not Cardy
branes and in the absence of an one-to-one correspondence between
the brane parameters and the representations of the chiral algebra,
there is no natural relation between the boundary three-point
couplings and the fusing matrices. In the absence of any ansatz, we
have to solve directly the constraints. Fortunately,
at least the couplings between open strings
living on the brane at $\chi=0$ are  simple. They are
given by \ba C^{000,r}_{s,t} &=& \d(s+t-r) \ , \nb \\
C^{000,s}_{(p.\jh_1),(-p,\jh_2)} &=& e^{\frac{s^2}{4}\s(p)} \ , \nb \\
C^{000,(p_1-p_2,\jh_1+\jh_2-n)}_{(p_1,\jh_1),(-p_2,\jh_2)} &=&
\frac{\sqrt{2}\pi^{1/4}}{n!!} \left [
\frac{\g(p_1)}{\g(p_2)\g(p_1-p_2)} \right
]^{\frac{n}{2}+\frac{1}{4}} \ ,
\hspace{0.6cm} n \in 2 \mathbb{N} \ , \nb \\
&=& 0  \ , \hspace{5.4cm} n \in 2 \mathbb{N}+1 \ , \ea and we see
that they are very similar to the square root of the bulk
couplings. The kinematical parts can be easily obtained from
$(\ref{kin1})$ and $(\ref{kin2})$. The fact that the last coupling
vanishes whenever $n \in 2\mathbb{N}+1$ is easy to understand if
we perform a semiclassical computation using the wave functions
(\ref{sm31}).

In order to find the solution for branes sitting at arbitrary
positions, we rely once more on the fact that the boundary vertex
operators in the continuous representations are very similar to
the standard tachyonic vertex operators in flat space. If we also
take into account the finite length of the open string stretched
between two branes at $\chi_i$ and $\chi_j$, we conclude that the
quantity that is conserved in the interactions of the
$\psi_{s,\jh}^{\chi_i \chi_j}$ vertex operators is not $s$ but \be
\tilde s \equiv {\rm sign}(s) \sqrt{s^2 - b^2\chi_{ij}^2} \ ,
\hspace{0.6cm} |s| \ge b|\chi_{ij}| \ , \hspace{0.6cm} \chi_{ij} =
\chi_i - \chi_j \ , \hspace{0.6cm}  b^2 = 1/2 \pi^2 \ . \ee As a
consequence, we expect that the general form of the coupling
between three vertex operators in the continuous representations
is \be C^{\chi_1 \chi_2 \chi_3}_{str} = \d(\tilde s + \tilde t +
\tilde r) \ . \label{sd214} \ee Consider now the correlator \be
\la \psi^{\chi_1\chi_2}_{(p,\jh_1)}(1)
\psi^{\chi_2\chi_3}_{(-p,\jh_2)}(2)
 \psi^{\chi_3\chi_4}_{(s,\jh_3)}(3) \psi^{\chi_4\chi_1}_{(t,\jh_4)}(4) \ra \ ,
\ee where for clarity the numbers in the round brackets stand for
both the charge variables and the worldsheet coordinates of the
corresponding vertex operators. This correlator factorizes on a
single block in the $s$-channel and leads to  the sewing
constraint \ba &&
C^{\chi_4\chi_2\chi_3}_{(p,-\jh_2-\jh_3+n),(-p,\jh_2),(s,\jh_3)}
C^{\chi_1\chi_2\chi_4}_{(p,\jh_1),(-p,\jh_2+\jh_3-n),(t,\jh_4)} =
\nb \\ && C^{\chi_1\chi_2\chi_3}_{(p,\jh_1),(-p,\jh_2),(-r,-\jh)}
e^{-\frac{st}{2}(\s(p) \cos \f - i \pi \cot \pi p \sin
\f)-i(n+\n)\f+i\z \n}  \ , \label{b27} \ea where \be s t e^{i \f}
= (\tilde s + i b \chi_{34})(\tilde t - i b \chi_{41}) \ ,
\hspace{1cm} r e^{i \z} = s + te^{i \f} \ . \ee Note that
according to (\ref{sd214}), when raising a continuous index $s$ we
must  multiply the coupling by $s/\tilde s$. From the previous
constraint, we may read the coupling between one continuous and
two discrete representations \ba C^{\chi_1 \chi_2
\chi_3,s}_{(p,\jh_1),(-p,\jh_2)} &=& \frac{s}{\tilde s} \left [
\frac{s}{\tilde s + i b \chi_{13}} \right ]^{\jh_1+\jh_2-\jh_3}
e^{- \frac{i \pi b \tilde s}{\sin \pi p}\chi_2 + i\frac{\pi}{2}
\cot \pi p  \, \tilde s b (\chi_1+\chi_3)
+\frac{s^2}{4}\s(p)} \ , \nb \\
C^{\chi_1 \chi_2 \chi_3,s}_{(-p,\jh_1),(p,\jh_2)} &=&
\frac{s}{\tilde s}\left [ \frac{s}{\tilde s - i b \chi_{13}}
\right ]^{\jh_1+\jh_2-\jh_3} e^{\frac{i \pi b \tilde s}{\sin \pi
p}\chi_2 - i\frac{\pi}{2} \cot \pi p  \,  \tilde s b
(\chi_1+\chi_3) +\frac{s^2}{4}\s(p)} \ , \label{b28} \ea where
$|s| \ge b|\chi_{13}|$. Actually, the phase proportional to
$\chi_2$ is not fixed by equation $(\ref{b27})$ but by the
constraint associated with the correlator \be \la
\psi^{\chi_1\chi_2}_{(p,\jh_1)}(1)
\psi^{\chi_2\chi_3}_{(-p,\jh_2)}(2)
 \psi^{\chi_3\chi_4}_{(p,\jh_3)}(3) \psi^{\chi_4\chi_1}_{(-p,\jh_4)}(4) \ra \ ,
\ee which reads \ba &&
C^{\chi_2\chi_3\chi_4,(s,\jh_s)}_{(-p,\jh_2),(p,\jh_3)}
C^{\chi_4\chi_1\chi_2}_{(-p,\jh_4),(p,\jh_1),(s,\jh_s)} + (-1)^\n
(s \rightarrow -s) = \int^\infty_{|\chi_{13}|} dt
\frac{\pi s}{\sin \pi p} e^{\frac{s^2-t^2}{2}\s(p)} \nb \\
&& J_{\n} \left ( \frac{\pi s t}{\sin \pi p} \right ) \left [
C^{\chi_1\chi_2\chi_3,(t,\jh_t)}_{(p,\jh_1),(-p,\jh_2)}
C^{\chi_3\chi_4\chi_1}_{(p,\jh_3),(-p,\jh_4),(t,\jh_t)} + (-1)^\n
(t \rightarrow -t) \right ]   \ . \ea In order to verify that the
couplings in $(\ref{b28})$ solve the previous constraint, one needs
the integral $(\ref{ui11})$. At this point the correlator \be \la
\psi^{\chi_1\chi_2}_{(p,\jh_1)}(1)
\psi^{\chi_2\chi_3}_{(-p,\jh_2)}(2)
 \psi^{\chi_3\chi_4}_{(q,\jh_3)}(3) \psi^{\chi_4\chi_1}_{(-q,\jh_4)}(4) \ra \ ,
\ee gives a constraint whose unknowns are  the couplings
between three discrete representations. The constraint is \ba &&
C^{\chi_2\chi_3\chi_4,(-(p-q),\jh_2+\jh_3+n)}_{(-p,\jh_2),(q,\jh_3)}
C^{\chi_1\chi_2\chi_4,(q,-\jh_4)}_{(p,\jh_1),(-(p-q),\jh_2+\jh_3+n)}
= \int^\infty_{|\chi_{13}|} ds \,
{\bf F}_{(s,\jh_s),(-(p-q),\jh_2+\jh_3+n)}  \small  \begin{bmatrix} (p,\jh_1) & (-p,\jh_2) \\
 (-q,\jh_4) & (q,\jh_3)
\end{bmatrix}
 \nb \\
&& \left [ C^{\chi_1\chi_2\chi_3,(s,\jh_s)}_{(p,\jh_1),(-p,\jh_2)}
C^{\chi_3\chi_4\chi_1}_{(q,\jh_3),(-q,\jh_4),(s,\jh_s)} + (-1)^\n
(s \rightarrow -s) \right ] \ , \ea and it involves the following
fusing matrix  \ba
&& {\bf F}_{(s,\jh_s),(-(p-q),\jh_2+\jh_3+n)}  \small  \begin{bmatrix} (p,\jh_1) & (-p,\jh_2) \\
 (-q,\jh_4) & (q,\jh_3)
\end{bmatrix} =  \frac{1}{n!} \left [ \frac{\G(p)\G(1-q)}{\G(p-q)} \right ]^{\n+1}
\left [ \frac{\g(p)}{\g(q)\g(p-q)} \right ]^{n-\n} \nb \\
&& \left [ \frac{s}{\sqrt{2}} \right ]^\n
e^{-\frac{s^2}{2}(\psi(p)+\psi(1-q) -2\psi(1))}L^\n_{n-\n}\left (
\frac{s^2 \pi \sin \pi (p-q)}{2 \sin \pi p \sin \pi q} \right ) \
. \ea The integral on the right hand side can be evaluated using
$(\ref{ui10})$. We obtain
\ba C^{\chi_1 \chi_2 \chi_3,
(-(p-q),\jh_1+\jh_2+n)}_{(-p,\jh_1),(q,\jh_2)} &=&
\frac{2^{\frac{1}{2}} \pi ^{\frac{1}{4}} i^n}{2^{\frac{n}{2}}n!}
\left [ \frac{\g(p)}{\g(q)\g(p-q)} \right
]^{\frac{n}{2}+\frac{1}{4}}
e^{-\frac{Q^2}{2}} H_n(-Q) \ , \nb \\
C^{\chi_1 \chi_2 \chi_3,
((p-q),\jh_1+\jh_2-n)}_{(p,\jh_1),(-q,\jh_2)} &=&
\frac{2^{\frac{1}{2}} \pi ^{\frac{1}{4}}i^n}{2^{\frac{n}{2}}n!}
\left [ \frac{\g(p)}{\g(q)\g(p-q)} \right
]^{\frac{n}{2}+\frac{1}{4}} e^{-\frac{Q^2}{2}}H_n(Q) \ ,
\label{d2o1} \ea where \be Q = \frac{(\chi_1 \sin \pi q + \chi_2
\sin \pi (p-q) - \chi_3 \sin \pi p)} {\sqrt{2 \pi \sin \pi p \sin
\pi (p-q) \sin \pi q}} \ . \label{sd2qa} \ee Similarly \ba
C^{\chi_1 \chi_2 \chi_3,
(p+q,\jh_1+\jh_2+n)}_{(p,\jh_1),(q,\jh_2)} &=&
\frac{2^{\frac{1}{2}} \pi ^{\frac{1}{4}} i^n}{2^{\frac{n}{2}}n!}
\left [ \frac{\g(p+q)}{\g(p)\g(q)} \right
]^{\frac{n}{2}+\frac{1}{4}}
e^{-\frac{Q^2}{2}} H_n(-Q) \ , \nb \\
C^{\chi_1 \chi_2 \chi_3,
(-p-q,\jh_1+\jh_2-n)}_{(-p,\jh_1),(-q,\jh_2)} &=&
\frac{2^{\frac{1}{2}} \pi ^{\frac{1}{4}}i^n}{2^{\frac{n}{2}}n!}
\left [ \frac{\g(p+q)}{\g(p)\g(q)} \right
]^{\frac{n}{2}+\frac{1}{4}} e^{-\frac{Q^2}{2}}H_n(Q) \ ,
\label{d2o2} \ea where \be Q = \frac{(\chi_1 \sin \pi q - \chi_2
\sin \pi (p+q) + \chi_3 \sin \pi p)} {\sqrt{2 \pi \sin \pi p \sin
\pi (p+q) \sin \pi q}} \ . \label{sd2qb} \ee The kinematical part
is similar to the one given in $(\ref{kin1})$. We did not check
the constraint $(\ref{1cl-2op})$ for this family of branes.

There are many similarities between the $H_4$ WZW model and the
orbifold CFT of a plane with points identified by a rotation,
stemming from the free field realization found in \cite{kk,kk2}.
It is therefore worth to compare the couplings derived in this section with
the couplings for intersecting branes in toroidal
compactifications, discussed in \cite{cim,cvetic,lust}. Actually,
we can start directly from branes at angles in flat space. The
boundary three-point couplings contains a quantum part that can be
computed using orbifold twist fields \cite{cvetic,lust} and that
coincides with $(\ref{d2o1})$ and $(\ref{d2o2})$ with $n=Q=0$.
They also receive contributions from disc world-sheet instantons
that behave as \be {\cal C}_{ijk} \sim e^{-\frac{A_{ijk}}{2\pi}} \
, \label{winston} \ee where $A_{ijk}$ is the area of the triangle
formed by the three intersecting branes (see fig. 1). Consider
first three branes intersecting at the origin. In this case
$A_{ijk}=0$. The couplings in the $H_4$ model in this case contain
no exponential contribution depending on the position of the
branes. We now move each brane parallel to itself a distance $d_i$
from the origin. If we call $\a_{ij}$ the angle between the brane
$i$ and the brane $j$, the area of the triangle is \be A_{ijk} =
\frac{\left [ d_1 \sin \a_{23} + d_2 \sin \a_{13} - d_3 \sin
\a_{12} \right ]^2 }{2\sin{\a_{12}}\sin{\a_{13}}\sin{\a_{23}}} \ .
\ee We recognize that the instanton contribution
$(\ref{winston})$  coincides with the exponential term in
$(\ref{sd2qa})$ and $(\ref{sd2qb})$ upon setting $d_i = \chi_i$
and identifying the angles $\a_{ij}$ with the light-cone momentum
carried by the vertex operators $\psi^{\chi_i\chi_j}_{\pm p,
\jh}$.

\begin{figure}[ht]
\begin{center}
\includegraphics[width=14cm]{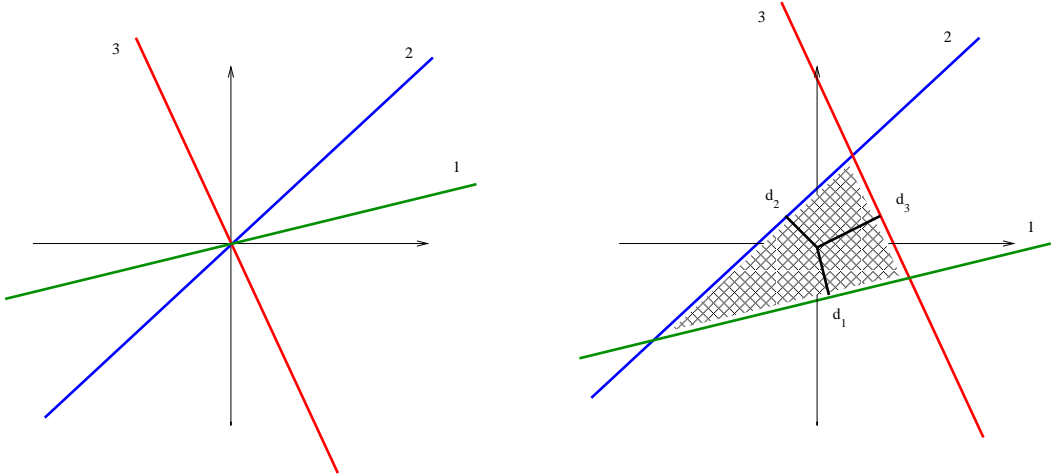}
\end{center}
\caption{World-sheet instanton contribution to the couplings for
intersecting brane models.}
\end{figure}

Finally there is an interesting limit to consider. The limit $\chi
\rightarrow \infty$ in the $H_4$ model  is the analogue of the limit $\psi
\rightarrow \infty$  in which an $AdS_2$ brane is moved towards the
boundary of $AdS_3$. In this limit, as discussed in \cite{oog1}, one
obtains a so called NCOS theory \cite{ncos,ncos2}, which is a theory of
open strings decoupled from the closed string sector. In our case however there is
an important difference: after the Penrose limit, the world-volume flux is
null and therefore there is no notion of a critical electric field. In
this respect, the world-volume theories of the $D2$ branes provide examples
of theories with light-like non-commutativity \cite{llnc}. We may also observe
 that in the limit $\chi
\rightarrow \infty$,  only the bulk continuous representations remain coupled to
the brane. The couplings in $(\ref{d2o1})$ and $(\ref{d2o2})$
between open strings in discrete representations are exponentially
suppressed. Thus the strings interact in this limit only through the exchange
of states in the continuous representations.

\subsection{The $S 1$ branes}
\setcounter{equation}{0}

In this section we provide a detailed solution for the structure
constants of the $S 1$ branes. We start with the boundary
two-point functions for open strings ending on two different
branes $a$ and $b$, sitting at different positions in the $u$
direction. As explained in section \ref{spectrum},  the vertex
operators for the open strings stretched between the two branes
belong to the $V^+_{p^{ab},\jh^{ab} - n}$ representations, $n \in
\mathbb{N}$, when $p_b > p_a$ and to the $V^-_{p^{ab},\jh^{ab} +
n}$ representations, $n \in \mathbb{N}$, when $p_b < p_a$. We will
use the shorthand notation $\psi^{ab}_{p^{ab},\jh^{ab}\mp n} =
\psi^{ab}_{p^{ab},n}$ where the sign in $\jh^{ab}\mp n$ is fixed
accordingly to the sign of $p^{ab}$. The two-point functions are
\be \la \psi^{ab}_{p^{ab},n_1}(t_1,x_1)
\psi^{ba}_{p^{ba},n_2}(t_2,x_2) \ra_a = C_{n_1;n_2}^{aba,1}\ib_a
e^{- |p^{ab}| x_1x_2} \d_{n_1,n_2} \ . \label{scb1} \ee When the
two branes are  at the same position in  the $u$ direction, the
open strings ending on them belong to the continuous
representations. In particular, when $a=b$ we have \be \la
\psi^{aa}_{s_1}(t_1,x_1) \psi^{aa}_{s_2}(t_2,x_2) \ra_a =
C_{s_1s_2}^{aaa,1}\ib_a 2 \pi \, \d(\a_1-\a_2-\pi)
\frac{\d(s_1-s_2)}{s_1} \ . \label{scb2} \ee

The Ward identities completely fix the dependence of the
bulk-boundary couplings on the charge variables  \ba \la
\F_{p,\jh}^{+}(z,x_1) \psi_s^{aa}(t,x_2) \ra_a &=& e^{-px_1\bar
x_1-\frac{s}{\sqrt{2}}\left ( \bar x_1 x_2 +\frac{x_1}{x_2} \right
)}
{}^aB^s_{p,\jh} \ib_a \ , \nb \\
\la \F_{p,\jh}^{-}(z,x_1) \psi_s^{aa}(t,x_2) \ra_a  &=&
e^{-px_1\bar x_1-\frac{s}{\sqrt{2}}\left ( x_1 x_2 +\frac{\bar
x_1}{x_2} \right )}
{}^aB^s_{-p,\jh} \ib_a \ ,  \\
\la  \F_{s_1,\jh}^{0}(z,x_1) \psi_{s_2}^{aa}(t,x_2) \ra_a  &=& 8
\pi^2 \d(\a_1-\bar \a_1 -2 \a_2+\pi)\d(\th+\a_1+\bar \a_1) \
{}^aB^{s_2}_{s_1,\jh} \ib_a \ . \nb \label{scb3} \ea The last
coupling  is non-zero only when $0 \le s_2 \le 2s_1$ and we set
\be s_2 = 2s_1\sin \frac{\th}{2}  \ . \label{scb4} \ee The bulk
one-point functions with the identity, are particular cases of the
previous expressions \ba \la \F_{p,\jh}^{\pm}(z,x) \ra_a &=& e^{-p
x\bar x} \ {}^aB_{\pm p, \jh}^1 \la 1 \ra_a \ ,
\nb \\
\la \F_{s,\jh}^{0}(z,x) \ra_a &=&  2 \pi \d(\a+\bar \a) \,
{}^aB^{1}_{s,\jh} \ib_a \ . \label{scb6} \ea We can fix all the
bulk-boundary structure constants by studying the factorization of
the three kinds of bulk two-point functions, namely $\la \F_p^+\F_q^+
\ra$, $\la \F_p^+\F_q^- \ra$ (where we have to distinguish between
$p>q$ and $p=q$) and $\la \F_p^+ \F_s^0 \ra$. Using the bulk
three-point couplings in $(\ref{qpppm}-\ref{cpmo2})$ and the
fusing matrices in appendix \ref{ff}, the sewing constraints can be
written as follows \ba {}^a B_{p,\jh_1}^s {}^a B_{q,\jh_2}^s &=&
\sqrt{\th_{p,q}}
e^{\frac{s^2}{2}(\psi(p)+\psi(q)-2\psi(1))}\sum_{n=0}^\infty
L_n\left (\th_{p,q} \frac{
s^2}{2}\right ) {}^aB^1_{p+q,\jh_1+\jh_2+n} \ ,  \\
{}^a B_{p,\jh_1}^s {}^a B_{-q,\jh_2}^s  &=& \sqrt{\th_{-p,q}}
 e^{\frac{s^2}{2}(\psi(q)+\psi(1-p)-2\psi(1))}\sum_{n=0}^\infty
L_n\left (\th_{-p,q} \frac{
s^2}{2}\right ) {}^aB^1_{p-q,\jh_1+\jh_2-n} \ ,  \nb \\
{}^a B_{p,\jh_1}^s {}^a B_{-p,\jh_2}^s  &=& \frac{\pi}{\sin \pi p
} e^{\frac{s^2}{2}(\psi(p)+\psi(1-p)-2\psi(1))} \int_0^\infty dt \
tJ_0\left (\frac{\pi
s t}{\sin \pi p} \right ) {}^a B_{t,\jh_1+\jh_2}^1 \ , \nb \\
{}^a B_{p,\jh_1}^{s_2} {}^a B_{s_1,\jh_2}^{s_2}  &=& \frac{e^{-
\frac{i \pi s_1^2\sin \th} {2 \tan (\pi p)}}}{\pi s_1^2 \sin
\th}e^{\frac{s_1^2 (1-\cos \th) }{2}(\psi(p)+\psi(1-p)-2\psi(1))}
\sum_{n \in \mathbb{Z}} e^{in \th}\ {}^a B^1_{p,\jh_1+\jh_2+n} \ ,
\nb \label{scb11} \ea where $\th_{p,q} = \pi \cot{\pi p}+\pi
\cot{\pi q}$ and as before $s_2^2 = 2s_1^2(1-\cos \th)$. The
solution is \ba {}^a B_{\pm p,\jh}^s &=& \sqrt{\frac{\pi}{\sin \pi
\m p }} \ \frac{e^{\pm 2 i p \h + i \jh u}}{1-e^{\pm i \mu u}}
e^{\frac{s^2}{4}[\psi(\m p)+\psi(1-\m p)-2 \psi(1)]  \mp
\frac{i\pi s^2}{4 \tan(\pi \m p) \tan \left(\frac{\m u}{2}\right
)}} \ , \nb \\
{}^a B_{s_1,\jh}^{s_2} &=& \frac{e^{i \jh u}}{\pi s_1^2 \sin \th}
2 \pi \d(\m u - \th) \ , \hspace{1cm} s_2^2 = 2s_1^2(1-\cos \th) \
. \label{bbc1m} \ea  Note the similarity between the first
coupling and a bulk three-point coupling of the form ${\cal
C}_{+-0}$ $(\ref{cpmo2})$, as expected on general grounds. The
second coupling can also be written as \be {}^a B_{s_1,\jh}^{s_2}
= e^{i \jh u} \frac{\d(s_2(u)-s_2)}{s_2} \ , \hspace{1cm} s_2^2(u)
= 4s_1^2 \sin^2 \frac{\m u}{2} \ . \label{bbc2m} \ee Finally the
one-point functions with the identity, relevant for the
construction of the boundary states, are a particular case of the
previous couplings and read \ba {}^a B_{\pm p,\jh}^{0} &=&
\sqrt{\frac{\pi}{\sin \pi \m p }} \frac{e^{\pm 2 i p \h + i \jh
u}}{1-e^{\pm i \mu u}} = \pm \frac{i}{2 \sin \frac{u}{2}}
\sqrt{\frac{\pi}{\sin \pi \m p }}e^{\pm i p \h + i \jh u \mp
\frac{i u}{2}}
\ , \nb \\
{}^a B_{s,\jh}^{0} &=& \frac{e^{i \jh u}}{4 \sin^2 \left (\frac{\m
u}{2} \right) } \frac{\d(s)}{s}  \ . \ea Whenever $\m u = \pi + 2
\pi n$ the couplings in $(\ref{bbc1m})$ simplify, since the brane
is  trivially embedded in the space-time. On the other hand,
whenever  $\m u = 2 \pi n$, the behavior of the couplings in
$(\ref{bbc1m})$ reflects the change in the geometry of the branes.
We have first to multiply the structure constants by the one-point
function of the identity $\ib_a = \sin(\m u_a/2)$. We then observe
that the first coupling in $(\ref{bbc1m})$ becomes non-trivial
only for $s=0$, as expected since only these representations live
on the $S(-1)$ brane world-volume. As for the couplings of the
continuous representations, they are now non-zero for every value
of $s_1$ while $s_2$ has to vanish. One can think of this process
as an interpolation between Neumann and Dirichlet boundary
conditions induced by the flux on the brane world-volume. In the
limit $\m p << 1$ these couplings reproduce the semi-classical
expressions derived in section \ref{semic}.

We determine now the boundary three-point couplings. Even though
the explicit form of these couplings is slightly more involved
than the couplings we computed in the previous section for the D2
branes, our task is in this case simplified since the $S 1$ branes
are the Cardy branes of the $H_4$ WZW model. The three-point
boundary couplings for Cardy boundary conditions in a RCFT, due to
the one-to-one correspondence between the boundary labels and the
representations of the chiral algebra, can be expressed using the
fusing matrices. This was first realized in the case of the Virasoro minimal
models \cite{ingo}.  Indeed, one can verify that setting \be
C^{abc,k}_{ij} \sim  {\bf F}_{\check{b}k} \small
\begin{bmatrix} i & j \\ a & \check{c}
\end{bmatrix} \ ,
\label{cf} \ee the constraint $(\ref{4open})$ is satisfied as a
consequence of the pentagon relation between the fusing matrices
\cite{ms}. The previous relation proved very useful in order to
study the effective field theory on the brane world-volume. Indeed,
the fusing matrices of a WZW model based on the group $G$
coincide with the Racah coefficients of the quantum group algebra
$U_q(G)$, where $q=e^{\frac{2 \pi i}{k+g}}$ with $k$ the level and
$g$ the dual Coxeter number. In the limit $q \rightarrow 1$, the
quantum Racah coefficients reduce to the classical ones, a fact
that has been exploited in \cite{volker,afs} to study the
non-commutative geometry on the brane world-volume. Similar
observations were made also for some non-compact CFTs, most
notably for the Euclidean version of $AdS_3$ \cite{schads} and the
Liouville model \cite{tbl}. We  now verify that the relation
between the fusing matrices and the three-point boundary couplings
(\ref{cf}) remains valid also for the non-compact $H_4$ WZW model.

In the following, we will repeatedly use the relation $(\ref{rel})$
between the brane parameters and the $H_4$ quantum numbers. As it
was the case for the bulk theory, we can distinguish between
various types of boundary three-point couplings. Consider three $S
1$ branes with labels $a$, $b$ and $c$. When $p_c
> p_b > p_a$, we have a boundary coupling similar to a bulk
${\cal C}_{++-}$ coupling \be \la \psi_{p^{ab},n_1}^{ab}(t_1,x_1)
\psi_{p^{bc},n_2}^{bc}(t_2,x_2) \psi_{p^{ca},n_3}^{ca}(t_3,x_3)
\ra_a = C^{abc,n_3}_{n_1 n_2} C^{aca,1}_{n_3 \check n_3} \ib_a
e^{-x_3(|p^{ab}|x_1+|p^{bc}|x_2)}(x_1-x_2)^\n \ , \ee where $\n =
n_1 + n_2 - n_3$. When $p_b > p_a = p_c$, we have a boundary
coupling similar to a bulk ${\cal C}_{+-0}$ coupling \be \la
\psi^{ab}_{p^{ab},n_1}(t_1,x_1) \psi^{bc}_{p^{bc},n_2}(t_2,x_2)
\psi^{ca}_{s,\jh^{ca}}(t_3,x_3) \ra_a  = C^{abc,s}_{n_1
n_2}C^{aca,1}_{ss} \ib_a e^{-|p^{ab}|
x_1x_2-\frac{s}{\sqrt{2}}\left ( x_2 x_3 + \frac{x_1}{x_3} \right
)}x_3^\n \ , \ee where $\n = n_1 - n_2$. Finally, when
$p_a=p_b=p_c$ we have a boundary coupling similar to a bulk ${\cal
C}_{000}$ coupling \be \la \psi^{ab}_{s_1}(t_1,x_1)
\psi^{bc}_{s_2}(t_2,x_2) \psi^{ca}_{s_3}(t_3,x_3) \ra_a  =
C^{abc,s_3}_{s_1 s_2}C^{aca,1}_{s_3 s_3} \ib_a (2\pi)^2
\d(\a_2-\a_3-\f)\d(\a_1-\a_2+\th) \ , \ee with \be s_3^2 =
s_1^2+s_2^2 - 2s_1s_2 \cos \th \ , \hspace{1cm} e^{i \f} =
\frac{s_1-s_2e^{-i\th}}{s_3} \ . \ee The other configurations can
be obtained by permuting the fields in the previous expressions.
It is also useful to remember that for the continuous
representations, raising an index is equivalent to $(x,\jh) \mapsto
-(x,\jh)$ while for the discrete representation it corresponds to
$(x,\jh) \mapsto -(x^*,\jh)$.

We adopt now the following strategy. We assume that the
three-point boundary couplings are given by the fusing matrices,
up to a choice of normalization for the boundary fields. We then
use the constraints pertaining to the correlators  $\la
\psi^{ab}\psi^{ba}\psi^{ab}\psi^{ba}\ra$ and  $\la
\psi^{ab}\psi^{ba}\psi^{aa}\psi^{aa}\ra$ to fix a convenient
normalization for both the discrete $\psi^{ab}_{p^{ab},n}$ and the
continuous $\psi^{aa}_s$ vertex operators. Finally we verify that
the couplings determined in this way solve the other constraints
as well.

Let us start with the three-point coupling between two open
strings stretched between a pair of $S 1$ branes with labels $a$
and $b$ and an open string that lives on the world-volume of one
of the two $S 1$ branes. This coupling involves two discrete and
one continuous representation and is related to the following
fusing matrix \be C^{aba,s}_{(p^{ab},n_1),(-p^{ab},n_2)} =
\w_{aba}(n_1,n_2) {\bf
F}_{(-p_b,\jh_2-\jh_a-n_2),s}  \small  \begin{bmatrix} (p^{ab},\jh_1) & (-p^{ab},\jh_2) \\
 (p_a,\jh_a) & (-p_a,-\jh_a)
\end{bmatrix} \ ,
\ee with $p_b > p_a$, $\jh_1 = \jh^{ab}-n_1$ and $\jh_2 =
\jh^{ba}+n_2$. Similarly, \be
C^{bab,s}_{(-p_{ab},n_1),(p_{ab},n_2)} = \w_{bab}(n_1,n_2)
 {\bf F}_{(-p_a,\jh_2-\jh_b+n_2),s}  \small  \begin{bmatrix} (-p_{ab},\jh_1) & (p_{ab},\jh_2) \\
 (p_b,\jh_b) & (-p_b,-\jh_b)
\end{bmatrix} \ ,
\ee with $\jh_1 = \jh_{ba}+n_1$ and $\jh_2 = \jh_{ab}-n_2$. Here
$\w_{aba}$ and $\w_{bab}$ are proportionality factors that can
depend on all the quantum numbers involved even though we only
emphasized their dependence on the labels $n_1$ and $n_2$. The
factorization constraint for $\la
\psi^{ab}\psi^{ba}\psi^{ab}\psi^{ba}\ra$ reads \ba && C^{bab,(s,\{
\jh^{ac} \})}_{(-p,\jh^{ba}+n_2),(p,\jh^{ab}-n_3)} C^{bab,(s,\{
\jh^{ac} \})}_{(-p,\jh^{ba}+n_4),(-p,\jh^{ab}-n_1)} \ib_b =
\int_0^\infty
dt \, t e^{\frac{s^2-t^2}{2} \s(p)} \nb \\
&& \frac{\pi}{\sin{\pi p}} J_{\n}\left ( \frac{\pi s t }{\sin{\pi
p}} \right ) C^{aba,(t,\{ \jh^{ac}
\})}_{(p,\jh^{ab}-n_1),(-p,\jh^{ba}+n_2)} C^{aba,(t,\{ \jh^{ac}
\})}_{(p,\jh^{ab}-n_3),(-p,\jh^{bc}+n_4)}  \ib_a \ , \ea where $p=
p_b-p_a>0$ and $\n = n_1+n_3-n_2-n_4$. Using the integral
$(\ref{ui5})$ one can verify that the constraint is satisfied if
the following relation holds between the proportionality factors
$\w_{aba}$ \be \ib_a \w_{aba}(n_1,n_2)\w_{aba}(n_3,n_4) \sin \pi
p_b = \ib_b \w_{bab}(n_2,n_3)\w_{bab}(n_4,n_1) \sin \pi p_a \ .
\label{ibrel} \ee Combining the previous equation with the
continuity of the two-point function on the disc \be \ib_a
\w_{aba}(n,n) \sin{\pi p_b} = \ib_b \w_{bab}(n,n) \sin{\pi p_a} \
, \ee we obtain $\w_{aba}(n,n) = \w_{bab}(n,n)$. A convenient
choice for the normalization is \ba \w_{aba}(n_1,n_2) &=&
\frac{1}{n_2!}\left [ \frac{\g(p_b)}{\g(p_a)\g(p)}
\right ]^{n_2}  \left [ \frac{\G(1-p_a)\G(p_b)}{\G(p)} \right ]^{n_1-n_2+1}  \ , \nb \\
\w_{bab}(n_1,n_2) &=&  \frac{1}{n_1!} \left [
\frac{\g(p_b)}{\g(p_a)\g(p)} \right ]^{n_1} \left [
\frac{\G(1-p_a)\G(p_b)}{\G(p)} \right ]^{n_2-n_1+1} \ . \ea Once
we make this choice, $(\ref{ibrel})$ implies the following
relation for the one-point function of the identity  \be \ib_a
\sin \pi p_b = \ib_b \sin \pi p_a \ , \ee which is satisfied if
$\ib_a \sim \sin \pi p_a$. This is indeed the case as we will se
in section \ref{annulus}. We may now write \ba
C^{aba,s}_{(p,n_1),(-p,n_2)} &=& \frac{\pi}{\sin \pi p_a} \left [
\frac{\pi \, s }{\sqrt{2} \sin \pi p_a} \right ]^{n_1-n_2}
e^{\frac{s^2}{2} \left [ \psi(p)-\psi(1) - \frac{\pi}{2}\cot{\pi
p_a} \right]} \nb \\ && L^{n_1-n_2}_{n_2} \left [ \frac{\pi
s^2}{2}( \cot \pi p + \cot \pi p_a ) \right ] \
, \label{b-3+-0}   \nb \\
C^{bab,s}_{(-p,n_1),(p,n_2)} &=& \frac{\pi}{\sin \pi p_b} \left [
\frac{\pi \, s}{\sqrt{2} \sin \pi p_b} \right ]^{n_2-n_1}
e^{\frac{s^2}{2} \left [ \psi(p)-\psi(1) + \frac{\pi}{2} \cot{\pi
p_b} \right]} \nb \\ && L^{n_2-n_1}_{n_1} \left [ \frac{\pi
s^2}{2}( \cot \pi p - \cot \pi p_b ) \right ] \ , \label{b-3-+0}
\ea with $p=p_b-p_a > 0$ and $L^m_n(x)$ a generalized Laguerre
polynomial. Note the different behavior of the first of these
couplings for $p_a \rightarrow 0$ and $p_b \rightarrow 0$. In the
first case it is non-zero only for $s=0$ and $n_1=n_2$  while it
remains essentially unchanged in the second case. This is as
expected since only the identity exists on the brane at $u_a=0$
while the open string spectrum for the brane at $\m u= 2 \pi p_b$
contains all the representations $\psi^{bb}_s$, $s \ge 0$.

Consider now the coupling between three open strings that live on
the same brane. This is a coupling between three continuous
representations. In terms of the fusing matrices we have \be
C^{aaa,r}_{s,t} = \w_{aaa}
 {\bf F}_{(-p_a,-\jh_a),r}  \small  \begin{bmatrix} s & t \\
 (p_a,\jh_a) & (-p_a,-\jh_a)
\end{bmatrix} \ .
\ee In this case we set $\w_{aaa}=1$ which is equivalent to
$C^{aaa,1}_{s,s}=1$. The constraint for $\la
\psi^{ab}\psi^{ba}\psi^{aa}\psi^{aa}\ra$ is \ba &&
C^{aba,r}_{(p,\jh^{ab}-n_1),(-p,-\jh^{ab}+n_2)} C^{aaa,r}_{s,t}  =
\frac{1}{\pi st \sin{\th} } e^{\frac{st}{2} - \left [\s(p)
\cos{\th} + i \sin{\th} \pi \cot{\pi p} \right ] - i  \f \n  } \nb \\
&& \sum_{n=0}^\infty e^{i(n_1-n)(\pi-\th)}
C^{baa,(-p,-\jh^{ab}+n)}_{(-p,-\jh^{ab}+n_2),s}
C^{aba,t}_{(p,\jh^{ab}-n_1),(-p,-\jh^{ab}+n)}  \ , \ea where $p =
p_b-p_a>0$, $\n = n_1-n_2$, $r^2=s^2+t^2-2st \cos{\th}$ and
$e^{i\f} = \frac{s-te^{-i\th}}{r}$. In order to verify that this
constraint is satisfied, one can use the series $(\ref{ui8})$. The
result is \be C^{aaa,r}_{t s} = \frac{1}{\pi st \sin \th}
e^{\frac{i \pi st \sin \th}{2 \tan \pi p_a}} \ , \label{b-3000}
\ee with $r^2 = s^2+t^2-2ts \cos \th$. For $r=0$, $\th=0$ the
three-point function reproduces the two-point function in
$(\ref{scb2})$.

At this point, we have completely fixed the normalization of the
boundary vertex operators. Thus, the factorization constraints
for the other four-point amplitudes represent a consistency check on the
solution.

The constraint for $\la \psi^{ab}\psi^{ba}\psi^{ad}\psi^{da}\ra$
is slightly more complicated and reads \ba && C^{bad,(-(p-q),
\jh^{bd}+n)}_{(-p,-\jh^{ba}+n_2),(q,\jh^{ad}-n_3)} C^{abd,(q,
\jh^{ad}-n_4 )}_{(p,\jh^{ab}-n_1),(-(p-q),\jh^{bd}+n)} =
\frac{1}{(n-n_2+n_3)!}\left [ \frac{\G(p)\G(1-q)}{\G(p-q)}\right
]^{\a+1} \nb \\ && \frac{\sin \pi p_d}{\pi}\left [
\frac{\g(q)}{\g(p)\g(q-p)} \right ]^{n+n_4-n_1}  \int_0^\infty ds
\, s \left [ \frac{s^2}{2} \right
]^{\n/2} e^{-\frac{s^2}{2}(\psi(p)+\psi(1-p)-2 \psi(1))}  \\
&& L^{\n}_{n+n_4-n_1}\left [ \frac{\pi s^2}{2} (\cot{\pi
q}-\cot{\pi p}) \right ]
C^{aba,s}_{(p,\jh^{ab}-n_1),(-p,-\jh^{ab}+n_2)}
C^{ada,s}_{(q,\jh^{ad}-n_3),(-q,-\jh^{ad}+n_4)} \ , \nb \ea where
$p= p_b-p_a > 0$, $q= p_d-p_a>0$ and  $\n = n_1+n_3-n_2-n_4$. In
order to verify that this constraint is satisfied one can use the
integral $(\ref{ui9})$. Consider now the couplings between states
in discrete representations. Using the relation $(\ref{cf})$ we
can write \be C^{abc,(p^{ac},\jh_3)}_{(p^{ab},\jh_1)
(p^{bc},\jh_2)} = {\bf F}_{(-p_b,\jh_2-\jh_c+n_2),(p^{ac},
\jh_1 +\jh_2+k )}  \small  \begin{bmatrix} (p^{ab},\jh_1) & (p^{bc},\jh_2) \\
 (p_a,\jh_a) & (-p_c,-\jh_c)
\end{bmatrix}  \ , \nb \\
 \label{b-3+++} \ee
where $\jh_1 = \jh^{ab}-n_1$, $\jh_2 = \jh^{bc}-n_2$, $\jh_3 =
\jh^{ac}-n_3$  and $k=n_1+n_2-n_3 \ge 0$. Similar expressions hold
for the other couplings between discrete representations and can
be found in appendix \ref{dsc}.

For completeness we display here the explicit form of one of these
couplings \ba C^{abc,(p-q,\jh_3)}_{(p,\jh_1) (-q,\jh_2)} &=&
\frac{(n_2+n_3)!}{n_3!(n_2+n_3-n_1)!} \frac{\sin \pi (p-q)}{\sin
\pi p_c} \left [\frac{\g(p)}{\g(q)\g(p-q)} \right
]^{\frac{k+1}{2}} \left [\frac{\pi \sin \pi q}{\sin \pi p \sin \pi
(p-q)}\right ]^{\frac{k+1}{2}} \nb \\ && \left [ \frac{\sin \pi
p_a}{\pi} \right ]^{k} F(-n_2,n_1-n_2-n_3,-n_2-n_3,\tau) \ ,
\label{bex} \ea where $\jh_1 = \jh^{ab}-n_1$, $\jh_2 =
\jh^{bc}+n_2$, $\jh_3 = \jh^{ac}-n_3$ and \be \tau =
\frac{\sin{\pi p}\sin{\pi p_c}}{\sin{\pi q}\sin{\pi p_a}} \ . \ee
Note that $k \equiv n_3 - n_1 + n_2 \ge 0$. There are other
constraints we should verify. We also checked the constraints
following from amplitudes with one bulk and two boundary operators
while we did not check the constraint related to the amplitude
$\la \psi^{ab}\psi^{bc}\psi^{cd}\psi^{da}\ra$.

It is interesting to compare the coupling in (\ref{b-3000}) with
the coupling between open tachyon vertex operators on a two-torus
with a magnetic field $B$ \cite{acny,swnc}. Consider two free
bosonic fields $X_1$ and $X_2$ with the boundary conditions \be
\p_\s X_1 + F \p_\t X_2 = 0 \ , \hspace{1cm} \p_\s X_2 - F \p_\t
X_1 = 0 \ , \label{scb40} \ee where $F = 2 \pi \a^{'} B$. In this
case the momenta are measured using the open string metric \be
G^{ij} = \frac{1}{1+F^2} \d^{ij} \ , \label{scb41} \ee and the
conformal dimension of a boundary tachyon vertex operator $e^{i
\vec p \vec X}$ is \be h = \frac{\a^{'} p^2}{1+F^2} \ .
\label{scb42} \ee Moreover in the OPE there is a momentum
dependent phase \be e^{i \vec p \vec X(t_1)} \,  e^{i \vec q \vec
X(t_2)} \sim (t_1-t_2)^{ 2 \a^{'} G^{ij}p_i q_j} \, e^{ i
\frac{\th^{ij}}{2} p_i q_j}  \, e^{i (\vec p+ \vec q) \vec X(t_2)}
\ , \label{scb43} \ee where the deformation parameter is \be
\th^{ij} = -\frac{2 \pi \a^{'} F}{1+F^2} \e^{ij} \ . \label{scb44}
\ee Comparing the conformal dimension of $\psi^{aa}_s$ with
(\ref{scb42}) we see that $p^2 = s^2(1+F^2)$. Thus comparing the
phase in (\ref{scb43}) with the one in (\ref{b-3000}) we can identify
\be F(u) = - \cot \frac{\m u}{2} \ , \label{scb45} \ee which is
the expected result. Note that the magnetic field vanishes for $u
= \pi + 2 \pi n$, which corresponds to the flat $S 1$ brane or
equivalently to Neumann boundary conditions on $X_1$ and $X_2$.
Changing the value of $u$ we get the mixed Neumann-Dirichlet
boundary conditions in $(\ref{scb40})$ until we reach $u = 2 \pi
n$, where the field-strength diverges and therefore the boundary
conditions become pure Dirichlet. In fact,  precisely for these
values of the coordinate $u$, the two-dimensional conjugacy classes
degenerate to points. According to the analogy with open strings
in a magnetic field, the strings that live on the brane
world-volume belong to the continuous representations since their
ends are both subject to the same magnetic field and then they
behave as free strings. On the other hand, the strings stretched
between two different branes feel generically different magnetic
fields and the corresponding vertex operators are therefore twist
fields or, in $H_4$ terminology, they belong to the discrete
representations. The twist for a string stretched from brane $b$
to brane $a$ is given by \cite{acny} \be \e^{ab} =
\frac{1}{\pi}\left [ {\rm arctan} F(u_b) - {\rm arctan} F(u_a)
\right ] = \frac{u_b-u_a}{2\pi} \ , \ee as expected. The effective
field theory on the world-volume of an $S 1$ brane is the limit of
the non-commutative field theory on the fuzzy sphere pertaining to
the $S^2$ branes in $S^3$: the volume and the magnetic flux are
both scaled to infinity in order to obtain the non-commutative
plane in the limit.

\section{Four-point amplitudes \label{fp}}
\renewcommand{\theequation}{\arabic{section}.\arabic{equation}}

In the previous section we derived all the structure constants for
the two families of boundary CFTs that describe the D2 and the $S
1$ branes of the $H_4$ WZW model. These are the essential ingredients for the solution
of the models. We may now compute arbitrary
correlation functions by sewing together the basic one, two and
three-point amplitudes. Here, we are going to discuss in
this section only those disc amplitudes that can be expressed in
terms of the four-point conformal blocks, namely amplitudes
containing either two bulk fields $\la \F \F \ra$ or one bulk and
two boundary fields $\la \F \psi \psi \ra$ or four boundary fields
$\la \psi \psi \psi \psi \ra$.

In the following, we will provide
some  examples for each type of amplitude, both for the
D2 and the $S 1$ branes. There are of course many other amplitudes
one could consider, besides the few we are going to discuss here.
In general, one can write for them a decomposition in terms of our
conformal blocks and structure constants but we expect that using
series and integrals more general than the ones we use in this
paper one should be able to find a closed form for many of them.
It will be  interesting to analyze their properties in detail,
both from the CFT and the string theory point of view.

 As in the
previous section, in order to avoid writing unnecessarily large
formulae, we denote with a single number in round brackets all the
variables a vertex operator depends on. This includes its insertion
point and the charge variables. The four-point conformal blocks we
will need in the following computations are displayed in appendix
\ref{bb}. We choose the gauge \be A(z_1,z_2,z_3,z_4) =
\prod_{j>i=1}^4 z_{ij}^{\frac{h}{3}-h_i-h_j} {\cal A}(z) \ ,
\hspace{1.4cm} z = \frac{z_{12}z_{34}}{z_{13}z_{24}} \ , \ee where
$h = \sum_{i=1}^4 h_i$, $z_{ij} = z_i - z_j$ and the $z_i$ can
represent the insertion points of either a bulk or a boundary
vertex operator. Finally we set $\n = - \sum_{i=1}^4 \jh_i$.

\subsection{The D2 branes}

\renewcommand{\theequation}{\arabic{section}.\arabic{subsection}.\arabic{equation}}
\setcounter{equation}{0}

For the $D2$ branes we only display a very simple amplitude \be A
= \la \psi^{\chi_1\chi_2}_{p,\jh_1}(1)
\psi^{\chi_2\chi_1}_{-p,\jh_2}(2) \psi^{\chi_1\chi_2}_{p,\jh_3}(3)
\psi^{\chi_2\chi_1}_{-p,\jh_4}(4) \ra \ . \ee In this case the
integral over the conformal blocks can be performed explicitly and
we obtain \ba  {\cal A}(z) &=& z^{\ka_{12}}(1-z)^{\ka_{14}}
e^{-p(x_1x_2+x_3x_4)}\left [\frac{x_1-x_3}{x_4-x_2}\right
]^{\frac{\n}{2}} \left [ \frac{2 \sin \pi p}{c_1(z)c_1(1-z)}
\right ]^{\frac{1}{2}}
 \\ && e^{xpz
-x z(1-z) \p \ln c_1(z) - \frac{x \sin \pi p}{2 \pi
c_1(z)c_1(1-z)}} I_{\n/2}\left ( \frac{x \sin \pi p}{2 \pi
c_1(z)c_1(1-z)} \right ) \nb \ , \ea where $c_1(z) =
F(p,1-p,1,z)$, $x = (x_1-x_3)(x_2-x_4)$ and $I_{\n}$ is a modified
Bessel function. Moreover \ba
\ka_{12} &=& h_1+h_2-\frac{h}{3}+p^2+(\jh_1-\jh_2)p-p \ ,  \nb \\
\ka_{14} &=& h_1+h_4-\frac{h}{3}+p^2+(\jh_1-\jh_4)p-p \ . \label{kad2} \ea
Even though this amplitude does not depend on the brane parameters
$\chi_i$, it is interesting since we can expand it in powers of the charge
variables and compare the correlator of the ground states of the $H_4$
representations with the correlator of open strings living at the
intersection of two branes at an angle \cite{cvetic,lust}, which can be
described by boundary twist fields. The two expressions indeed coincide
upon identifying the light-cone momentum $p$ with the angle formed by the
two branes, as expected given the relationship between the primary vertex
operators in the Nappi-Witten background and the orbifold twist fields
\cite{kk,kk2,dk}. Our open string amplitudes in the gravitational
wave can be thought of as generating functions for the correlators of
arbitrarily excited boundary twist fields in orbifold models or
equivalently of open strings living at the intersection of a configuration of
branes at angles.

\subsection{The $S 1$ branes}
\setcounter{equation}{0}

For the $S 1$ branes we provide several explicit examples of
four-point amplitudes. As we mentioned in the introduction, the
space-time interpretation of these correlation functions deserves
further investigation, in particular those involving on-shell open
string states stretched between branes localized at different
positions in the $u$ direction.

We start from the bulk two-point functions on the disc. This  gives
the first correction to the propagation of the closed strings due
to the presence of the S-brane. Using the results of section
\ref{sc} and the conformal blocks in appendix \ref{bb} we obtain
\be \la \F_{p,\jh_1}^+ \F_{q,\jh_2}^- \ra_a = \ib_a e^{-px_1\bar
x_1-qx_2\bar x_2}
\frac{z^{\ka_{12}}(1-z)^{\ka_{14}}}{\sqrt{\g(p)\g(q)}}
\frac{e^{i(p-q)\h_a+i(\jh_1+\jh_2)u_a}}{1-e^{iu_a}}
\frac{e^{-x(1-z)q -xz(1-z)\p \ln B(z)}}{B(z)} \ , \ee where
$(u_a,\h_a)$ are the brane parameters, \be B(z) =
\frac{F(q,1-p,1-p+q,z)}{\g(p)\G(1-p+q)} - e^{-iu_a} z^{p-q}\frac{
F(p,1-q,1-q+p,z)}{\g(q)\G(1-q+p)}
 \ , \ee and \be z =
\frac{|z_1-z_2|^2}{|z_1-\bar z_2|^2} \ , \hspace{1cm} x =
(x_1-\bar x_2)(x_2 - \bar x_1) \ . \ee Here \be \ka_{12} =
h_1+h_2-\frac{h}{3}+p \, q-\jh_2p+\jh_1q-q \ ,  \hspace{0.6cm}
\ka_{14} = 2h_1-\frac{h}{3}+p^2+2\jh_1p-p \ . \label{aa5ffs1} \ee
When $p=q$ only the identity couples to the brane in the closed
channel and the correlator simplifies  \be \la \F_{p,\jh_1}^+
\F_{p,\jh_2}^- \ra_a = \ib_a e^{-px_1\bar x_1-qx_2\bar x_2}
\frac{e^{i(\jh_1+\jh_2)u_a}}{4\sin^2 \frac{u_a}{2}}
\frac{z^{\ka_{12}}(1-z)^{\ka_{14}}}{c_1(z)}e^{-x(1-z)p -xz(1-z)\p
\ln c_1(z)} \ , \ee where \be c_1(z) = F(p,1-p,1,z) \ . \ee We
also display the closely related amplitude \be \la \F_{p,\jh_1}^+
\F_{q,\jh_2}^+ \ra_a = \ib_a e^{-px_1\bar x_1-qx_2\bar x_2}
\frac{z^{\l_{12}}(1-z)^{\l_{14}}}{\sqrt{\g(1-p)\g(q)}}
\frac{e^{i(p+q)\h+i(\jh_1+\jh_2)u_a}}{1-e^{iu_a}} \frac{e^{
-x(1-z)\p \ln D(z)}}{D(z)} \ , \ee where \be D(z) =
\frac{F(p,q,p+q,z)}{\G(p+q)\g(1-p)} - e^{iu_a} z^{1-p-q}
\frac{F(1-p,1-q,2-p-q,z)}{\G(-p-q)\g(q)}  \ , \ee and \be z =
\frac{|z_1-z_2|^2}{|z_1-\bar z_2|^2} \ , \hspace{1cm} x = (x_1-
x_2)(\bar x_1 - \bar x_2) \ . \ee In this case \be \l_{12} =
-h_1+h_2-\frac{h}{3}+(1-p-\jh_1)(p+q)-(\jh_1+\jh_2)p-q \ ,
\hspace{0.6cm} \l_{14} = 2h_1-\frac{h}{3}+p \, q + (\jh_1+\jh_2)p
-p \ . \label{aa68ffs1} \ee We consider finally \ba \la
\F^+_{p,\jh_1} \F^0_{s,\jh_2} \ra_a &=& \ib_a\sqrt{\frac{\pi}{\sin
\pi p}} \frac{e^{i p \h_a + i(\jh_1+\jh_2)u}}{1-e^{iu_a}}
e^{-px_1\bar x_1-\frac{s}{\sqrt{2}}\left ( x_1\bar x_2 + \bar x_1
x_2 +
\frac{x_1}{x_2}+\frac{\bar x_1}{\bar x_2} \right )}  \\
&&
z^{h_1-\frac{h}{3}-p\jh_2}(1-z)^{s^2-\frac{h}{3}}e^{\frac{s^2}{2}
(\psi(p)+\psi(1-p)-2\psi(1))}
e^{-x a(z) - \frac{b(z)}{x}} \sum_{n \in \mathbb{Z}} \left [ x
z^{-p} e^{iu_a} \right ]^n  \ , \nb \ea where $x = x_2 \bar x_2$
and \be a(z) = \frac{s^2}{2p} F(p,1,1+p,z) \ , \hspace{1cm} b(z) =
\frac{s^2}{2(1-p)} F(1-p,1,2-p,z) \ . \ee

We now turn to the
four-point open string amplitudes. The cross ratio in this case is
$z = \frac{t_{12}t_{34}}{t_{13}t_{24}}$.  The first amplitude we
consider is \be A_{n_1,n_2,s_1,s_2} \equiv \la
\psi^{ab}_{(p,\jh^{ab}-n_1)}(1)\psi^{ba}_{(-p,\jh^{ba}+n_2)}(2)
\psi^{aa}_{s_1}(3) \psi^{aa}_{s_2}(4) \ra \ , \hspace{0.6cm} p =
p_b-p_a > 0 \ .  \ee It describes the correlation between two
open strings stretched between the branes $a$ and $b$ and two open
strings ending on the brane $a$. We obtain  \ba && {\cal
A}_{n_1,n_2,s_1,s_2}  = \ib_a
e^{-px_1x_2-\frac{x_1}{\sqrt{2}}\left(\frac{s_1}{x_3}+\frac{s_2}
{x_4}\right)-\frac{x_2}{\sqrt{2}}(s_1x_3+s_2x_4)}
x_3^{n_1-n_2} (1-w)^{\ka_{12}}w^{\ka_{14}} \nb
\\ &&
\left [ \frac{s_2+s_1xw^{-p}}{\sqrt{2}} \right ]^{n_1-n_2}
e^{\frac{s_1^2+s_2^2}{4}[2\psi(p)-2\psi(1)-\pi \cot \pi
p_a]-\frac{s_1s_2}{2} \left [ \frac{x a(w)}{p} +\frac{w
b(w)}{x(1-p)}+\frac{\pi(\cot \pi p +\cot \pi p_a)w^p}{x} \right]}
 \nb \\ &&  \left [ \frac{\pi }{\sin \pi p_a} \right
]^{n_1-n_2+1} L^{n_1-n_2}_{n_2} \left [ \frac{\pi }{2}( \cot \pi p
+ \cot \pi p_a ) \left ( s_1^2+s_2^2+ \frac{s_1s_2
x}{w^{p}}+\frac{s_1s_2 u^p}{x} \right ) \right ] \ , \ea where $w
= 1 - z$ and  \be a(w) = F(p,1,1+p,w) \ , \hspace{1cm} b(w) =
F(1-p,1,2-p,w) \ . \ee Moreover $x = \frac{x_4}{x_3}$ and \be
\ka_{12} = \frac{s_1^2+s_2^2}{2}-\frac{h}{3} \ , \hspace{1cm}
\ka_{14} = h_1-\frac{h}{3}-p\jh_4 \ . \ee Another interesting
amplitude is \be A_{n_1,n_2,n_3,n_4} \equiv \la
\psi^{ab}_{(p,\jh^{ab}-n_1)}(1)\psi^{ba}_{(-p,\jh^{ba}+n_2)}(2)
\psi^{ab}_{(p,\jh^{ab}-n_3)}(3) \psi^{ba}_{(-p,\jh^{ba}+n_4)}(4)
\ra \ , \ee  which computes the correlation between four open
strings stretched between the $a$ and the $b$ branes. When
$n_1=n_3=n$, $n_2=n_4=m$ we can use the integral $(\ref{ui6})$ and
the result is \ba && {\cal A}_{n,m,n,m} =
e^{-p(x_1x_2+x_3x_4)}(x_1-x_3)^{2(n-m)}
z^{\ka_{12}}(1-z)^{\ka_{14}} \frac{n!}{m!} \frac{ \ib_a }{\sin \pi
p_a}
 \nb \\ && \left [ \frac{\sin \pi p }{R_-(z)}\right ]^{2(n-m)+1}
e^{xpz -\frac{xp}{2} -xz(1-z) \p \log c_1(z)+\frac{ x \sin \pi p
\sin \pi p_a}{\pi c_1(z)R_-(z)}} \label{anmnm}
\\
&& \sum_{l=0}^{m} (-1)^l
\frac{\G(m-l+1/2)\G(l+1/2)}{\G(l+1+n-m)(m-l)!} \left [
\frac{R_+(z)}{R_-(z)} \right ]^{2l} L_{2l}^{2(n-m)} \left [ -
\frac{ x \sin \pi p \sin \pi p_a \sin \pi p_b}{\pi R_+(z)R_-(z)}
\right ] \ , \nb \ea where \be R_{\pm}(z) = c_1(z) \sin \pi p_b
\pm c_1(1-z) \sin \pi p_a  \ , \hspace{1cm} c_1(z) = F(p,1-p,1,z)
\ . \ee Moreover $x = (x_1-x_3)(x_2-x_4)$ and \ba
\ka_{12} &=& h_1+h_2-\frac{h}{3}+p^2+(2\jh^{ab}-n-m)p-p \ ,  \nb \\
\ka_{14} &=& h_1+h_4-\frac{h}{3}+p^2+(2\jh^{ab}-n-m)p-p \ .
\label{kas1} \ea

For the $D2$ branes we showed that the $H_4$
boundary amplitudes can be considered as generating functions for
the open string amplitudes in models with intersecting branes. In
the case of the $S 1$ branes one can show in exactly the same way
that the boundary amplitudes are generating functions for the open
string amplitudes in models with magnetized branes. For instance,
the amplitude between the ground states of the $\hat {\cal H}_4$
representations, which can be easily extracted from
$(\ref{anmnm})$, coincides when $p_a=n=m=0$ with the amplitude
computed in \cite{narain}. This result is not surprising since the
magnetized and the intersecting branes in toroidal
compactifications are related by the operation of charge
conjugation, as it is also the case for the $S 1$ and the $D2$
branes in the Nappi-Witten gravitational wave.

For the sake of completeness, we also present a correlator with one
bulk and two boundary vertex operators. There are four types of
such correlators \be \la \F^+ \psi^{ab}\psi^{ba} \ra \ ,
\hspace{1cm} \la \F^+ \psi^{aa}\psi^{aa} \ra \ , \hspace{1cm} \la
\F^0 \psi^{ab}\psi^{ba} \ra \ , \hspace{1cm} \la \F^0
\psi^{aa}\psi^{aa} \ra \ , \ee and the example we chose is \ba &&
\la \F^+_{p,\jh}
\psi^{ab}_{q,\jh^{ab}-n_1}\psi^{ba}_{-q,\jh^{ba}+n_2} \ra = \ib_a
\sqrt{\frac{\pi}{\sin{\pi p}}} \left [ \frac{\pi}{\sin \pi p_a}
\right ]^{\n+1} \frac{e^{ip \h_a +i \jh u_a}}{1-e^{i \m u_a}}
e^{-px_1\bar x_1 - q x_3x_4} (x_1-x_3)^{\n}
\nb \\
&& z^{\ka_{12}}(1-z)^{\ka_{14}}e^{x \left [
g_1(z)+\frac{(1-z)^{p-q}}{c_1(z) B(z)}\right ]}
\frac{D(z)^{n_2}}{B(z)^{n_1+1}} L^{\n}_{n_2} \left [ \frac{\pi
\sin{\pi p_b} \, x(1-z)^{p-q}}{\sin{\pi q}\sin{\pi p_a}B(z)D(z)}
\right ] \ . \ea The cross-ratio in this case is $z =
\frac{(z_1-\bar z_1}{t_3-t_4}$ and we have also defined \ba B(z)
&=& \left [ \psi(1-p_a)-\psi(q) -\frac{\s(p)}{2} +\frac{i \pi
}{2}\cot{\pi p }\cot{\pi p_a} \right ] c_1(z)-c_2(z)
\ , \nb \\
D(z) &=& \left [ \psi(p_a)-\psi(1-q) -\frac{\s(p)}{2} +\frac{i \pi
}{2}\cot{\pi p }\cot{\pi p_a} \right ] c_1(z)-c_2(z) \ , \ea and
\be \ka_{12} = 2h_1-\frac{h}{3}+p^2+2\jh_1 p - p \ ,
\hspace{0.4cm} \ka_{14} = h_1+h_4-\frac{h}{3} + pq-(\jh^{ba}+n_2)p
+\jh_1 q - q \ . \ee

\section{Annulus amplitudes \label{annulus}}
\renewcommand{\theequation}{\arabic{section}.\arabic{equation}}

All the amplitudes we discussed so far, were defined on the disc.
The consistency conditions of a boundary CFT impose a constraint
also on the one-point functions of the boundary fields on the
annulus. When the boundary field is the identity, this additional
constraint reduces to the Cardy constraint, which interchanges the
two equivalent interpretations of the annulus diagram. The first is the
partition function of the boundary CFT, when time is running along
the boundary. The other is  as a tree-level amplitude for the propagation of
the bulk states when time is running
perpendicular to the boundary. In this second case the boundary conditions are
specified by the introduction of two boundary states.  Passing from one description to the
other requires an S modular transformation $t \mapsto 1/t$, where
$t$ is the modulus of the annulus.

The Cardy constraint has been
at the origin of many important insights concerning the operator
content of a rational boundary CFT \cite{Cardy,pss,zuber}.
It has
also been exploited for the investigation of some non-compact
models \cite{kutads,oog2,schads,schsyl,fnp}. The best way to
analyze the annulus constraint is to introduce characters for
the representations of the chiral algebra and study their modular
transformations. This is a non trivial problem for conformal
$\s$-models describing non-compact curved space-times or time-dependent branes
\cite{schsyl,fnp,malstrom,sugawara,characters}.

The characters of a generic representation $\a$ of the affine
$\hat {\cal H}_4$ algebra are defined as follows \be \z_\a(z,v|\t)
= {\rm tr}_{{\cal H}_\a} \left [ q^{L_0-\frac{c}{24}}e^{-2 \pi(z
J_0+vK_0)} \right ] \ . \ee For the $\Sigma_w[\hat V_{\pm
p,\jh}^{\pm}]$ representations we have \be
\z_{p,\jh;w}^{\pm}(z,v|\t) = \mp \frac{i
q^{h^{\pm}_{p,\jh;w}+\frac{w}{2}(1+w)}}{\h(\t)\th_1(z|\t)} e^{-2
\pi i z \left(\jh  \mp w \pm \frac{1}{2}\right) \mp i \pi w
\pm(p+w)v} \ . \ee Note that $\z^+_{p,\jh;-1-w} =
\z^-_{1-p,\jh;w}$, as required by (\ref{relpm}). It is therefore
convenient to express everything in terms of the $\z^+_{p,\jh;w}$
characters, letting $w \in \mathbb{Z}$. It is useful to define \be
r = p+w \ , \hspace{1cm} t = \jh - \frac{1-r}{2}-w \ , \ee  and
rewrite the character as \be \z_{p,\jh;w}^{+}(z,v|\t) = - \frac{i
e^{- i \pi w}}{\h(\t)\th_1(z|\t)} e^{- 2 \pi i \t r t} e^{- 2\pi i
r v -2 \pi i z \left ( t - \frac{r}{2} \right ) } \ . \ee We can
derive the following S modular transformation \be
\z^{+}_{p_1,\jh_1,w_1}\left(\frac{z}{\t},\frac{v}{\t}
|-\frac{1}{\t} \right ) = e^{-2 \pi i \frac{vz}{\t}}\sum_{w_2 \in
\mathbb{Z}} \int_0^1 dp_2 \int_{-\infty}^\infty d \jh_2 \
S_{(p_1,\jh_1,w_1);(p_2,\jh_2,w_2)} \
\z^{+}_{p_2,\jh_2,w_2}(z,v,\t) \ , \ee where \ba
S_{(p_1,\jh_1,w_1);(p_2,\jh_2,w_2)} &=& e^{-i \pi (w_1 - w_2-1)}
e^{ 2 \pi i (p_2+w_2)\left (\jh_1 -\frac{1-p_1-w_1}{2} - w_1
\right ) + 2 \pi i (p_1+w_1) \left ( \jh_2 - \frac{1-p_2-w_2}{2} -
w_2 \right )} \nb \\
&=& e^{-i \pi(w_1- w_2-1)}e^{2 \pi i r_1 t_2 + 2 \pi i t_1r_2} \ .
\label{dsm} \ea The characters of the $\Sigma_w[\hat V_{s,\jh}^0]$
representations are \be \z_{s,\jh;w}^{0}(z|\t) =
\frac{q^{\frac{s^2}{2}}}{\h^4(\t)} \sum_{k \in \mathbb{Z}} e^{2
\pi i k \jh} \d(z+w\t+k)  \ , \ee and they have the following
modular transformation properties \be
\z_{s_1,\jh_1;w_1}^{0}\left({z\over \t}|-{1\over \t}\right)
=\sum_{w_2\in Z}\int_{-1/2}^{1/2}d\jh_2~\int_{0}^{\infty}s_2~ds_2~
S^0_{(s_1,\jh_1;w_1);(s_2,\jh_2;w_2)}~\z_{s_2,\jh_2;w_2}^{0}(z|\t)
\ , \ee where \be S^0_{(s_1,\jh_1;w_1);(s_2,\jh_2;w_2)}=2\pi
i~e^{2\pi i(w_2\jh_1+w_1\jh_2)}~J_0(2\pi s_1s_2) \ . \label{csm}
\ee Note that when $s$ = $0$, all the representations with $\jh \in
\mathbb{R}$ are inequivalent and their base is one-dimensional. The
corresponding characters are \be \z_{0,\jh;w}^{0}(z|\t) =
\frac{e^{-2 \pi i \jh (z+w \tau)}}{\h^4(\t)}   \ . \ee

The torus vacuum amplitude of the Nappi-Witten gravitational wave
\cite{kk2,trusso} can be expressed in terms of the $\hat {\cal H}_4$
characters. The discrete series contribution to the closed string
partition function is given by \ba &&
Z^{+-}(\tau,z;\bar{\tau},\bar{z}) = {\rm Tr} \left [
 q^{L_0-\frac{c}{24}}e^{-2 \pi z J_0}
\bar{q}^{L_0-\frac{c}{24}}e^{-2 \pi \bar{z} J_0} \right ] = \int d
\jh \int_0^1 dp \sum_{w \in \mathbb{Z}}^\infty \left
|\z^+_{p,\jh;w}(z,\tau) \right|^{2}
 \nb \\
&=& \frac{1}{\left |\eta  \, \theta_1 \right |^2} \int d \jh
\int_0^1 dp \sum_{w \in \mathbb{Z}}^\infty e^{2 \pi \tau_2\left [
\left( p+\jh-\frac{1}{2}\right )^2-\left (\jh-\frac{1}{2}-w\right
)^2 \right ] +4 \pi {\rm Im}z \left ( \jh - w - \frac{1}{2} \right
)} \ . \ea Changing variable in each term of the sum $\hat l = \jh
- w - \frac{1}{2} $ and setting $\tilde p = p + w$ we obtain \be
Z^{+-} = \frac{1}{\left |\eta \, \theta_1 \right |^2} \int d \hat
l \int_{-\infty}^\infty d \tilde p \ e^{2 \pi \tau_2 [(\tilde
p+\jh)^2-\jh^2]+4 \pi {\rm Im} z \jh} =  \frac{i e^{2 \pi
\frac{({\rm Im}z)^2}{{\tau_2}}}}{2 \, \tau_2 \left | \eta \,
\theta_1 \right |^2} \ , \ee where we performed the rotation
$\tilde p \rightarrow i \tilde p$ in order to evaluate the
gaussian integral. Similarly the contribution of the type-0
characters is \be Z^0= V_2 \sum_{w\in
\mathbb{Z}}\int_{-1/2}^{1/2}d\jh~\int_0^{\infty} ds s \, \left |
\z_{s,\jh;w}^{0}(z|\t) \right|^{2}={V_2 \over 4 \pi
\tau_2\eta^4\bar \eta^4}\sum_{w,k\in
\mathbb{Z}}|\delta(z+w\tau+k)|^2 \ , \ee where $V_2$ is the volume
of the transverse plane. This additional volume factor is a
consequence of the fact that the states that belong to the
continuous representations can move freely in the transverse plane
and their wave functions are only delta function normalizable. On
the other hand the discrete states are confined around the origin
of the transverse plane and have normalizable wave functions.

We turn now to the annulus amplitudes. In the closed channel they
can be constructed using suitable boundary states. In the open
channel they  encode the spectrum of the open strings ending on
the two given branes. We will display the amplitudes in both
channels and investigate how they are related by the $S$ modular
transformation. In the following we will make several assumptions
and formal manipulations and the results we obtain even though
apparently consistent are not rigorous and we think that the
modular properties of these amplitudes deserve further study. We
will use the short-hand notation $\z_\a(z|\t) \equiv \z_\a$ and
$\z_\a(z/\t|-1/\t) \equiv \tilde \z_\a$. Boundary states for the
$H_4$ WZW model \footnote{Boundary states for the $H_4$ WZW model
were also considered in the recent paper \cite{hikida}.} can be
easily constructed using the bulk-boundary couplings derived in
section \ref{sc}.  The boundary state for a D2 brane localized at
$\chi$ is \ba |\chi \ra\ra &=& \ib_\chi
V_2^{1/4}\sum_{\jh=0,1/2}\int_0^\infty ds s \ {}^\chi B_{s,\jh}
|s,\jh;0\ra \nb \\ &=& \ib_\chi V_2^{1/4}\int_0^\infty ds \ \left
[ \cos (\sqrt{2} \chi s) |s,0;0\ra + i \sin (\sqrt{2} \chi s)
|s,1/2;0\ra \right ] \ .
 \ea
Here we have a factor of $V_2^{1/4}$, since the boundary continuous
representations correspond to one-dimensional waves. The annulus
amplitude in the closed string channel for two branes localized at
$\chi_1$ and $\chi_2$ then reads \ba \tilde{\cal A}_{\chi_1
\chi_2} &=& \ib_{\chi_1}\ib_{\chi_2} V_2^{1/2}\int_0^\infty ds
\sum_{\jh=0,1/2} \ {}^{\chi_1}B_{s,\jh}
\ {}^{\chi_2}B^*_{s,\jh} \ \tilde \z^0_{s,\jh}  \\
&=& \ib_{\chi_1}\ib_{\chi_2} V_2^{1/2}\int_0^\infty ds \left [\cos
(\sqrt{2}\chi_1 s) \cos (\sqrt{2} \chi_2 s) \tilde \z^0_{s,0;0}
+\right.\\
&&
+\left.
\sin (\sqrt{2} \chi_1 s)\sin (\sqrt{2} \chi_2 s) \tilde
\z^0_{s,1/2;0} \right ] \ . \label{tad2} \nb \ea On the other
hand, since the spectrum of the BCFT contains all the $\hat {\cal
H }_4$ representations, the annulus amplitude in the open channel
reads \be {\cal A}_{\chi_1 \chi_2} = \sum_{w=0}^\infty \int_0^1 dp
\int_{-\infty}^\infty d \jh \ \z^+_{p,\jh;w}+ 2 V_2^{1/2} \sum_{w
\in \mathbb{Z}} \int_{0}^\infty d \tilde s(w) \int_{-1/2}^{1/2} d
\jh \ \z^0_{s,\jh;w} \ , \ee where ${\tilde s}^2(w) = s^2 + \left
[ \frac{\sqrt{2} \s(w)}{2 \pi} \right ]^2$. We introduced the
quantity $\s(w) = |\chi_1-e^{i\pi w}\chi_2|$ in order to specify
the domain of the integral in $s$. The contribution of the
continuous representation can also be written as \be {\cal
A}_{\chi_1\chi_2} = \ib_{\chi_1}\ib_{\chi_2} V_2^{1/2} \sum_{w \in
\mathbb{Z}} \int_{\frac{\sqrt{2} \s(w)}{2\pi}}^\infty dt
\int_{-1/2}^{1/2} d \jh \frac{t}{\sqrt{t^2-\left[\frac{\sqrt{2}
\s(w)}{2\pi}\right]^2}} \z^0_{t,\jh;w} \ . \ee Note that the even
and odd spectral-flowed continuous representations appear with
different ranges of integration in the partition function, a
remnant after the Penrose limit of the different density of the
corresponding states for the $AdS_2$ branes in $AdS_3$
\cite{oog1}.

If we compute the modular transformation of the transverse annulus
using the $S$ matrix in $(\ref{csm})$, we correctly reproduce the
spectrum of the continuous representations in the direct annulus.
It is less clear how the discrete contribution should appear.
By comparing the normalization of the annulus amplitude
in the direct and in transverse channel we can fix the one-point
function of the identity. We have \be \ib_\chi = \sqrt{2} \ . \ee
The discussion for the $S1$ branes is similar. The boundary state
for an $S 1$ brane labeled by $(u,\h)$ is \be |u,\h\rangle\rangle
= \ib_{(u,\h)} \sum_{w\in\mathbb{Z}}^\infty \left [ \int_0^1 dp
\int_{-\infty}^\infty d \jh  \ {}^{u,\h}B_{p,\jh;w}|p,\jh,w\rangle
+ V_2  \int_{-1/2}^{1/2}d\jh \int_0^\infty ds s \
{}^{u,\h}B_{s,\jh;w}|s,\jh,w \rangle \right ] \ . \ee The
bulk-boundary couplings with the identity are \be
{}^{u,\h}B_{p,\jh;w} = \sqrt{\frac{\pi}{\sin \pi p}} \
\frac{e^{2i(p+w)\h+iu(\jh-w)+i \pi w}}{1-e^{iu}} \ , \ee and we
recall the relations  $u = 2 \pi(q+a)$ and $2 \h = \pi(2q+2 \hat
l-1)$. The annulus amplitude in the closed channel reads \ba
\tilde {\cal A}_{12} &=& \ib_a\ib_b \sum_{w \in \mathbb{Z}}
\int_0^1 dp \int_{-\infty}^\infty d\jh \ \frac{\pi}{\sin \pi p} \
\frac{e^{2i(p+w)(\h_1-\h_2)+i(\jh-w)(u_1-u_2)}}{(1-e^{iu_1})(1-e^{-iu_2})}
\tilde \z^+_{p,\jh;w} \nb \\
&+& \ib_a\ib_b V_2 \sum_{w \in \mathbb{Z}} \int_{-\infty}^{\infty}
d \jh \ \frac{e^{i(\jh-w)(u_1-u_2)+2 i w
(\h_1-\h_2)}}{16\sin^2(u_1/2)\sin^2(u_2/2)} \tilde \z^0_{0,\jh;w}
\ , \label{s1at} \ea where we used the fact that  $\jh \in
\mathbb{R}$ also for the continuous representations when $s=0$. In
the previous expression we separated the contribution of the
discrete and the continuous representations, which are weighted by
different volume factors. In fact the first term has to be
considered as a regularized term without the divergences due to
the almost delocalized states in the transverse plane with $p \sim
w$, $ w \in \mathbb{Z}$ and the second term as the term containing
all these divergences, since the continuous representations
capture the  behavior of the discrete representations when $p$
approaches an integer value.
What this means is that when manipulating the term
containing the discrete representations,
whenever a constraint arises forcing to evaluate it at the
boundary of the interval $0 < p <1$, the corresponding
contribution should be discarded and only the contribution coming
from the continuous representations in the second term of (\ref{s1at})
retained. Equivalently, we
could keep only the first term in $(\ref{s1at})$ and take into
account the divergent behavior of the integrand when $p$ becomes
an integer. We will show in the following that both points of view
lead to the same result.

Let us transform the amplitude $(\ref{s1at})$ to the open string
channel using the $S$ matrix in $(\ref{dsm})$, $(\ref{csm})$
writing for instance \be \tilde \z^+_{p,\jh;w} = \sum_{a \in
\mathbb{Z}} \int_0^1 dp \int d \jh \ S_{(p,\jh;w);(q,\hat l;a)} \
\z^+_{q,\hat l;a} \ . \ee We perform first the integration over
$\jh$ which gives the constraint \be q + a = \frac{u_2-u_1}{2\pi}
\ . \label{annel1} \ee Let us suppose for the moment that $ u_2 -
u_1 \notin 2 \pi \mathbb{Z}$ so that only the discrete
representations contribute. We can recombine the integral over $p$
and the sum over $w$ in a single integral over $\tilde p = p + w$
\be
 {\cal A}_{12} = - \frac{\ib_a\ib_b
 e^{\frac{i (u_{1}-u_2)}{2}}}{(1-e^{i u_1})(1-e^{-i u_2})}
\int d \hat l \int_{- \infty}^\infty d \tilde p \frac{\pi}{\sin
\pi \tilde p} e^{i \tilde p(2 \h_{12}+2 \pi (l + q) - \pi) }
\z^+_{q,\hat l;a} \ . \ee We now need a prescription to perform
the integral over $\tilde p$. The prescription that reproduces in
the open channel the spectrum of the boundary operators is the
following \be \tilde p \rightarrow \tilde p + (-1)^a i \e \ . \ee
Consider for instance the case $a$ even. We can expand \be
\frac{\pi}{\sin \pi( \tilde p  + i \e)} = - 2 \pi i
\sum_{n=0}^\infty e^{2 \pi i p \left ( n + \frac{1}{2} \right )} \
, \ee and then perform the integral over $\tilde p$ that gives the
constraint \be \hat l = - \jh_1 + \jh_2 - n \ , \hspace{1cm} n \in
\mathbb{N} \ . \ee Therefore \be {\cal A}_{12} = \frac{2 \pi i
\ib_a\ib_b e^{\frac{i (u_{1}-u_2)}{2}}}{(1-e^{i u_1})(1-e^{-i
u_2})} \sum_{n=0}^\infty \z^+_{q,-\jh_1+\jh_2-n;a} \ .
\label{ann40} \ee This is precisely the expected result for the
annulus amplitude in the open string channel. The reason is that when $u_2-u_1 =
2\pi(q+ a)$ with $0 < q < 1$ and $\h_2-\h_1 = \pi(q+\jh)$, the open
string spectrum only contains discrete spectral-flowed
representations \be {\cal A}_{12} = \sum_{n=0}^\infty
\z^+_{q,\jh-n;a} \ . \label{ann40bis} \ee From the comparison of
the amplitudes in the open and closed string channel, we can fix
the last structure constants required for the complete solution of
the boundary $H_4$ model, namely the one-point functions of the identity.
We obtain \be \ib_{u,\h} = \sqrt{\frac{2}{\pi}}\sin \left
(\frac{u}{2} \right ) \ . \ee

When $u_2 - u_1 = 2 \pi k$ with $k \in \mathbb{Z}$, the constraint
$(\ref{annel1})$ has two solutions, $q = 0$, $a = k$ and $q=1$, $a
= k-1$. We have two options. The first is to think as proposed
before, that the term that contains the discrete representations
is a regularized term. In this case, it does not contribute while
the term containing the continuous representations after the
modular transformation gives \be {\cal A}_{12} = - \frac{2 \pi i
\ib_a\ib_b e^{\frac{i (u_{1}-u_2)}{2}}}{(1-e^{i u_1})(1-e^{-i
u_2})} \ \frac{V_2}{\sin^2 \frac{u}{2}} \int_0^\infty ds s \
\z^0_{s,\jh_2-\jh_1;k} \ . \label{s1cont} \ee This is the expected
result, since whenever $u_2-u_1=2 \pi k$ and $\h_2-\h_1 = \pi \jh$
the open string states belong to the continuous spectral-flowed
representations \be {\cal A}_{12} = \frac{V_2}{\sin^2 \frac{u}{2}}
\int_0^\infty ds s \ \z^0_{s,\jh;k} \ . \ee

The second option is to think of the continuous representations as
already included in the divergent behavior of the discrete
representations for $p$ close to an integer. It is instructive to
derive $(\ref{s1cont})$ once more, adopting this point of view,
and  to show explicitly its equivalence with the previous one. In
this case, we have to extract $(\ref{s1cont})$ from the integral
over the discrete representations when the integrand is evaluated
at the extrema of the interval. Performing the same steps as
before and keeping both contributions $q = 0$, $a = k$ and $q=1$,
$a = k-1$, we obtain \be {\cal A}_{12} = \frac{2 \pi i \ib_a\ib_b
e^{\frac{i (u_{1}-u_2)}{2}}}{(1-e^{i u_1})(1-e^{-i u_2})}
\sum_{n=0}^\infty (\z^+_{0,-\jh_1+\jh_2-n;k} -
\z^+_{1,-\jh_1+\jh_2+n;k-1}) \ . \ee Using the explicit form of
the $\z^+$ characters and sending $k \rightarrow k + i \e$, the
previous expression can be rewritten as follows for $\e \sim 0$
\be {\cal A}_{12} = \frac{2 \pi i \ib_a\ib_b e^{\frac{i
(u_{1}-u_2)}{2}}}{(1-e^{i u_1})(1-e^{-i u_2})} \frac{1}{\e \tau}
\frac{1}{\h^4(\tau)}\sum_{m \in \mathbb{Z}} e^{2\pi
i(\jh_2-\jh_1)m} \d(z+k \tau +m) \ . \ee We should now interpret
the divergence $1/\e$ as due to the infinite volume of the
transverse plane (consistently with the power we have of the
modular parameter which is the one commonly associated with two
non-compact directions)  \be \lim_{\e \rightarrow 0} \frac{1}{\e}
= \frac{V_2}{\sin^2 \frac{u}{2}} = V_2^{{\rm open}}\ , \ee which
is the volume measured using the open string metric \cite{swnc} \be
G_{op}^{-1} \equiv (g_{cl}+{\cal F})^{-1}g_{cl}(g_{cl}-{\cal
F})^{-1} \ . \ee In this way, we obtain again the amplitude
displayed in $(\ref{s1cont})$.

As we mentioned earlier, our aim in this section is not to provide a rigorous
discussion of the modular transformation properties of the annulus
amplitudes, but rather to give a plausible  suggestion about how the
standard open-closed duality should work for the branes of the
$H_4$ model.

\subsection{Contraction of the BCFT}
\renewcommand{\theequation}{\arabic{section}.\arabic{subsection}.\arabic{equation}}
\setcounter{equation}{0}

The Penrose limit that connects the Nappi-Witten gravitational
wave and $\mathbb{R}\times S^3$ or $AdS_3\times S^1$ can also be
extended to the branes contained in these space-times, as
described in section \ref{a}. In \cite{dk} we gave a detailed
description of the world-sheet equivalent of the Penrose limit in
 space-time. This is the contraction of the $\mathbb{R} \times
SU(2)_k$ WZW model to the $H_4$ WZW model. It is interesting to
perform the same contraction for the boundary CFT. Here we
shall comment on how the BCFT describing the $H_4$ branes arises as
the limit of the BCFT describing the $S^2$ branes in $S^3$. We
first recall how to derive the affine $\hat {\cal H}_4$ characters
from the contraction of the $U(1) \times SU(2)_k$ characters
\cite{dk}. It is convenient to write the latter as follows \be
\chi^k_l = \sum_{n \in \mathbb{Z}} \left ( \chi^+_{l,n}+
\chi^-_{l,n} \right ) \ , \ee where $l$ is the spin of the
representations and \be \chi^\pm_{l,n}(z|\tau) = \mp \frac{1}{i
\theta_1} e^{2 \pi i (k+2) \left [ \left
(n+\frac{2l+1}{2(k+2)}\right )^2\tau \mp \left( n
+\frac{2l+1}{2(k+2)}\right )z \right ]} \ . \ee The $U(1)$
characters are \be
 \psi_Q(z|\tau) =
\frac{e^{-2 \pi i \tau \frac{Q^2}{k}+2 \pi i z Q}}{\eta} \ . \ee

The characters of the original CFT become the $H_4$ characters if
we send the level $k$ to infinity and scale simultaneously the
spin $l$  and the charge $Q$ in a correlated way \be
\psi_Q(-z-2v/k|\tau) \chi^\pm_{l,n}(z|\tau) \rightarrow
\z_\a(z,v|\tau) \ . \ee More precisely for the discrete
representations we obtain \ba \psi_{-k(n+\frac{p}{2})}\chi^+_{l,n}
&\rightarrow& \z^-_{p,-\jh;2n}\ ,
\hspace{2.2cm} n \ge 0 \ , \nb \\
\psi_{-k(-m+\frac{p}{2})}\chi^+_{l,-m} &\rightarrow&
\z^+_{1-p,-\jh;2m-1} \ ,
\hspace{1.3cm} m \ge 1 \ , \nb \\
\psi_{k(n+\frac{p}{2})}\chi^-_{l,n} &\rightarrow& \z^+_{p,\jh;2n}
\ ,
\hspace{2.5cm} n \ge 0 \ , \nb \\
\psi_{k(-m+\frac{p}{2})}\chi^-_{l,-m} &\rightarrow&
\z^-_{1-p,\jh;2m-1} \ , \hspace{1.6cm} m \ge 1 \ ,  \label{d28}
\ea with $l = \frac{k}{2}p-\jh$. The characters for the continuous
representations require a different scaling, namely  \ba
\psi_{-kn+\jh}\left ( \chi^+_{l,n}+\chi^-_{l,-n} \right )
&\rightarrow& \z^0_{s,\jh;-2n}
\ , \hspace{1.4cm} n \in \mathbb{Z} \ , \hspace{1cm} l = \sqrt{\frac{k}{2}}s
\label{d2828} \\
\psi_{-k(n+1/2)+\jh}\left ( \chi^+_{l,n}+\chi^-_{l,-n-1} \right )
&\rightarrow& \z^0_{s,\jh;-2n-1} \ , \hspace{1cm} n \in \mathbb{Z}
\ , \hspace{1cm} l = \frac{k}{2} - \sqrt{\frac{k}{2}}s\nb \ . \ea
In the following, we will only discuss the contraction
involving the $SU(2)_k$ WZW model. Similar relations however, can be
written for the $SL(2,\mathbb{R})_k$ characters. The boundary
state for an $S^2$ brane in $S^3$ reads \be | i, R \rangle \rangle
= \sum_{l=0}^{\frac{k}{2}}
\frac{S_{il}}{\sqrt{S_{0l}}}D^l_{m,n}(\a,\b,\g)| l,m,n \rangle  \
. \label{d10} \ee Here, we are considering the general case where
the gluing condition (\ref{a04}) also involves the adjoint action
of a group element $R(\a,\b,\g) \in SU(2)$. The
$D^l_{mn}(\a,\b,\g)$ are the matrix elements of $R$ in the spin
$l$ representation and can be expressed in terms of the Jacobi
polynomials \be D^l_{mn}(\a,\b,\g) = i^{m-n}e^{-i(n+m)\a
-i(n-m)\b}P^l_{mn}(\cos \g) \ . \ee Finally $S_{il}$ is the
modular S matrix of the affine $SU(2)_k$ algebra \be S_{il} =
\sqrt{\frac{2}{k+2}} \sin \left [ \frac{\pi}{k+2}(2i+1)(2l+1)
\right ] \ . \ee  Using the fact that for $R=1$ the label $i$ of
the branes is related to the coordinate $\psi$ by \be \psi_i =
\frac{\pi (2i+1)}{k+2} \ , \hspace{1cm} i = 0, ..., \frac{k}{2} \
, \label{d11} \ee we can derive the relation between the $S 1$
brane parameters and the quantum numbers of the $\hat{\cal H}_4$
representations that is inherited from the original relations
between the $S^2$ brane parameters and the spin of $SU(2)$. The
results are summarized in the following table. In the first
column we listed the discrete $H_4$ representations. In the second
and in the third column we show the labels of the corresponding branes.
Finally,
in the fourth column, we list the $U(1) \times SU(2)_k$ representations they
originate from in the Penrose limit \ba &\z^+_{p,\jh,2n}&
\hspace{0.4cm} \m u = 2 \pi(\m p +2 n ) \hspace{1.4cm} 2 \h =
\pi(2 \m p +2 \jh -1) \hspace{0.6cm}
\psi_{k(n+\frac{p}{2})} \, \chi_{\frac{k}{2}p-\jh} \ , \nb \\
&\z^+_{p,\jh,2m-1}& \hspace{0.4cm} \m u = 2 \pi(\m p +2 m-1 )
\hspace{0.6cm} 2 \h = - \pi(2 \m p +2 \jh -1)  \hspace{0.4cm}
\psi_{k\left(m-\frac{1-p}{2}\right)} \, \chi_{\frac{k}{2}(1-p)+\jh} \ , \nb \\
&\z^-_{p,\jh,2n}& \hspace{0.4cm} \m u = - 2 \pi(\m p +2 n )
\hspace{1.1cm} 2 \h = - \pi(2 \m p -2 \jh -1)  \hspace{0.4cm}
\psi_{-k(n+\frac{p}{2})} \, \chi_{\frac{k}{2}p+\jh} \ , \nb \\
&\z^-_{p,\jh,2m-1}& \hspace{0.4cm} \m u = -2 \pi(\m p +2 m-1 )
\hspace{0.4cm} 2 \h = \pi(2 \m p -2 \jh -1)  \hspace{0.6cm}
\psi_{-k\left(m-\frac{1-p}{2}\right)} \,
\chi_{\frac{k}{2}(1-p)-\jh} \ . \nb \label{bpqnr} \ea In a similar
fashion, one may show that the label $\chi$ of the $D2$ branes is
related to the label $l$ of the original $S^2$ brane by \be l =
\frac{k}{4} + \sqrt{k} \frac{\chi}{2 \pi} \ . \ee We may now write
down the annulus amplitudes for the brane configurations in
$\mathbb{R} \times S^3,$ whose Penrose limit is one of the branes
in $H_4$, as explained in section \ref{a}. We will show that in
the limit, the direct and the transverse annulus amplitudes become
the corresponding amplitudes for the $H_4$ model we discussed in
the previous section. In order to do this, we have to first
understand for each amplitude, how we can scale the quantum numbers
of the original representations and then restrict our attention to
states that have finite charges and conformal dimension in the
limit.

For the D2 branes, we start with a brane with Neumann boundary
conditions along the time direction. The original annulus
amplitude in the open string channel is \be {\cal A}_{l_1,l_2} =
\int_{-\infty}^\infty dQ \sum_{l=|l_1-l_2|}^{{\rm
min}(l_1+l_2,k-l_1-l_2)} \psi_Q \, \chi_l \ . \label{d30} \ee The
brane labels have to be scaled in the limit as  $l_i =
\frac{k}{4}+\frac{\sqrt{k}}{2 \pi} \, \chi_i$, $i=1,2$ and
therefore the range of the possible $SU(2)_k$ representations is
\be \frac{\sqrt{k}}{2 \pi} \, |\chi_1-\chi_2| \le l \le
\frac{k}{2} - \frac{\sqrt{k}}{2 \pi} \, |\chi_1+\chi_2| \ . \ee
Since there is a lower bound in the range of $l$, we have to
slightly change the way we scale the spin in the Penrose limit.
For the representations $\z^+_{p,\jh;2a}$ we simply set \be l=
\frac{k}{2}p+\sqrt{k}\frac{|\chi_1-\chi_2|}{2\pi} - \jh \ ,
\hspace{1cm} Q = - k\left(w
+\frac{p}{2}\right)+\sqrt{k}\frac{|\chi_1-\chi_2|}{2\pi} \ , \ee
and similarly for all the other discrete representations. For the
continuous representations, the lower bound in $l$ shifts the
lower bound for the integral in $s$. It is easy to se that \be s
\ge \frac{|\chi_1-\chi_2|}{\sqrt{2}\pi} \ , \hspace{0.6cm} {\rm
for} \ \ \z^0_{s,\jh;2w} \ , \hspace{0.8cm} s \ge
\frac{|\chi_1+\chi_2|}{\sqrt{2}\pi} \ , \hspace{0.6cm} {\rm for} \
\ \z^0_{s,\jh;2w+1} \ . \ee We observe that the different behavior
of the even and odd spectral-flowed continuous representations,
arises also in a very transparent way from the contraction of
$SU(2)_k$. Note that the minimal conformal dimension for the
vertex operators $\psi^{\chi_2\chi_1}_{s,\jh;0}$ is $h =
\frac{(\chi_1-\chi_2)^2}{2\pi^2}$, which can be ascribed to the
tension of the string stretched between the two branes, as
expected. Since the original amplitude contains arbitrary $U(1)$
charges, in the limit we obtain an amplitude that contains all
possible $H_4$ representations \be {\cal A}_{\chi_1 \chi_2} \sim
\sum_{w=0}^\infty \int^1_0 dp \int_{-\infty}^\infty d \jh \
[\z^+_{p,\jh;w} + \z^-_{-p,\jh;w}] + \sqrt{2}V_2^{1/2}\sum_{w \in
\mathbb{Z}} \int_0^\infty d\tilde s \int_{-1/2}^{1/2} d \jh \
\z^0_{s,\jh;w} \ . \ee In the transverse channel we can reason in
the same way. The original amplitude is \be \tilde{\cal A}_{l_1
l_2} = \tilde\psi_0 \sum_{l=0}^{k/2}
\frac{S_{l_1l}S_{l_2l}}{S_{0l}} \tilde\chi_l \ . \ee Since now all
the states have zero $U(1)$ charge, according to
(\ref{d28}-\ref{d2828}) we can only obtain in the limit the
highest-weight continuous representations.

The discussion for the $S 1$ branes is similar. We label the
branes with their position in time $u$ and with the $SU(2)$ spin
$l$. In the open string channel, the original amplitude is \be
{\cal A}_{(u_1,l_1)(u_2,l_2)} = \psi_{\frac{u_2-u_1}{4 \pi}}
\sum_{l=|l_1-l_2|}^{{\rm min}(l_1+l_2,k-l_1-l_2)}  \chi_l \ . \ee
We scale $l_i = \frac{k}{2}p_i - \jh_i$, $i=1,2$. As before we
have to distinguish  two cases. When $u_2-u_1$ is not an integer
multiple of $2 \pi $, we may write \be \frac{u_2-u_1}{4 \pi} =
k\left ( w +\frac{p}{2} \right ) \ , \ee with $0<p<1$. We then
have to scale $l$ as $l = \frac{k}{2}(p_2-p_1) -\jh$ and the
possible values of $\jh$ follow from the original range of $l$ \be
\jh_2-\jh_1 \ge \jh \ge -\infty \ , \ee in integer steps and
therefore $\jh = \jh_2-\jh_1-n$, $n \in \mathbb{N}$, as expected.
On the other hand, when $u_2-u_1=2 \pi (kw-\jh)$, we may use the
relations $ (\ref{d2828})$. In the limit, we obtain an annulus
amplitude that only involves the continuous representations.

In the closed string channel, the original amplitude is \be \tilde
{\cal A}_{(u_1,l_1)(u_2,l_2)} =  \int^\infty_{-\infty} dQ \, e^{i
Q(u_2-u_1)} \tilde \psi_Q \sum_{l=0}^{k/2}
\frac{S_{l_1l}S_{l_2l}}{S_{0l}} \tilde \chi_l \ . \ee As in the
amplitude (\ref{d30}), we have again arbitrary $U(1)$ charges. We
thus obtain in the limit all possible $H_4$ representations.
Therefore, the annulus amplitudes, both in the closed and in the
open string channel, reproduce in the limit the results we expect
for the $D2$ and the $S 1$ branes. It would be very interesting to
pursue this line of thinking, in order to gain a more detailed
understanding of the contraction of the boundary CFT.

\section{The DBI approach \label{dbi}}

\renewcommand{\theequation}{\arabic{section}.\arabic{equation}}

We will study here the DBI approach for the branes described in
this paper. This approach although in most cases approximate, has
the advantage of an obvious geometric interpretation. We will also
be able to provide an independent confirmation of the spectrum of
fluctuations for the $H_4$ branes and justify some of the
assumptions made during the solution of the BCFT. In the bosonic
case the lowest state is the tachyon. In anticipation of the
supersymmetric case we will use the bosonic part of the
supersymmetric DBI action that describes the dynamics of the
``massless" modes. To simplify the formulae, we use here the
background (\ref{a01},\ref{a01a}) with $\mu=2$  so that the metric
and antisymmetric tensor read \be
ds^2=-2dudv-r^2du^2+dr^2+r^2d\theta^2\sp B_{r\theta}=2ur \ . \ee
We can put back $\mu$ by rescaling $u\to \mu u/2$, $v\to 2v/\mu$.

\subsection{The $S 1$ branes and the spectrum of their fluctuations}
\renewcommand{\theequation}{\arabic{section}.\arabic{subsection}.\arabic{equation}}
\setcounter{equation}{0}

We will find a class of solutions to the DBI equations that will
contain
 as special cases the   $S 1$ branes discussed in this paper.
The $S 1$  will have Dirichlet boundary conditions
on the $u,v$ coordinates. We choose a static gauge where the brane
world-volume is parameterized by $r,\theta$. The induced metric
and antisymmetric tensor is \be \hat g_{rr}=1-2u'v'-r^2u'^2\sp
\hat g_{\theta\theta}=r^2-r^2\dot u^2-2\dot u\dot v \ , \ee \be
 \hat g_{r\theta}=-u'\dot v-\dot u v'-r^2u'\dot u \sp \hat B_{r\theta}=2ur
\ . \ee In the formulae above, a dot stands for a $\theta$
derivative and a prime for an $r$ derivative.

We can directly evaluate the Nambu-Goto-Dirac-Born-Infeld Lagrangian as
\be L=\sqrt{\det(\hat g+\hat B+F)}=\sqrt{\hat g_{rr}\hat
g_{\theta\theta}-\hat g_{r\theta}^2+(2ur+F_{r\theta})^2} \ ,
\label{ll} \ee where $F_{r\theta}$ is the world-volume gauge field
strength. The equations of motion for the gauge field can be
integrated to \be {2ur+F_{r\theta}\over L}={E\over 2}\to
2ur+F_{r\theta}={Er\over \sqrt{4-E^2}}\sqrt{1-2u'v'-r^2u'^2} \ ,
\ee where E is a constant (the ``electric field"). The $u,v$
equations are \be \p_{\theta}{(r^2\dot u+\dot v)\hat
g_{rr}-(r^2u'+v')\hat g_{r\theta}\over L}+\partial_r{-(r^2\dot
u+\dot v)\hat g_{r\theta} +(r^2u'+v')\hat g_{\theta\theta}\over
L}=-Er \ , \ee \be
\partial_{\theta}{\dot u\hat g_{rr}-u'\hat g_{r\theta}\over
L}-\p_r {\dot u\hat g_{r\theta}-u'\hat g_{\theta\theta}\over L}=0
\ . \ee We will from now on  consider a rotationally invariant
ansatz. Dropping $\theta$-derivatives the equations simplify to
\be (r^2u'+v')\hat g_{\theta\theta}=(-{1\over 2}Er^2+A)L\sp u'\hat
g_{\theta\theta}={B\over 2}L \ , \label{sol0}\ee with $A,B$
integration constants and $L$ from (\ref{ll}) given by \be
L={2r\over \sqrt{4-E^2}}\sqrt{1-2u'v'-r^2u'^2} \ . \label{sol1}\ee
Massaging (\ref{sol0},\ref{sol1})  we obtain \be r^2u'^2={B^2\over
4-E^2}(1-2u'v'-r^2u'^2)\sp r^2+{v'\over u'}={-Er^2+2A\over B} \ ,
\label{sol2}\ee \be v'={2A-(B+E)r^2\over B}u'\sp u'^2={B^2\over
(4-(B+E)^2)r^2+4AB} \ . \label{sol3}\ee We will  look here for
solutions where $u$ is a constant corresponding to the symmetric
$S 1$ branes discussed in this paper. $u=$constant implies that
$B=0$ and \be v'=\left(-{1\over 2}Er+{A\over r}\right){2\over
\sqrt{4-E^2}}\Rightarrow v=v_0-{Er^2\over 2\sqrt{4-E^2}}+{2A\over
\sqrt{4-E^2}}\log r \ . \ee Since the class variable is
$\xi=2v\sin
u-r^2\cos u$,  we learn that the symmetric $S 1$ branes have also
$A=0$. Comparison with (\ref{radius}) and (\ref{sfluxb}) gives
 \be \cot u_0=-{E\over \sqrt{4-E^2}}\sp \xi=2v_0\sin u_0
\ . \label{electricu}\ee The fluctuations around the classical
embedding are in one-to-one correspondence with on-shell open
marginal deformations. Although for a single $S 1$ brane this is
not very rich, it is still useful to verify it explicitly. The
richer case of two branes at a non-zero distance in light-cone is
much harder to analyse and we will not do it here.

In appendix (\ref{adbi}) we analyse the action for the
fluctuations $u,V,F$ around the S1 solution that turns out to be
\be L_2={\sqrt{4-E^2}\over 16r}\left[-(4-E^2)\left(F+{8ur\over
4-E^2}\right)^2+
 8(\dot
u\dot V+r^2u'V')+32{r^2u^2\over 4-E^2} \right] \ . \ee The
equations of motion that ensue are \be \Box u=0\Rightarrow {1\over
r}(ru')'+{1\over r^2}\ddot u=0\label{laplace} \ , \ee \be \Box
V={1\over r}(rV')'+{1\over r^2}\ddot V=-{4\over 4-E^2}(2u+C) \ ,
\label{laplace2} \ee with the gauge field satisfying \be
F+2ur-{E^2\over 4-E^2}r^2u'={2C\over (4-E^2)}r \ee with $C$ a
constant.

The regular solution of (\ref{laplace}) is $u=u_0$ constant. On
the other hand,
 the solution of  (\ref{laplace2}) is \be
V=V_0-{2u_0+C\over 4-E^2}r^2 \ee In order to be regular at
$r=\infty$, the electric field fluctuation and the $u$ fluctuation
must be related by \be C=-2u_0\;\;. \label{c}\ee This is indeed
implied by (\ref{electricu}). In the non-symmetric case where
(\ref{electricu}) is no longer valid, (\ref{c}) must still be in
effect for the fluctuations to be continuum normalizable and thus
physical states of the theory.

 The
two physical states obtained
  correspond to $K_{-1}|s=0>$ and
$J_{-1}|s=0>$ in the bosonic case and $\psi^K_{-{1\over 2}}|s=0>$
and $\psi^J_{-{1\over 2}}|s=0>$ in the supersymmetric case \cite{book}
in
accordance with the BCFT discussion. Note that here, unlike the
D-brane case, including the contribution of the additional
coordinates of the ten-dimensional string theory does not change
our results. The reason is that the world-sheet is Euclidean and
the physical states conditions are very restrictive, implying the
vanishing of all momenta.

\subsection{The D2 branes and the spectrum of their fluctuations}
\renewcommand{\theequation}{\arabic{section}.\arabic{subsection}.\arabic{equation}}
\setcounter{equation}{0}

The cartesian coordinates on the plane $(x,y)$ are more convenient
here. The metric and antisymmetric tensor read \be
ds^2=-2dudv-(x^2+y^2)du^2+dx^2+dy^2\sp B_{xy}=2u \ . \ee Putting
Dirichlet boundary conditions on $y$ we obtain the following
induced metric \be d\hat
s^2=(-x^2-y^2+y_u^2)du^2+y_v^2dv^2-2(1-y_uy_v)dudv+
2y_uy_xdudx+2y_vy_xdvdx+(1+y_x^2)dx^2
\ , \ee while \be \hat B=2u(y_u dx\wedge du+y_vdx\wedge dv) \ .
\ee The action  is \be S_{D2}=\int dxdudv\sqrt{1+L_2} \ , \ee with
\be L_2=-2y_uy_v+(x^2+y^2)(y_v^2+(-2uy_v+F_{vx})^2)+y_x^2-
2(-2uy_u+F_{ux})(-2uy_v+F_{vx}) \ee
$$
-F_{vx}^2y_u^2-F_{uv}^2(1+y_v^2)+2F_{ux}F_{vx}y_uy_v+
F_{ux}^2y_v^2+2F_{uv}F_{ux}y_vy_x-2F_{uv}F_{vx}y_uy_x
\ . $$ We will now search for solutions where the D2 brane is
sitting at $y=y_0$ constant that contain the  symmetric D2
solutions studied in this paper.

Setting $y=y_0$ we obtain the following equations to be solved \be
\p_u{F_{uv}\over L}+\p_x \left [(y_0^2+x^2){F_{vx}\over
L}-{F_{ux}\over L} \right ]=0 \ , \label{f1}\ee \be
\p_u{F_{vx}\over L}-\p_v\left [(y_0^2+x^2){F_{vx}\over
L}-{F_{ux}\over L} \right ]=0 \ , \label{f2}\ee \be
\p_v{F_{uv}\over L}+\p_x{F_{vx}\over L}=0 \ , \label{f3}\ee with
\be L=\sqrt{1-F_{uv}^2-2F_{ux}F_{vx}+(x^2+y_0^2)F_{vx}^2} \ , \ee
while the y equation gives\footnote{There is another possibility
here, namely $F_{vx}=-2/y_0\ $, but this does not correspond to a
symmetric solution.} \be F_{vx}=0 \ . \label{f4} \ee Equations
(\ref{f1}-\ref{f3}) are then solved by \be
F_{uv}=f_{uv}=constant\sp F_{ux}=-y_0+f_{ux}=constant \ . \ee The
symmetric solution corresponds to $f_{ux}=f_{uv}=0$. The gauge
field can be dualized to a scalar here as follows \be {F_{uv}\over
L}=\p_x A\sp {F_{ux}\over L}=\p_u A-(y_0^2+x^2)\p_vA\sp
{F_{vx}\over L}=-\p_vA \ . \ee Solving for the gauge field
strength we obtain \be F_{uv}={\p_x A\over \hat L}\sp
F_{ux}={\p_uA-(x^2+y_0^2)\p_vA\over \hat L}\sp F_{vx}=-{\p_vA\over
\hat L} \ . \ee Such expressions solve equations
(\ref{f1})-(\ref{f2}) but now the Bianchi identity gives \be
\p_x{\p_xA\over \hat L}-\p_u{\p_v A\over \hat
L}+\p_v{-\p_uA+(x^2+y_0^2)\p_vA\over \hat L}=0 \ , \label{bia}\ee
where \be \hat
L=\sqrt{1-2\p_uA\p_vA+(\p_xA)^2+(x^2+y_0^2)(\p_vA)^2} \ . \ee
Using (\ref{f4}), (\ref{bia}) becomes \be \p_x{\p_xA\over \hat
L}=0\sp \p_{v}A=0 \ . \ee Thus the  $A$ corresponding to our
previous solution is \be A={f_{uv}~x+(f_{ux}-y_0)u\over
\sqrt{1-f_{uv}^2}} \ . \ee We will now study the spectrum of
fluctuations around  the simplest solution $y=y_0$,
$F_{vx}=F_{uv}=0$, $F_{ux}=-y_0$. Setting $ y\to y_0+y$,
$F_{vx}\to F_{vx}$, $F_{uv}\to F_{uv}$, $F_{ux}\to -y_0+F_{ux}$
and expanding the action to quadratic order we obtain \be
S_2={1\over 2}\int \left[-2\p_{u}y\p_v y+(\p_x y)^2+x^2(\p_v
y)^2-F_{uv}^2+x^2(F_{vx}-2uy_v)^2-\right. \ee
$$
\left.-2(F_{ux}-2uy_u)(F_{vx}-2uy_v)\right] \ .
$$
The ensuing equations of motion are \be 2\p_u\p_v y-\p_x^2
y-x^2\p_v^2 y-2(F_{vx}-2uy_v)=0 \ , \ee \be
\p_uF_{uv}+\p_x[x^2(F_{vx}-2uy_v)-F_{ux}+2uy_u]=0 \ , \ee \be
\p_u(F_{vx}-2uy_v)-\p_v[x^2(F_{vx}-2uy_v)-F_{ux}+2uy_u]=0 \ , \ee
\be \p_vF_{uv}+\p_x(F_{vx}-2uy_v)=0 \ . \ee Introducing a dual
scalar field A by \be F_{uv}=\p_x A\sp F_{ux}=2uy_u+\p_u
A-x^2\p_vA\sp F_{vx}=-\p_vA+2uy_v \ , \ee the equations read \be
\Box A=2y_v\sp \Box y=-2 \p_v A \ , \ee where \be \Box
=2\p_u\p_v-\p_x^2-x^2\p_v^2 \ . \ee In terms of the dual variable,
the quadratic action can be written as \be S_{2}=\int dudvdx
\left[{1\over 2}A\square A+{1\over 2}y\square y-2A\partial_v
y\right] \ . \ee Defining a new  complex scalar field as
$\Phi=(A+iy)e^{-iu}$ we find \be S_{2}=\int dudvdx \left[{1\over
4}\Phi^*\square \Phi+{1\over 4}\Phi\square \Phi^* \right] \ . \ee
Thus, $\Phi$ is a massless scalar. Its solutions are in one-to-one
correspondence with the discrete and the continuous
representations in accordance with the BCFT discussion.

Since here the world-volume has Minkowski signature it is the
eigenvalues of the Laplacians that are relevant when we include 6
extra flat coordinates in order to study strung theory in the $H_4
\times \mathbb{R}^6$ background. Thus we need to solve \be \square
\Phi=E\Phi \ . \ee We parameterize, \be \Phi= e^{ip_-u+ip_+v}z \ ,
\ee and $z$ satisfies the harmonic oscillator equation \be
\left[-\p_x^2+p_+^2 x^2\right]z=(2p_+p_+E )z \ , \ee with
quantized values of $p_-$. For $p_+=0$ the equation becomes \be
\p_x^2 z=-E z \ , \ee and the solutions for $z$ are plane waves in
one dimension.

Thus, the spectrum is in agreement with the BCFT findings in
section \ref{spectrum}.

\section{Conclusions and generalizations \label{f} }
\renewcommand{\theequation}{\arabic{section}.\arabic{equation}}

In this paper we provided the complete solution for the BCFT pertaining
to the two classes of symmetric branes of the $H_4$ model, the $D2$ and
the $S1$ branes. In both cases we solved the consistency BCFT conditions
\cite{lew,lew2} and obtained the BCFT data, namely the bulk-boundary and
the three-point boundary couplings.

The bulk-boundary couplings for the
$D2$ branes can be found in Eq. $(\ref{d2bb1})$ and $(\ref{d2bb2})$ while
the three-point boundary couplings are in Eq. $(\ref{sd214})$,
$(\ref{b28})$, $(\ref{d2o1})$ and $(\ref{d2o2})$. The bulk-boundary
couplings for the $S 1$ branes are in Eq. $(\ref{bbc1m})$ while the
boundary three-point couplings can be found in Eq. $(\ref{b-3-+0})$
$(\ref{b-3000})$ and $(\ref{bex})$. To our knowledge, with the notable
exception of the Liouville model \cite{zzbl1,zzbl2,hoso,pons,tbl}, this is
the first {\em complete} tree-level solution of D-brane dynamics in a
curved non-compact background.

Our solution of the $H_4$ model with and
without a boundary should help to clarify the properties of the
non-compact WZW models and the closed and open string dynamics in curved
space-times. Among other results we provided the first example of
structure constants for twisted symmetric branes in a WZW model (the D2
branes) and of open four-point functions in a curved background.

There are two aspects of our work we think deserve further study. The
first is to perform a more detailed analysis of the four-point amplitudes
and the second to clarify the relation between the open and closed string
channel of the annulus amplitudes. There are also several other issues it
would be worth pursuing and we mention here a few. One is the
study of the symmetric branes of the other WZW models based on the Heisenberg
groups $H_{2+2n}$, $n \ge 2$. Their generators satisfy the following
commutation relations \be [P_i^+,P_i^-] = -2i \m_i K \ , \hspace{1cm}
[J,P^\pm_i] = \mp i \m_i P^{\pm}_i \ , \label{high} \ee with $i = 1, ...,
n$. It will be interesting to generalize our results to the higher
dimensional analogues of the $D2$ and the $S1$ branes as well as to extend
them to encompass other classes of symmetric branes. In fact whenever two
or more of the $\m_i$ parameters in (\ref{high}) coincide, the higher
dimensional Heisenberg algebras have additional outer automorphisms which
permute the corresponding pairs of $P^{\pm}_i$ generators. The existence
of additional outer automorphisms parallels the enhancement of the
isometry group of these pp-wave backgrounds when some of the $\m_i$
parameters coincide \cite{bdkz}. As a consequence these models display a
richer set of symmetric branes, some of them similar to the oblique branes
discussed in \cite{hy,gr3,sz}.

It should also be possible to study less
symmetric branes which can be obtained by performing a $T$-duality along
the Cartan torus, following \cite{mms}. Also the supersymmetric $H_{2n+2}$
WZW models should be analyzed and brane configurations preserving some or
none of the bulk supersymmetries. An interesting brane is the $H_4$ brane
in the $H_6$ gravitational wave \cite{dbr,dbr2}, the Penrose limit of the
$AdS_2 \times S^2$ brane in $AdS_3 \times S^3$. The dynamics of the open
strings ending on this brane should be described by a direct
generalization of our results for the $D2$ branes.

\section{Acknowledgments}

The authors are grateful to  Costas Bachas and Volker Schomerus
for several discussions. G.D. would like to thank the Erwin
Schr\"odinger International Institute in Vienna for the warm
hospitality provided to him during the program {\it String theory
in curved backgrounds and BCFT} and the workshop {\it String
theory on non-compact and time-dependent backgrounds}, June
$7$-$18$ $2004$. He is particularly grateful to the organizers of
the program for the brilliant atmosphere they created and for
giving him the opportunity to present this work. Most of the
present work was done while G. D. was supported by an European
Commission Marie Curie Individual Postdoctoral Fellowship at the LPTHE,
Paris, contract HMPF CT 2002-01908. G.D. also acknowledge support
by the PPARC grant PPA/G/O/2002/00475 and by the EU network
HPRN-CT-2000-00122. The work of E. K. was partially supported by
INTAS grant, 03-51-6346, RTN contracts MRTN-CT-2004-005104 and
MRTN-CT-2004-503369 and by a European Union Excellence Grant,
MEXT-CT-2003-509661.

\newpage

\appendix

\vskip 10mm
 \renewcommand{\theequation}{\thesection.\arabic{equation}}
\centerline{\Large\bf Appendices}

\section{$\hat {\cal H}_4$ representations \label{rep}}

The Heisenberg group $H_4$ has three types of unitary
representations.

1) Lowest-weight representations $ V_{p,\jh}^+$, where $p > 0$.
They are constructed starting from a state $|p,\jh\rangle$ which
satisfies $P^+|p,\jh \rangle=0$, $K|p,\jh\rangle = i p
|p,\jh\rangle$ and $J|p,\jh\rangle = i \jh |p,\jh\rangle$. The
spectrum of $J$ is given by $\{ \jh +n \}$, $n \in \mathbb{N}$ and
the value of the Casimir is $\mathcal{C}=-2p\jh+ p$ .

2) Highest-weight representations $ V_{p,\jh}^-$, where $p > 0$.
They are constructed starting from a state $|p,\jh\rangle$ which
satisfies $P^-|p,\jh \rangle=0$, $K|p,\jh\rangle = - i p
|p,\jh\rangle$ and $J|p,\jh\rangle = i \jh |p,\jh\rangle$. The
spectrum of $J$ is given by $\{ \jh - n \}$, $n \in \mathbb{N}$
and the value of the Casimir is $\mathcal{C}= 2p\jh+p$. The
representation $V^-_{p,-\jh}$ is the representation conjugate to
$V^+_{p,\jh}$.

3) Continuous representations $V _{s,\jh}^0$ with $p=0$. These
representations are characterized by $K |s,\jh \rangle = 0$, $J
|s,\jh \rangle = i \jh|s,\jh \rangle $ and $P^\pm|s,\jh
\rangle\neq 0$. The spectrum of $J$ is then given by $\{ \jh
+n\}$, with $n \in \mathbb{Z}$ and $| \jh | \le \frac{1}{2}$. The
value of the Casimir is ${\cal C} = s^2$. The one dimensional
representation can be considered as a particular continuous
representation, where the charges $s$ and $\jh$ are zero.

For the study of the $H_4$ WZW model, three types of highest-weight
representations of the affine $\hat {\cal H}_4$ algebra will be
relevant. Affine representations $\hat V^\pm_{p,\jh}$ based on
$V^\pm_{p,\jh}$ representations of the horizontal algebra, with
conformal dimension \be h = \mp p \jh + \frac{p}{2}(1-p) \ , \ee
and affine representations $\hat V^0_{s\jh}$  based on
$V^0_{s\jh}$ representations, with conformal dimension \be h =
\frac{s^2}{2}  \ . \ee Highest-weight representations of the
current algebra lead to a string spectrum free from negative norm
states only if they satisfy the constraint \be 0 < p < 1 \ .
\label{bound} \ee States with larger values of $p$ belong to new
representations resulting from spectral flow of the original
highest-weight representations \cite{moog1,kk}. In fact the
spectral-flowed representations are highest-weight representations of an
isomorphic algebra whose modes are related to the original ones by
\ba \tilde{P}^+_{n} &=& P^+_{ n - w} \ , \hspace{1cm}
\tilde{P}^{-}_{n} = P^{- }_{n + w} \ , \hspace{1cm}
\tilde{J}_n = J_n \ , \nb \\
\tilde{K}_n &=& K_n - i w \d_{n,0} \ ,  \hspace{1cm} \tilde{L}_n =
L_n - i w J_n \ . \ea An important piece of information for
understanding the structure of the three-point couplings is
provided by the decomposition of the tensor products of the $H_4$
representations  \ba V^+_{p_1,\jh_1} \otimes V^+_{p_2,\jh_2} &=&
\sum_{n=0}^\infty V^+_{p_1+p_2,\jh_1+\jh_2+n}
\ , \nb \\
V^+_{p_1,\jh_1} \otimes V^-_{p_2,\jh_2} &=& \sum_{n=0}^\infty
V^+_{p_1+p_2,\jh_1+\jh_2-n}
\ , \hspace{0.5cm}  p_1 > p_2  \ , \nb \\
V^+_{p_1,\jh_1} \otimes V^-_{p_2,\jh_2} &=& \sum_{n=0}^\infty
V^-_{p_1+p_2,\jh_1+\jh_2+n} \ , \hspace{0.5cm} p_1 < p_2 \ , \nb
\\
V^+_{p \, ,\jh_1} \otimes V^-_{p \, ,\jh_2} &=& \int_0^\infty s ds
V^0_{s,\jh_1+\jh_2} \ , \nb \\
V^+_{p_1,\jh_1} \otimes V^0_{s,\jh_2} &=& \sum_{n=-\infty}^\infty
V^+_{p_1+p_2,\jh_1+\jh_2+n} \ . \label{tensorH4} \ea
The fusion
rules for the primary vertex operators of the $H_4$ model can be
obtained from the previous tensor products. When the
representations involved are spectral-flowed representations, one
has to use the relation \cite{gab-ts} \be \Sigma_{w_1}[\F_{\a_1}]
\otimes \Sigma_{w_2}[\F_{\a_2}] = \Sigma_{w_1+w_2}[\F_{\a_1}
\otimes \F_{\a_2}] \ . \ee

\section{Fusing matrices \label{ff}}

Consider the correlator $\la \f_i(z_1)\f_j(z_2)\f_k(z_3)\f_l(z_4)
\ra$ and let ${\cal F}_p^{ijkl}(z)$ denote the conformal blocks in
the $s$-channel $z_1 \!\! \sim \!\! z_2 $ and ${\cal
F}_q^{lijk}(1-z)$ the conformal blocks in the $u$-channel $z_1
\!\!\sim \!\!z_4$, where $z = \frac{z_{12}z_{34}}{z_{13}z_{24}}$.
We use the following convention for the fusing matrices \be {\cal
F}^{ijkl}_p(z) =  \sum_q {\bf F}_{pq}  \small
\begin{bmatrix} j & k \\ i & l
\end{bmatrix} \,
{\cal F}^{lijk}_q(1-z) \ .
\ee
${\cal F}_{pq}$ defines a linear transformation
\be
 {\bf F}_{pq}  \small  \begin{bmatrix} j & k \\ i & l
\end{bmatrix} : V_{jp}^i \otimes V_{kl}^p \rightarrow V_{ql}^i \otimes V_{jk}^q \ ,
\ee
where $V_{jk}^i$ is the space of the three-point couplings.
Moreover,
\be
\sum_q {\bf F}_{pq}  \small  \begin{bmatrix} j & k \\ i & l
\end{bmatrix} \,  {\bf F}_{qr}  \small  \begin{bmatrix} l & k \\ i &
j
\end{bmatrix} = \d_{r,s} \ .
\ee Since in our non-compact CFT the conformal blocks are labeled
either by discrete or continuous indexes, in the previous
expressions we will have a sum or an integral, according to the
case. The following are the fusing matrices we used in section
\ref{sc} to compute the structure constants. We set  $\n = -
\sum_{i=1}^4 \jh_i$. For correlators of the form $\la +++- \ra$ we
have

\ba && {\bf F}_{(p_1+p_2,\jh_1+\jh_2+n),(p_1-p_4,\jh_1+\jh_4+m)}
\small  \begin{bmatrix} (-p_4,\jh_4) & (p_1,\jh_1)
\\ (p_3,\jh_3) & (p_2,\jh_2)
\end{bmatrix} = \nb \\ && \frac{(\n-n)!}{m!\G(\n-n-m+1)}
 \left [ \frac{\G(p_2+p_3)\G(p_1+p_2)}{\G(p_2)\G(p_4)}\right
]^{\n}  \left [
\frac{\G(p_2)\G(p_4)}{\g(p_1+p_2)\G(1-p_1)\G(p_3)}\right ]^n \nb
\\ && \left [\frac{\G(p_2)\G(p_4)}{\g(p_2+p_3)\G(p_3)\G(1-p_1)}\right
]^m F(-n,-m,\n-n-m+1,-\th) \ , \ea where \be \th = \frac{\sin{\pi
p_2}\sin{\pi p_4}}{\sin{\pi p_1}\sin{\pi p_3}} \ . \ee

For correlators of the form $\la +-+- \ra$ we have

\ba && {\bf F}_{(p_1-p_2,\jh_1+\jh_2-n),(p_1-p_4,\jh_1+\jh_4-m)}
\small  \begin{bmatrix} (-p_4,\jh_4) & (p_1,\jh_1)
\\ (p_3,\jh_3) & (-p_2,\jh_2)
\end{bmatrix} = \nb \\ && \frac{(m+n+\n)!}{m!(m+\n)!}
\left [ \frac{\g(p_1-p_2)\G(p_2)\G(1-p_4)}{\G(p_1)\G(1-p_3)}\right
]^n \left
[\frac{\G(p_2)\G(1-p_4)}{\g(p_2-p_3)\G(p_3)\G(1-p_1)}\right ]^m
\nb \\ && \left [
\frac{\G(p_2)\G(1-p_4)}{\G(1-p_1+p_2)\G(p_2-p_3)}\right ]^{\n+1}
F(-n,-m,-n-m-\n,\th) \ , \ea where \be \th = \frac{\sin{\pi
p_2}\sin{\pi p_4}}{\sin{\pi p_1}\sin{\pi p_3}} \ . \ee
 \ba &&  {\bf
F}_{(s,\{\jh_1+\jh_2\}),(p-q,\jh_1+\jh_4-n)} \small
\begin{bmatrix} (-q,\jh_4) & (p,\jh_1)
\\ (q,\jh_3) & (-p,\jh_2)
\end{bmatrix} = \frac{1}{(n+\n)!}\left [ \frac{\G(p)\G(1-q)}{\G(p-q)} \right ]^{\n+1} \nb \\
&& \left [ \frac{\g(p)}{\g(q)\g(p-q)} \right ]^n \left [
\frac{s^2}{2} \right
]^{\frac{\n}{2}}e^{-\frac{s^2}{2}(\psi(p)+\psi(1-q)-2\psi(1))}
L_n^\n \left [ \frac{s^2}{2}(\pi {\rm ctg} \pi q - \pi {\rm ctg}
\pi p) \right ] \ . \ea

\ba && {\bf F}_{(p-q,\jh_1+\jh_4-n),(s,\{\jh_3+\jh_4\})} \small
\begin{bmatrix} (-p,\jh_2) & (p,\jh_1)
\\ (q,\jh_3) & (-q,\jh_4)
\end{bmatrix} = n! \left [ \frac{\G(q)\G(1-p)}{\G(1-p+q)} \right ]^{\n+1} \nb \\
&& \left [ \frac{\g(q)\g(p-q)}{\g(p)} \right ]^n \left [
\frac{s^2}{2} \right
]^{\frac{\n}{2}}e^{\frac{s^2}{2}(\psi(q)+\psi(1-p)-2\psi(1))}
L_n^\n \left [ \frac{s^2}{2}(\pi {\rm ctg} \pi q - \pi {\rm ctg}
\pi p) \right ] \ . \ea

\be
{\bf F}_{(s,\{\jh_1+\jh_2\}),(t,\{\jh_1+\jh_4\})}
\small  \begin{bmatrix} (-p,\jh_4) & (p,\jh_1)
\\ (p,\jh_3) & (-p,\jh_2)
\end{bmatrix} = \frac{\pi}{\sin \pi p} e^{\frac{t^2-s^2}{2}
\left ( \psi(p)+\psi(1-p)-2 \psi(1) \right )} J_\n \left (
\frac{\pi s t}{\sin \pi p} \right ) \ . \ee

Similar expressions hold for correlators of the form $\la ++--
\ra$.

\ba && {\bf F}_{(p_1+p_2,\jh_1+\jh_2+n),(p_1-p_4,\jh_1+\jh_4-m)}
\small  \begin{bmatrix} (-p_4,\jh_4) & (p_1,\jh_1)
\\ (-p_3,\jh_3) & (p_2,\jh_2)
\end{bmatrix} = \nb \\ && \frac{(m+n+\n)!}{m!(m+\n)!}\left [ \frac{\G(p_1)\G(p_3)}{\g(p_1+p_2)
\G(1-p_2)\G(1-p_4)}\right ]^{n-\n} \nb \left
[\frac{\G(p_1)\G(p_3)}{\g(p_3-p_2)\G(p_2)\G(p_4)}\right ]^m \nb
\\ && \left [ \frac{\G(p_1)\G(p_3)}{\G(p_1+p_2)\G(p_3-p_2)}\right
]^{\n+1} F(-n+\n,-m,-n-m,\th) \ , \ea where \be \th =
\frac{\sin{\pi p_1}\sin{\pi p_3}}{\sin{\pi p_2}\sin{\pi p_4}} \ .
\ee

\ba &&  {\bf
F}_{(s,\{\jh_1+\jh_2\}),(p+q,\jh_1+\jh_4+n)} \small
\begin{bmatrix} (-p,\jh_4) & (p,\jh_1)
\\ (-q,\jh_3) & (q,\jh_2)
\end{bmatrix} = (n-\n)!\left [ \frac{\G(p)\G(q)}{\G(p+q)} \right ]^{\n+1}  \\
&& \left [ \frac{\g(p)}{\g(q)\g(p+q)} \right ]^{n-\n} \left [
\frac{s^2}{2} \right
]^{\frac{\n}{2}}e^{\frac{s^2}{2}(\psi(p)+\psi(q)-2\psi(1))} L_n^\n
\left [ \frac{s^2}{2}(\pi {\rm ctg} \pi p + \pi {\rm ctg} \pi q)
\right ] \ . \nb \ea

\ba && {\bf F}_{(p+q,\jh_1+\jh_2+n),(s,\{\jh_3+\jh_4\})} \small
\begin{bmatrix} (q,\jh_2) & (p,\jh_1)
\\ (-q,\jh_3) & (-p,\jh_4)
\end{bmatrix} = \frac{1}{n!} \left [ \frac{\G(1-q)\G(1-p)}{\G(1-p-q)} \right ]^{\n+1}  \\
&& \left [ \frac{\g(p+q)}{\g(p)\g(q)} \right ]^{n-\n} \left [
\frac{s^2}{2} \right
]^{\frac{\n}{2}}e^{-\frac{s^2}{2}(\psi(1-q)+\psi(1-p)-2\psi(1))}
L_{n-\n}^\n \left [ \frac{s^2}{2}(\pi {\rm ctg} \pi p + \pi {\rm
ctg} \pi q) \right ] \ . \nb \ea

For correlators of the form $\la +-00 \ra$ we have

\be
{\bf F}_{(s,\{\jh_1+\jh_2\}),(p,\jh_1+\jh_4+n)}
\small  \begin{bmatrix} (s_4,\jh_4) & (p,\jh_1)
\\ (s_3,\jh_3) & (-p,\jh_2)
\end{bmatrix} = e^{-\frac{s_3s_4}{2} \cos \f \,
\s(p)+\frac{i \pi s_3s_4 \sin \f }{2 \tan \pi p}-i n \f +i \h \n}
\ , \ee \be {\bf F}_{(p,\jh_1+\jh_4+n)(s,\{\jh_1+\jh_2\})} \small
\begin{bmatrix} (-p,\jh_2) & (p,\jh_1)
\\ (s_3,\jh_3) & (s_4,\jh_4)
\end{bmatrix} = \frac{1}{\pi s_3s_4 \sin \f} e^{\frac{s_3s_4}{2} \cos \f
\, \s(p)-\frac{i \pi s_3s_4 \sin \f }{2 \tan \pi p}+i n \f -i \h
\n} \ , \ee where $\s(p) = \psi(p)+\psi(1-p)-2 \psi(1)$, $s^2 =
s_3^2+s_4^2+2 s_3s_4 \cos \f$ and $e^{i \h} =
\frac{s_3+s_4e^{i\f}}{s}$.

\section{Bases of conformal blocks \label{bb}}

In this appendix we collect various bases of conformal blocks for
correlators of the form $\langle +-+- \rangle$, $\la ++-- \ra$ and
$\la +-00 \ra$. Using the global conformal and $H_4$ symmetries,
the four-point amplitudes can be written as follows \be
A_4(z_i,x_i;\bar z_i,\bar x_i) = \prod_{j>i=1}^4 \left
|z_{ij}\right |^{2h_i+2h_j-\frac{2h}{3}} {\cal K}(x_i) \bar {\cal
K}(\bar x_i) \sum_n {\cal F}_n(z) \bar {\cal F}_n(\bar z) \ . \ee
The kinematical functions ${\cal K}$ and $\bar {\cal K}$ are
completely fixed by the Ward identities of the left and right $\hat
{\cal H}_4$ algebras and we chose the standard gauge for the
global conformal transformations. The conformal blocks ${\cal
F}_n(z,x)$ thus depend only on the cross-ratio $z =
\frac{z_{12}z_{34}}{z_{13}z_{24}}$ and a suitable combination $x$
of the four charge variables. In the following $\n = -
\sum_{i=1}^4 \jh_i$. Consider first the correlator \be \la
\F^+_{p_1,\jh_1}(z_1,\bar{z}_1,x_1,\bar{x}_1)
\F^+_{p_2,\jh_2}(z_2,\bar{z}_2,x_2,\bar{x}_2)
\F^+_{p_3,\jh_3}(z_3,\bar{z}_3,x_3,\bar{x}_3)
\F^-_{p_4,\jh_4}(z_4,\bar{z}_4,x_4,\bar{x}_4) \ra \ . \label{aa1}
\ee The $H_4$ Ward identities require \be p_1+p_2+p_3=p_4 \ ,
\label{aa2} \ee and give the function ${\cal K}$ \be {\cal K}(x_i)
= e^{-x_4(p_1x_1+p_2x_2+p_3x_3)} (x_3-x_1)^{\n}  \ , \label{aa3}
\ee as well as the invariant combination  \be x =
\frac{x_2-x_1}{x_3-x_1} \ . \label{aa4} \ee We set ${\cal F} =
z^{\ka_{12}}(1-z)^{\ka_{14}}F(z,x)$ where \ba
\ka_{12} &=& h_1+h_2-\frac{h}{3}-p_1p_2-\jh_2p_1-\jh_1p_2 \ ,  \nb \\
\ka_{14} &=&
h_1+h_4-\frac{h}{3}+p_1p_4-\jh_4p_1+\jh_1p_4-p_1+\n(p_2+p_3) \ .
\label{aa5} \ea We then arrive at the following form for the KZ
equation \ba  \p_z F(z,x) &=& \frac{1}{z}\left [ -\left (p_1x
+p_2x(1-x)\p_x \right ) -\n p_2 x  \right ] F(z,x) \nb \\
&-& \frac{1}{1-z} \left [ (1-x)(p_2x+p_3) \p_x -\n p_2(1-x) \right
] F(z,x) \ . \label{aa6} \ea The correlator vanishes when $\n <
0$. In the $s$-channel we have the $\F^+_{p_1+p_2,\jh_1+\jh_2+n}$
representations with $0 \le n \le \n$. The conformal blocks are
\be F_{n}(z,x) = f^n(z,x)(g(z,x))^{\n-n} \ , \ee where \ba f(z,x)
&=& \frac{p_3}{1-p_1-p_2}z^{1-p_1-p_2}\f_0(z) -x z^{-p_1-p_2}
\f_1(z) \ , \nb \\
g(z,x) &=& \g_0(z) -\frac{xp_2}{p_1+p_2}\g_1(z) \ , \ea and \ba
\f_0(z) &=& F(1-p_1,1+p_3,2-p_1-p_2,z) \ , \nb \\
\f_1(z) &=& F(1-p_1,p_3,1-p_1-p_2,z) \ , \nb \\
\g_0(z) &=& F(p_2,p_4,p_1+p_2,z) \ , \nb \\
\g_1(z) &=& F(1+p_2,p_4,1+p_1+p_2,z) \ . \ea In the $u$-channel we
have the representations $\F^-_{p_4-p_1,\jh_1+\jh_2-m}$, $0 \le m
\le \n$ and their conformal blocks read \be F_{m}(u,x) = \tilde
f^{\n-m}(u,x)(\tilde g(u,x))^{m} \ , \ee where \ba \tilde f(u,x)
&=& u^{-p_2-p_3}(\tilde \f_0(u) -x \tilde
\f_1(u)) \ , \nb \\
\tilde g(z,x) &=& \frac{p_3}{p_2+p_3} \tilde \g_0(u)
+\frac{xp_2}{p_2+p_3}\tilde \g_1(u) \ , \ea and \ba
\tilde \f_0(u) &=& F(p_1,-p_3,1-p_2-p_3,u) \ , \nb \\
\tilde \f_1(u) &=& F(1-p_3,p_1,1-p_2-p_3,u) \ , \nb \\
\tilde \g_0(u) &=& F(p_2,p_4,1+p_2+p_3,u) \ , \nb \\
\tilde \g_1(u) &=& F(1+p_2,p_4,1+p_2+p_3,u) \ . \ea Consider now the
correlator \be \la \F^+_{p_1,\jh_1}(z_1,\bar{z}_1,x_1,\bar{x}_1)
\F^-_{p_2,\jh_2}(z_2,\bar{z}_2,x_2,\bar{x}_2)
\F^+_{p_3,\jh_3}(z_3,\bar{z}_3,x_3,\bar{x}_3)
\F^-_{p_4,\jh_4}(z_4,\bar{z}_4,x_4,\bar{x}_4) \ra \ . \label{baa1} \ee The
$H_4$ Ward identities require \be p_1+p_3=p_2+p_4 \ , \label{baa2} \ee and
give the function ${\cal K}$ \be {\cal K}(x_i) =
e^{-p_2x_1x_2-p_3x_3x_4-(p_1-p_2)x_1x_4} (x_1-x_3)^{\n}  \ . \label{baa3}
\ee In this case the invariant combination is \be x = (x_1-x_3)(x_2-x_4) \
. \label{baa4} \ee We set ${\cal F} = z^{\ka_{12}}(1-z)^{\ka_{14}}F(z,x)$
where \ba
\ka_{12} &=& h_1+h_2-\frac{h}{3}+p_1p_2-\jh_2p_1+\jh_1p_2-p_2 \ ,  \nb \\
\ka_{14} &=&  h_1+h_4-\frac{h}{3}+p_1p_4-\jh_4p_1+\jh_1p_4-p_4 \ .
\label{baa5} \ea We then arrive at the following form for the KZ equation
\ba z(1-z) \p_z F(z,x) &=& \left [ x\p^2_x +\left (
(p_1-p_2)x+1+\n \right )\p_x \right ] F(z,x) \nb \\
&+& z \left [ -(p_1+p_3)x \p_x + x p_2p_3-(1+\n)p_3 \right ] F(z,x) \ .
\label{baa6} \ea When $p_1 > p_2$ in the $s$-channel we have
$\F^+_{p_1-p_2,\jh_1+\jh_2-n}$ with $n \ge 0$ for $\n \ge 0$ and $n=m-\n$
with $m \ge 0$ for $\n \le 0$. The conformal blocks are \be F_n(z,x) =
\n_n \frac{e^{xg_1(z)}}{(f_1(z))^{1+\n}} L^{\n}_n(x \g_{\psi}(z))
\psi(z)^n \ , \hspace{1cm}  n \in \mathbb{N} \ , \label{baa11} \ee where
$L^{\n}_n$ is the n-th generalized Laguerre polynomial, \ba \psi(z) &=&
\frac{f_2(z)}{f_1(z)}  \ , \hspace{1cm}  \g_{\psi}(z) =
-z(1-z)\p \ln{\psi} \ ,  \hspace{1cm} \n_n = \frac{n!}{(p_1-p_2)^n} \ , \nb \\
g_1(z) &=& z p_3-z(1-z)\p \ln{f_1} \ , \label{aa12} \ea and \ba
f_1(z) &=& F(p_3,1-p_1,1-p_1+p_2,z) \ , \nb \\
f_2(z) &=& z^{p_1-p_2}F(p_4,1-p_2,1-p_2+p_1,z) \ . \label{aa13} \ea When
$p_1 < p_2$, the intermediate states belong to the
$\F^-_{p_2-p_1,\jh_1+\jh_2+n}$ representation with $n=m+\n$ with $m \ge 0$
for $\n \ge 0$ and $n \ge 0$ for $\n \le 0$. The conformal blocks are very
similar. Finally when $p_1=p_2=p$ and $p_3=p_4=q$ the intermediate
representation is $\F^0_{s,\jh}$ and the conformal blocks read \be
F_{s}(z,x) = \frac{e^{xg_1(z)}}{(c_1(z))^{1+\n}} e^{\frac{s^2}{2} \r(z)}
(-x z(1-z)\p \r)^{-\frac{\n}{2}} J_{\n}(s\sqrt{2x\g}) \ , \label{baa27}
\ee where \be \r(z) = \frac{c_2(z)}{c_1(z)}  \ , \hspace{1cm} \g =
-z(1-z)\p \r(z)  \ , \label{baa28} \ee and \ba
c_1(z) &=& F(q,1-p,1,z) \ , \nb \\
c_2(z) &=& [ \ln{z}+2\psi(1)-\psi(q)-\psi(1-p)]  c_1(z) \nb \\
&+& \sum_{n=0}^{\infty} \frac{(q)_n(1-p)_n}{n!^2} [\psi(q+n)+\psi(1-p+n) -
2 \psi(n+1)] z^n \ , \label{baa29} \ea where \be (a)_n \equiv
\frac{\G(a+n)}{\G(a)} \ . \label{baa30} \ee Moreover \be g_1(z) =qz
-z(1-z) \p_z \ln{c_1} \ . \label{baa31} \ee In the $u$-channel when $p_1 >
p_4$ we have the representations $\F^+_{p_1-p_4,\jh_1+\jh_4-n}$ with $n
\in \mathbb{N}$ for $\n \ge 0$ and $n=m-\n$, $m \ge 0$ for $\n \le 0$. The
conformal blocks are \be F_n(u,x) = \n_n
\frac{e^{xg_1(u)}}{(f_1(u))^{1+\n}} L^{\n}_n(x \g_{\psi}(u)) \psi(u)^n \ ,
\hspace{1cm}  n \in \mathbb{N} \ , \label{aa24} \ee where \ba \psi(u) &=&
\frac{f_2(u)}{f_1(u)}  \ , \hspace{1cm} \g_{\psi}(u) =
u(1-u)\p \ln{\psi} \ ,  \hspace{1cm} \n_n = \frac{n!}{(p_2-p_3)^n} \ , \nb \\
g_1(u) &=& (1-u)p_3+u(1-u)\p \ln{f_1} \ , \label{aa25} \ea and \ba
f_1(u) &=& F(p_3,1-p_1,1-p_2+p_3,u) \ , \nb \\
f_2(u) &=& u^{p_2-p_3}F(p_2,1-p_4,1+p_2-p_3,u) \ . \label{aa26}
\ea When $p_1 < p_4$, the intermediate states belong to the
$\F^-_{p_4-p_1,\jh_1+\jh_4+n}$ representation with $n=m+\n$ with
$m \ge 0$ for $\n \ge 0$ and $n \ge 0$ for $\n \le 0$. The
conformal blocks are very similar. Finally, when $p_1=p_4=p$ and
$p_2=p_3=q$ the intermediate representation is $\F^0_{s,\jh}$ and
the conformal blocks read \be F_{s}(u,x) =
\frac{e^{xg_1(u)}}{(c_1(u))^{1+\n}} e^{\frac{s^2}{2} \r(u)} (x
u(1-u)\p \r)^{-\frac{\n}{2}} J_{\n}(s\sqrt{2 x \g}) \ ,
\label{aa27} \ee where \be \r(u) = \frac{c_2(u)}{c_1(u)}  \ ,
\hspace{1cm} \g = u(1-u)\p \r(u)  \ , \label{aa28} \ee and \ba
c_1(u) &=& F(q,1-p,1,u) \ , \nb \\
c_2(u) &=& [ \ln{u}+2\psi(1)-\psi(q)-\psi(1-p)]  c_1(u) \nb \\
&+& \sum_{n=0}^{\infty} \frac{(q)_n(1-p)_n}{n!^2}
[\psi(q+n)+\psi(1-p+n) - 2 \psi(n+1)] u^n \ .\label{aa29} \ea
Moreover \be g_1(u) = (1 - q) u  + u(1-u) \p_u \ln{c_1} \ .
\label{aa31} \ee We will also need correlators of the form  \be
\la\F^+_{p_1,\jh_1}(z_1,\bar{z}_1,x_1,\bar{x}_1)
\F^+_{p_2,\jh_2}(z_2,\bar{z}_2,x_2,\bar{x}_2)
\F^-_{p_3,\jh_3}(z_3,\bar{z}_3,x_3,\bar{x}_3)
\F^-_{p_4,\jh_4}(z_4,\bar{z}_4,x_4,\bar{x}_4)\ra \ . \label{aa1b}
\ee In this case the $H_4$ symmetry requires \be p_1+p_2=p_3+p_4 \
, \label{aa32} \ee and gives \be {\cal K}(x_i) =
e^{-p_3x_1x_3-p_2x_2x_4-(p_1-p_3)x_1x_4} (x_1-x_2)^{\n} \ ,
\label{aa33} \ee as well as $x = (x_1-x_2)(x_3-x_4)$. Proceeding
as before we pass to the conformal blocks and we set ${\cal F} =
z^{\l_{12}}(1-z)^{\l_{14}}F(z,x)$ where \ba \l_{12} &=&
-\ka_{12}-\ka_{14}-(1+\n)p_2 \nb \\ &=&
-h_1+h_2-\frac{h}{3}+(1-p_1)(p_1+p_2)+p_1(\jh_3+\jh_4)
-\jh_1(p_1+p_2)-(1+\n)p_2 \ ,  \nb \\
\l_{14} &=&  \ka_{14} =
h_1+h_4-\frac{h}{3}+p_1p_4-\jh_4p_1+\jh_1p_4-p_4 \ . \label{aa35}
\ea We then arrive at the following form for the KZ equation \ba
z(1-z) \p_z F_n(z,x) &=& z \left [ x\p^2_x +\left (
(p_1-p_3)x+1+\n \right )\p_x
-(1+\n)p_2 \right ] F_n(z,x) \nb \\
&+& \left [ -(p_1+p_2)x \p_x + xp_2p_3 \right ] F_n(z,x) \ .
\label{aa36} \ea
In the $s$-channel, the representations
$\F^+_{p_1+p_2,\jh_1+\jh_2+n}$ with $n - \n \in \mathbb{N}$ when
$\n \ge 0$ and $n \in \mathbb{N}$ when $\n \le 0$. In the first
case with $m = n - \n$ the conformal blocks are \be F_m(z,x) =
\n_m \frac{e^{xg_1(z)}}{(f_1(z))^{1+\n}} L^{\n}_m(x \g_{\psi}(z))
\psi(z)^m \ , \hspace{1cm}  m \in \mathbb{N} \ , \label{aa37} \ee
where  \ba \psi(z) &=& \frac{f_2(z)}{f_1(z)} \ , \hspace{1cm}
\g_{\psi}(z) =
-(1-z)\p \ln{\psi} \ ,  \hspace{1cm} \n_m = \frac{m!}{(1-p_1-p_2)^m} \ , \nb \\
g_1(z) &=& p_2-(1-z)\p \ln{f_1} \ , \label{aa38} \ea and \ba
f_1(z) &=& F(p_2,p_4,p_1+p_2,z) \ , \nb \\
f_2(z) &=& z^{1-p_1-p_2}F(1-p_1,1-p_3,2-p_1-p_2,z) \ .
\label{aa39} \ea When $\n \le 0$ the conformal blocks are given by
the same expression except that now $n \ge 0$. Using \be
L^\n_{n-\n}(x) = \frac{n!}{(n-\n)!}(-x)^{-\n} L_n^{-\n}(x) \ , \ee
and the wronskian \be W(f_1,f_2) = (1-c)z^{-c}(1-z)^{c-a-b-1} \ ,
\ee they can be rewritten as \be F_n(z,x) =
z^{\n(p_1+p_2)}(1-z)^{-\n(p_1+p_4)}x^{-\n} \n_n
\frac{e^{xg_1(z)}}{(f_1(z))^{1+|\n|}} L^{|\n|}_n(x \g_{\psi}(z))
\psi(z)^n \ , \hspace{1cm}  n \in \mathbb{N} \ . \label{aa40} \ee
In the $u$-channel, when $p_1 > p_4$, we have  the representations
$\F^+_{p_1-p_4,\jh_1+\jh_4-n}$. The conformal blocks are  \be
F_n(z,x) = \n_n \frac{e^{xb_1(z)}}{(a_1(z))^{1+\n}} L^{\n}_n(x
\g_{\chi}(z)) \chi(z)^n \ , \hspace{1cm}  n \in \mathbb{N} \ ,
\label{aa41} \ee where  \ba \chi(z) &=& \frac{a_2(z)}{a_1(z)} \ ,
\hspace{1cm} \g_{\chi}(z) =
u \p \ln{\chi} \ ,  \hspace{1cm} \n_n = \frac{n!}{(p_3-p_2)^n} \ , \nb \\
b_1(z) &=& p_2+u \p \ln{a_1} \ , \label{aa42} \ea and \ba
a_1(z) &=& F(p_2,p_4,1+p_2-p_3,u) \ , \nb \\
a_2(z) &=& u^{p_3-p_2}F(p_3,p_1,1-p_2+p_3,u) \ . \label{aa43} \ea
Here $n \ge 0$ when $ \n \ge 0$ and $n = m - \n$ with $m \ge 0$
when $\n \le 0$. When $p_1 < p_4$ we have the representations
$\F^-_{p_4-p_1,\jh_1+\jh_4+n}$ with $n \ge 0$ when $ \n \le 0$ and
$n = m + \n$ with $m \ge 0$ when $\n \ge 0$. The conformal blocks
are similar to the ones already displayed. Finally when
$p_1=p_4=p$ and $p_2=p_3=q$ the intermediate states belong to the
continuous representations $\F^0_{s,\{\jh_1+\jh_4 \}}$. Let us now
turn to correlators of the form
 \be
\la \F^+_{p,\jh_1}(z_1,\bar{z}_1,x_1,\bar{x}_1)
\F^-_{p,\jh_2}(z_2,\bar{z}_2,x_2,\bar{x}_2)
\F^0_{s,\jh_3}(z_3,\bar{z}_3,x_3,\bar{x}_3)
\F^0_{t,\jh_4}(z_4,\bar{z}_4,x_4,\bar{x}_4) \ra \ . \label{aa1blo}
\ee In this case \be {\cal K}(x_i) =
e^{-px_1x_2-\frac{x_1}{\sqrt{2}}\left(\frac{s}{x_3}+\frac{t}{x_4}\right)-\frac{x_2}{\sqrt{2}}(sx_3+tx_4)}
x_3^{n_1-n_2} \ , \ee and $x = \frac{x_4}{x_3}$. The conformal
blocks corresponding to the propagation of
$\F^+_{(p,\jh_1+\jh_4+n)}$ in the $u$-channel are ${\cal F} =
u^{\ka_{14}}(1-u)^{\ka_{12}}F(u,x)$ where \be \ka_{12} =
\frac{s^2+t^2}{2}-\frac{h}{3} \ , \hspace{1cm} \ka_{14} =
h_1-\frac{h}{3}-p\jh_4 \ . \ee They solve the following KZ
equation \be \p_u F_n(u,x) = -\frac{1}{u}\left ( px \p_x
+\frac{stx}{2}\right )F_n(u,x) -\frac{1}{1-u}
\frac{st}{2}\left(x+\frac{1}{x} \right ) F_n(u,x) \ . \ee Their
explicit form is \be F_n(u,x) = x^n u^{-n p} e^{-\frac{st}{2}
\left ( \frac{x a(u)}{p} + \frac{u b(u)}{x(1-p)} \right ) } \ ,
\ee where \be a(u) = F(p,1,1+p,u) \ , \hspace{1cm} b(u) =
F(1-p,1,2-p,u) \ . \ee Similarly the blocks pertaining to
$\F^-_{p,\jh_2+\jh_3-m}$ are given by $F_{\n+m}$. In the
$s$-channel the blocks for the representation $\F^0_{r}$ with \be
r^2 = s^2+t^2+2st \cos{\f} \ , \hspace{1cm} e^{i \h} =
\frac{s+te^{i \f}}{r} \ , \hspace{1cm} \f \in [0,2\pi)  \ , \ee
are \be {\cal F}_r(z,x) = e^{-\frac{st}{2}[\cos{\f} \s(p) -i
\sin{\f} \pi \cot \pi p]+i \h \n}\sum_{n \in \mathbb{Z}} e^{-i n
\f} {\cal F}_n(u,x) \ . \ee

\section{Sewing constraints \label{dsc}}

In this appendix we outline with an example the main steps that
are necessary in order to verify that the structure constants
given in section \ref{sc} solve the sewing constraints. We
consider the bulk-boundary couplings for the $S 1$ branes and
study the factorization of the following bulk two-point functions:
$\la \F_p^+\F_q^+ \ra$, $\la \F_p^+\F_q^- \ra$ and $\la \F_p^+
\F_s^0 \ra$. The first correlator gives \be {}^a B_{p,\jh_1}^s
{}^a B_{q,\jh_2}^s C^{aaa,1}_{s s} = \sum_{n=0}^\infty {\it
C}_{(p,\jh_1);(q,\jh_2)}^{(p+q,\jh_1+\jh_2+n)} \ {}^a
B^1_{p+q,\jh_1+\jh_2+n} {\bf F}_{(p+q,\jh_1+\jh_2+n),s}  \small
\begin{bmatrix} (-p,-\jh_1) & (p,\jh_1) \\ (-q,-\jh_2) & (q,\jh_2)
\end{bmatrix} \ . \label{scb7a} \ee
The second correlator gives
 \be {}^a B_{p,\jh_1}^s {}^a
B_{-q,\jh_2}^s C^{aaa,1}_{s s} = \sum_{n=0}^\infty {\it
C}_{(p,\jh_1);(-q,\jh_2)}^{(p-q,\jh_1+\jh_2-n)} \ {}^a
B^1_{p-q,\jh_1+\jh_2-n} {\bf F}_{(p-q,\jh_1+\jh_2-n),s} \small
\begin{bmatrix} (-p,-\jh_1) & (p,\jh_1) \\ (q,-\jh_2) & (-q,\jh_2)
\end{bmatrix}
\ , \label{scb7b} \ee when $p>q$ and \be {}^a B_{p,\jh_1}^s {}^a
B_{-p,\jh_2}^s C^{aaa,1}_{s s} = \int_{0}^\infty dt \, t {\it
C}_{(p,\jh_1);(-p,\jh_2)}^{(t,\{\jh_1+\jh_2\})} \ {}^a
B^1_{t,\{\jh_1+\jh_2\}} {\bf F}_{t,(s,\{\jh_1+\jh_2\})}  \small
\begin{bmatrix} (-p,-\jh_1) & (p,\jh_1) \\ (p,-\jh_2) & (-p,\jh_2)
\end{bmatrix} \ , \label{scb7c} \ee
when $p=q$. Finally the third correlator gives \be {}^a
B_{p,\jh_1}^{s} {}^a B_{s_2,\jh_2}^{s} C^{aaa,1}_{s s} =
\sum_{n=0}^\infty {\it
C}_{(p,\jh_1);(s_2,\jh_2)}^{(p,\jh_1+\jh_2+n)} \ {}^a
B^1_{p,\jh_1+\jh_2+n} {\bf F}_{(p,\jh_1+\jh_2+n),s}  \small
\begin{bmatrix} (-p,-\jh_1) & (p,\jh_1) \\ (s_2,-\jh_2) & (s_2,\jh_2)
\end{bmatrix} \ . \label{scb7d} \ee
Using the bulk three-point couplings in
$(\ref{qpppm}-\ref{cpmo2})$ and the fusing matrices in appendix
\ref{ff} the previous equations become \ba {}^a B_{p,\jh_1}^s {}^a
B_{q,\jh_2}^s &=& \sqrt{\th_{p,q}}
e^{\frac{s^2}{2}(\psi(p)+\psi(q)-2\psi(1))}\sum_{n=0}^\infty
L_n\left (\th_{p,q} \frac{
s^2}{2}\right ) {}^aB^1_{p+q,\jh_1+\jh_2+n} \ ,  \\
{}^a B_{p,\jh_1}^s {}^a B_{-q,\jh_2}^s  &=& \sqrt{\th_{-p,q}}
 e^{\frac{s^2}{2}(\psi(q)+\psi(1-p)-2\psi(1))}\sum_{n=0}^\infty
L_n\left (\th_{-p,q} \frac{
s^2}{2}\right ) {}^aB^1_{p-q,\jh_1+\jh_2-n} \ ,  \nb \\
{}^a B_{p,\jh_1}^s {}^a B_{-p,\jh_2}^s  &=& \frac{\pi}{\sin \pi p
} e^{\frac{s^2}{2}(\psi(p)+\psi(1-p)-2\psi(1))} \int_0^\infty dt \
tJ_0\left (\frac{\pi
s t}{\sin \pi p} \right ) {}^a B_{t,\jh_1+\jh_2}^1 \ , \nb \\
{}^a B_{p,\jh_1}^{s_2} {}^a B_{s_1,\jh_2}^{s_2}  &=& \frac{e^{-
\frac{i \pi s_1^2\sin \th} {2 \tan (\pi p)}}}{\pi s_1^2 \sin
\th}e^{\frac{s_1^2 (1-\cos \th) }{2}(\psi(p)+\psi(1-p)-2\psi(1))}
\sum_{n \in \mathbb{Z}} e^{in \th}\ {}^a B^1_{p,\jh_1+\jh_2+n} \ ,
\nb \label{scb11a} \ea where $\th_{p,q} = \pi \cot{\pi p}+\pi
\cot{\pi q}$ and $s_2^2 = 2s_1^2(1-\cos \th)$.

We make the following ansatz \be {}^a B_{\pm p,\jh}^s =
\sqrt{\frac{\pi}{\sin \pi \m p }} \ e^{ \pm 2 i p \h +
\frac{s^2}{4} [\psi(\m p)+\psi(1-\m p)-2 \psi(1)]} b^s_{\pm
p,\jh}(u) \ , \ee with $b^s_{\pm p, \jh+n}(u) =e^{i n \m u}
b^s_{\pm p,\jh}(u) $. The constraints then simplify \ba
b^s_{p,\jh_1}(u)b^s_{-q,\jh_2}(u) &=& e^{-\frac{i\pi [\cot(\pi \m
p)-\cot(\pi \m q)] s^2} {4 \tan \left(\frac{\m u}{2}\right )}}
\frac{b^1_{p-q,\jh_1+\jh_2}(u)}{1-e^{-i \m u}} \ , \nb \\
b_{p,\jh_1}^s(u) b_{-p,\jh_2}^s(u) &=& \int_0^\infty dt \ t
J_0\left (\frac{\pi
s t}{\sin \pi p} \right ) {}^a B_{t,\jh_1+\jh_2}^1 \ , \nb \\
b_{p,\jh_1}^{s_2}(u)  {}^a B_{s_1,\jh_2}^{s_2} &=&  \frac{e^{-
\frac{i \pi s_1^2\sin \th} {2 \tan (\pi p)}}}{\pi s_1^2 \sin
\th}\sum_{n \in \mathbb{Z}} e^{i n \g} b^1_{p,\jh_1+\jh_2+n}(u) \
, \ea and are solved by  $b_{\pm p,\jh}^s(u) = e^{i \jh \, u \mp
\frac{i\pi s^2}{4 \tan(\pi \m p) \tan \left(\frac{\m u}{2}\right
)}}$.

Therefore \ba {}^a B_{\pm p,\jh}^s &=& \sqrt{\frac{\pi}{\sin \pi
\m p }} \ \frac{e^{\pm 2 i p \h + i \jh u}}{1-e^{\pm i \mu u}}
e^{\frac{s^2}{4}[\psi(\m p)+\psi(1-\m p)-2 \psi(1)]  \mp
\frac{i\pi s^2}{4 \tan(\pi \m p) \tan \left(\frac{\m u}{2}\right
)}} \ , \nb \\
{}^a B_{s,\jh}^{t} &=& \frac{e^{i \jh u}}{\pi s^2 \sin \th} 2 \pi
\d(\m u - \th) \ , \hspace{1cm} t^2 = 2s^2(1+\cos \th) \ .
\label{bbc1} \ea

Here we show some explicit examples of the relation $(\ref{cf})$
for the $S1$ branes of the $H_4$ model. We have \be
C^{abc,(p+q,\jh_3)}_{(p,\jh_1) (q,\jh_2)} = {\bf
F}_{(-p_b,\jh_2-\jh_c+n_2),(p+q,
\jh_1 +\jh_2+k )}  \small  \begin{bmatrix} (p,\jh_1) & (q,\jh_2) \\
 (p_a,\jh_a) & (-p_c,-\jh_c)
\end{bmatrix}  \ , \nb \\
 \label{b-3+++b} \ee
where $\jh_1 = \jh_{ab}-n_1$, $\jh_2 = \jh_{bc}-n_2$, $\jh_3 =
\jh_{ac}-n_3$  and $k=n_1+n_2-n_3 \ge 0$. Similarly \be
C^{abc,(p-q,\jh_3)}_{(p,\jh_1) (-q,\jh_2)} = {\bf
F}_{(-p_b,\jh_2-\jh_c-n_2),(p-q,
\jh_1 +\jh_2-k )}  \small  \begin{bmatrix} (p,\jh_1) & (-q,\jh_2) \\
 (p_a,\jh_a) & (-p_c,-\jh_c)
\end{bmatrix} \ , \hspace{0.3cm} p > q \ , \nb \\
 \label{b-3++-} \ee
where $\jh_1 = \jh_{ab}-n_1$, $\jh_2 = \jh_{bc}+n_2$, $\jh_3 =
\jh_{ac}-n_3$  and $k=n_3+n_2-n_1 \ge 0$. We also have
 \be
C^{abc,(-(q-p),\jh_3)}_{(p,\jh_1) (-q,\jh_2)} = {\bf
F}_{(-p_b,\jh_2-\jh_c-n_2),(p-q,
\jh_1 +\jh_2+k )}  \small  \begin{bmatrix} (p,\jh_1) & (-q,\jh_2) \\
 (p_a,\jh_a) & (-p_c,-\jh_c)
\end{bmatrix} \ , \hspace{0.3cm} p < q \ , \nb \\
 \label{bbb-3++-} \ee
where $\jh_1 = \jh_{ab}-n_1$, $\jh_2 = \jh_{bc}+n_2$, $\jh_3 =
\jh_{ac}+n_3$ and $k=n_3+n_1-n_2 \ge 0$.

\section{Penrose limit of the $SU(2)$ and $SL(2,\mathbb{R})$ branes \label{plb}}

In this appendix we discuss the Penrose limit of the symmetric
branes in $S^3$ and in $AdS_3$ using coordinate systems adapted to
their world-volume. For $S^3$ we use spherical coordinates \be
ds^2 = k \left [ -dt^2 + d\psi^2 + \sin^2 \psi ( d\th^2 + \sin^2
\th d \varphi^2 ) \right ] \ , \hspace{1cm} H_{\psi \th \f} = 2 k
\sin^2 \psi \sin \th \ . \label{d8} \ee The symmetric branes sit
at $\psi_n = \pi n/k$. The integer $n$, $0 < n < k$,
parameterizes a uniform world-volume flux $F = -n/2 \sin \th$,
which stabilizes the brane \cite{bd}. When $n=0$ or $n=k$, the
brane world-volume degenerates to a point. In order to describe
the $S 1$ branes we make the following change of variables \be t =
\frac{\m x^+}{2} + \frac{x^-}{\m k}\ , \hspace{1cm} \psi =
\frac{\m x^+}{2} - \frac{x^-}{\m k} \ , \hspace{1cm} \th =
\frac{\r}{\sqrt{k}} \ . \label{d7b} \ee This leads in the limit
$k \rightarrow \infty$ to the Nappi-Witten wave in Rosen
coordinates. We can easily see that the flux  on the brane
world-volume becomes \be {\cal F} \equiv B + 2 \pi F =
-\frac{1}{2} \sin x^+ \r d \r \wedge d \varphi \ , \ee as
expected. Moreover we can exploit the relation between the brane
location $\psi$ and the spin of the $SU(2)$ representations \be
\psi_j = \frac{\pi (2j+1)}{k+2} \ , \hspace{1cm} j = 0, ...,
\frac{k}{2} \ , \label{d11bbb} \ee to derive an analogous relation
between the labels of the $H_4$ conjugacy classes $(u,\h)$ and the
quantum numbers of the $H_4$ representations. If we scale the spin
of $SU(2)$ in such a way as to obtain a discrete representation
$V_{\pm p, \jh}$ \cite{dk} \be j = \frac{k}{2}p \mp \jh \ ,
\hspace{1cm} p > 0 \ , \ee we obtain \be \m u = \pm 2 \pi (\m p +
n) \ , \hspace{1cm} 2 \h = \pi(2\jh \pm 2p \mp 1) \ . \ee For the
$D2$ we set \be t = \frac{\m u}{2} \ , \hspace{1cm} \f = \frac{\m
u}{2} - \frac{2v}{\m k} \ , \hspace{1cm} \psi =
\frac{\chi}{\sqrt{k}} +\frac{\pi}{2} \ , \hspace{1cm} \th =
\frac{\xi}{\sqrt{k}} + \frac{\pi}{2}\ , \label{d7bbb} \ee and take
the limit $k \rightarrow \infty$, which leads to the Nappi-Witten
wave in Brinkman coordinates. In this case we focus on $S^2$
branes very close to the equator of $S^3$ and scale the $SU(2)$
spin as \be j = \frac{k}{4} + \sqrt{k} \, \frac{\chi}{2 \pi} \ .
\ee As expected, the twisted-branes are in one-to-one
correspondence with the representations invariant under the action
of the external automorphism $\W$, $V^0_{s,0}$ and
$V^0_{s,\frac{1}{2}}$. Note that in the first case, the null
geodesic used to take the limit intersects the brane world-volume
while in the second case it is contained within the brane
world-volume.

The limit of the $AdS_2$ branes is better described using the
following coordinate system for $AdS_3 \times S^1$ \be d s^2 = k d
\psi^2 + k \cosh ^2 \psi \left ( d \w^2 - \cosh^2 \w d \t^2 \right
) + k d x^2 \ , \hspace{1cm} H_{\psi \w \t} = 2 k \cosh^2 \psi
\cosh \w \ . \ee The $AdS_2$ branes are surfaces with constant
$\psi$ and a world-volume flux $F_{\w \tau} = - \frac{k \psi}{2
\pi} \cosh \w$. The Penrose limit is  \be \tau = \frac{\m u}{2} +
\frac{2v}{\m k} \ , \hspace{1cm} x = \frac{\m u}{2} \ ,
\hspace{1cm} \psi = \frac{\chi}{\sqrt{k}} \ , \hspace{1cm} \w =
\frac{\xi}{\sqrt{k}} \ , \label{d2b} \ee with  $k \rightarrow
\infty$. In the process, the $AdS_2$ brane at constant $\psi$ (with
Neumann boundary condition along $S^1$) becomes the $D2$ brane at
constant $\chi$, with a null world-volume flux $F_{u \xi} =
\frac{\m \chi}{2}$, as expected. Similarly, the limit of the $H_2$
branes is more easily described if we use hyperbolic coordinates
for $AdS_3$ writing \be ds^2 = -k d\tilde \tau^2 +k \sin^2 \tilde
\tau (d \l^2+\sinh^2 \l d \phi^2) + k dx^2 \ , \hspace{1cm}
H_{\tilde \tau \phi \l} = 2 k \sin^2 \tilde \tau \sinh \l \ , \ee
with $\tilde \tau \in [-\pi,\pi]$, $\l \ge 0$. The $H_2$ branes
are surfaces with constant $\tilde \tau$ and a world-volume flux
$F_{\phi \l} = - \frac{k \tilde \tau}{2 \pi} \sinh \l$. In the
limit $k \rightarrow \infty$ the change of coordinates \be \tilde
\tau = \frac{\m x^+}{2} \ , \hspace{1cm} x = \frac{\m x^+}{2} -
\frac{2x^-}{\m k} \ , \hspace{1cm} \l = \frac{\r}{\sqrt{k}} \ ,
\hspace{1cm} \phi = -\f \ , \label{ddd2b} \ee leads to the
Nappi-Witten wave in Rosen coordinates (here $\r^2 =
y_1^2+y_2^2$). The $H_2$ branes with Dirichlet boundary conditions
along $S^1$ become the $S 1$ branes with the flux given in
$(\ref{sfluxr})$.

\section{The DBI approach \label{adbi}}

In this appendix we will study the more general class of
rotationally invariant solutions found in section \ref{dbi}. We
 consider $B\not= 0$. It is convenient to distinguish the
following cases:

(I) $|B+E|<2$. In this case the brane embedding can be written as
\be u={|B|\over \sqrt{4-(B+E)^2}}\log\left[r+\sqrt{r^2+{4AB\over
4-(B+E)^2}}\right]+u_0 \ee

\be v=v_0+{2A}{4-BE-E^2\over (4-(B+E)^2)^{3\over
2}}\log\left[r+\sqrt{r^2+{4AB\over 4-(B+E)^2}}\right]- \ee
$$-
{B+E\over 2\sqrt{4-(B+E)^2}}r\sqrt{r^2+{4AB\over 4-(B+E)^2}}
$$

(II) $|B+E|=2$. Here the embedding simplifies to

\be u=\sqrt{B\over 4A}r+u_0 \ee \be v=v_0+\sqrt{B\over A}~ {3Ar\pm
r^3\over 3B} \ee

(III) $|B+E|>2$. In this cases we obtain a trigonometric embedding

\be u=u_0+{|B|\over
\sqrt{(B+E)^2-4}}\arcsin\left[{r\sqrt{(B+E)^2-4}\over\sqrt{4AB}}\right]
\ee \be v=v_0-{|B|\over
2\sqrt{(B+E)^2-4}}r\sqrt{-r^2+{4AB\over(B+E)^2-4}}+ \ee
$$
+{2A(E^2+BE-4)\over ((B+E)^2-4)^{3\over
2}}\arcsin\left[{r\sqrt{(B+E)^2-4}\over\sqrt{4AB}}\right]
$$

Solving for r and substituting we obtain that the points on the brane are
on the curve
\be
[(B+E)^2-4](v-v_0)-2A{E^2+BE-4\over \sqrt{(B+E)^2-4}}(u-u_0)=
\ee
$$=
-{1\over
2}\sqrt{|AB^3|}~\sin\left[\sqrt{(B+E)^2-4}{2(u-u_0)\over |B|}\right]
$$

Using the embedding equations (\ref{sol2},\ref{sol3}) we can
calculate  the induced metric as \be d\hat s^2={4-E^2\over
B^2}r^2u'^2dr^2+r^2d\th^2=r^2\left[{4-E^2\over
B^2}du^2+d\th^2\right] \ee while the antisymmetric tensor is
$B_{r\theta}=2ur$. The induced two-dimensional curvature is \be
R\sim u'+ru'' \ee The induced metric is flat when $A=0$, when the
solution is

\be u={|B|\over \sqrt{4-(B+E)^2}}\log r+u_0\sp v=v_0- {B+E\over
2\sqrt{4-(B+E)^2}}r^2 \ee Our symmetric branes are a special case
of the flat branes with $B=0$.

The open string metric is the induced metric rescaled by
$(detg+B)/detg$. We find \be d s^2_{\rm
open}={4B^2u^2+(4-E^2)r^2u'^2\over
(4-E^2)r^2u'^2}\left[{4-E^2\over B^2}r^2u'^2dr^2+r^2d\th^2\right]
\ee
$$
=\left(r^2+{4B^2u^2\over (4-E^2)u'^2}\right)\left[{4-E^2\over
B^2}du^2+d\th^2\right]
$$

For the symmetric branes this is again flat.

\bigskip

The critical electric field case $E=\pm 2$ is a bit special and we
will discuss it here, separately.
 The gauge field equation implies in this case
\be r^2u'^2+2u'v'=1 \ee while the others \be 2r^2u'=B|2ur+F|\sp
r^2u'+v'=\left(\mp 1+{A\over r^2}\right)|2ur+F| \ee
The previous equations can be massaged into  \be u'=\sqrt{B\over 4A}{1\over
\sqrt{1-{B\pm 4\over 4A}r^2}}\Rightarrow r=R\sin\left[\sqrt{B\pm
4\over B}(u-u_0)\right] \ee \be v=v_0+{R\over \sqrt{B(B\pm
4)}}\left[{B\pm 2\over 2}r\sqrt{1-{r^2\over R^2}}\pm R\arcsin
{r\over R}\right] \ee This class of solutions describes an
embedding with a compact support $0<r<R$ with $R=\sqrt{4A\over
B\pm 4}$. The induced metric here is degenerate \be d\hat
s^2=r^2d\theta^2 \ee

These are the null branes mentioned (but not analyzed in detail) in the main
body of this paper.

\subsection{$S 1$ fluctuations}

Expanding around the classical solution $u_*$,$v_*$, $F_*$
satisfying (\ref{sol1})-(\ref{sol3}) \be u\to u_*+u\sp v\to
v_*+v\sp F\to F_*+F \ee we obtain to quadratic order \be L=L_*+L_2
+{\cal O}(u^3,v^3, F^3) \ee \be L_2= {1\over
2L_*^{3}}\left[(1-2u_*'v_*'-r^2u_*'^2)r^2(2ur+F)^2+\right. \ee
$$
+2r^2(2u_*r+F_*)[(v_*'+r^2u_*')u'+u_*'v'](2ur+F)-L_*^2[u_*'^2\dot
v^2+(r^2+v_*'^2)\dot u^2+2(1-u_*'v_*')\dot u\dot v]
$$
$$
\left.-r^4u_*'^2v'^2-r^4[r^2+v_*'^2+(2u_*r+F_*)^2]u'^2
-2r^2[(2u_*r+F_*)^2+r^2(1-u_*'v_*')]\right]u'v'$$

{}From equations (\ref{sol1})-(\ref{sol3}) \be L_*={2\over
|B|}r^2u_*'\sp 1-2u_*'v_*'-r^2u_*'^2={4-E^2\over B^2}r^2u_*'^2\sp
(2u_*r+F_*)^2={E^2\over B^2}r^2u_*'^2 \ee follow and we can
rewrite $L_2$ as

\be L_2={(4-E^2)\sqrt{4AB+(4-(B+E)^2)r^2}\over
16r^2}\left(F+2ur+{E [(2A-Er^2)u'+B v']\over 4-E^2}\right)^2- \ee
$$
-{4(A^2-AEr^2+r^4)\dot u^2+B^2\dot v^2+2(2AB-[(E(B+E)-4]r^2)\dot
u\dot v\over 4r^2\sqrt{4AB+(4-(B+E)^2)r^2}}
$$
$$
-\sqrt{4AB+(4-(B+E)^2)r^2}~~\times
$$
$$\times
{4(A^2-AEr^2+r^4)u'^2+B^2 v'^2+2(2AB-[(E(B+E)-4]r^2) u' v'\over
4(4-E^2)r^2}
$$

Specializing to the symmetric solutions $A=B=0$ \be u_*=const.\sp
v_*\to v_0-{Er^2\over 2\sqrt{4-E^2}}\sp L_*={2r\over
\sqrt{4-E^2}}\sp 2u_*r+F_*={Er\over \sqrt{4-E^2}} \ee we obtain.
\be L_2={\sqrt{4-E^2}\over
16r}\left[-(4-E^2)(2ur+F)^2+2E^2r^2u'(2ur+F)+{16r^2\over
4-E^2}\dot u^2+8\dot u\dot v+ \right. \ee
$$\left.+
(4+E^2)r^4u'^2+8r^2u'v'\right]
$$

We redefine \be V=v+{2r^2\over 4-E^2}u\sp A_{\theta}\to
A_{\theta}+{E^2r^2u\over 4-E^2} \ee and rewrite the action (after
performing an integration by parts) as \be L_2={\sqrt{4-E^2}\over
16r}\left[-(4-E^2)\left(F+{8ur\over 4-E^2}\right)^2+
 8(\dot
u\dot V+r^2u'V')+32{r^2u^2\over 4-E^2} \right] \ee

It is obvious from the Lagrangian above that $u$ satisfies the
flat two-dimensional Laplace equation. \be \Box u=0\Rightarrow
{1\over r}(ru')'+{1\over r^2}\ddot u=0\label{flaplace} \ee The
solution to the equation for the gauge fluctuation  is \be
F+2ur-{E^2\over 4-E^2}r^2u'={2C\over (4-E^2)}r \ee with $C$ a
constant.

Finally the  $u$ equation reads

\be \Box V={1\over r}(rV')'+{1\over r^2}\ddot V=-{4\over
4-E^2}(2u+C) \label{flaplace2} \ee Thus the V field is a free
field subject to a source linear in u and C.

The regular solution of (\ref{flaplace}) is $u=u_0$ constant. On
the other hand in order that  (\ref{flaplace2}) has a regular
solution we must have $C=-2u_0$ and then $V=V_0$ constant.

In the critical case , we take $E\to 2$ and rescale $F\to
F/\sqrt{4-E^2}$, $(u,v)\to (u,v)\sqrt{4-E^2}$ to obtain \be
S_2(E=2)={\sqrt{4AB+(4-B^2)r^2}\over
16r^2}\left(F+2ur+2[2(A-r^2)u'+B v']\right)^2- \ee
$$
-{\sqrt{4AB+(4-B^2)r^2}\over 4r^2}\left[2(A-r^2)u'+B v'\right]^2
$$
This system is degenerate. The solution to its equation of motion
is \be F+2ur=-{4Cr^2\over \sqrt{4AB+(4-B^2)r^2}} \sp 2(A-r^2)u'+B
v'={10Cr^2\over \sqrt{4AB+(4-B^2)r^2}} \ee

\section{Bulk one-point couplings from the DBI action}

We can compute the coupling to the bulk metric from \be
S^{\m\n}\equiv {\delta S_{DBI}\over \delta G_{\m\n}} \ee

\be S^{\m\n}={1 \over 4}\sqrt{-\det(\hat G+\hat
B+F)}\left[\left(\hat G+\hat B+F\right)^{\a\b}+\left(\hat G-\hat
B-F\right)^{\a\b}\right]\p_{\a}x^{\m}\p_{\b}x^{\n} \ee

For the D2-branes at $y=y_0$ we obtain the following coupling

\be S^{uv}= -{1\over 2\sqrt{1-f_{uv}^2}}\sp S^{vv}= {-f_{ux}^2 +
x^2 +
  2y_0f_{ux}\over 2\sqrt{1-f_{uv}^2}}
\ee
$$
 S^{vx}= {f_{uv}(f_{ux}-y_0)\over 2\sqrt{1-f_{uv}^2}}\sp S^{xx}={\sqrt{1-f_{uv}^2}\over 2}
 $$
 all others being zero. In summary,
 \be
S=\left(%
\begin{array}{cccc}
   0 & -{1\over 2\sqrt{1-f_{uv}^2}} & 0 & 0 \\
  -{1\over 2\sqrt{1-f_{uv}^2}} & {-f_{ux}^2 +  x^2 +
  2y_0f_{ux}\over 2\sqrt{1-f_{uv}^2}} & {f_{uv}(f_{ux}-y_0)\over 2\sqrt{1-f_{uv}^2}} & 0 \\
   0 &  {f_{uv}(f_{ux}-y_0)\over 2\sqrt{1-f_{uv}^2}} & {\sqrt{1-f_{uv}^2}\over 2} & 0\\
   0 & 0 & 0 & 0 \\
\end{array}%
\right) \ee At the symmetric point \be
S={1\over 2}\left(%
\begin{array}{cccc}
   0 & -1 & 0 & 0 \\
  -1 & x^2 & 0 & 0 \\
   0 &  0& 1 & 0\\
   0 & 0 & 0 & 0 \\
\end{array}%
\right) \ee

For the D1 branes \be S^{\mu\nu}_{D1}={B\over
4u'}r^2\p_{r}x^{\mu}\p_{r}x^{\nu}+{4-E^2\over
4B}u'\p_{\theta}x^{\mu}\p_{\theta}x^{\nu} \ee For the symmetric
configurations  we obtain \be S^{vv}_{D1}={E^2 r^3\over
4\sqrt{4-E^2}}\sp S^{vr}_{D1}=-{E\over 4}r^2\sp
S^{rr}_{D1}=\sqrt{4-E^2}~{r\over 4}\sp
S^{\theta\theta}_{D1}={\sqrt{4-E^2}\over 4r} \ee

The coupling to the antisymmetric tensor is given by \be
A^{\m\n}\equiv {\delta S_{DBI}\over \delta B_{\m\n}} \ee

\be A^{\m\n}={1 \over 4}\sqrt{-\det(\hat G+\hat
B+F)}\left[\left(\hat G+\hat B+F\right)^{\a\b}-\left(\hat G-\hat
B-F\right)^{\a\b}\right]\p_{\a}x^{\m}\p_{\b}x^{\n} \ee By direct
calculation for the D1 case we obtain that the only non-zero
components are \be A^{\theta u}={E\over 4}u'\sp A^{\theta
v}={E\over 4}v'\sp A^{\theta r}={E\over 4} \ee For the particular
case of symmetric D1 branes we have \be A^{\theta u}=0\sp
A^{\theta v}=-{E^2r\over 4\sqrt{4-E^2}}\sp A^{\theta r}={E\over 4}
\ee

For the D2 branes we obtain

\be
A={1\over 2}\left(%
\begin{array}{cccc}
   0 & {f_{uv}\over \sqrt{1-f_{uv}^2}} & 0 & 0 \\
  -{f_{uv}\over \sqrt{1-f_{uv}^2}} & 0 & {f_{ux}-y_0\over \sqrt{1-f_{uv}^2}} & 0 \\
   0 &  -{f_{ux}-y_0\over \sqrt{1-f_{uv}^2}}& 0 & 0\\
   0 & 0 & 0 & 0 \\
\end{array}%
\right)~~~{\Longrightarrow\atop\lim_{f\to 0}}~~~
{1\over 2}\left(%
\begin{array}{cccc}
   0 & 0 & 0 & 0 \\
 0 & 0 & -y_0 & 0 \\
   0 &  y_0& 0 & 0\\
   0 & 0 & 0 & 0 \\
\end{array}%
\right) \ee

The one-point coupling to the dilaton is given by \be F\equiv
{\delta S_{DBI}\over \delta \Phi}=-\sqrt{-\det(\hat G+\hat B+F)}
\ee We obtain \be F_{D1}=-{2r^2u'\over B}=-{2r^2\over
\sqrt{(4-(B+E)^2)r^2+4AB}}~~~~{\Longrightarrow\atop{\rm
symmetric}}~~~~~-{2r\over \sqrt{4-E^2}} \ee \be
F_{D2}=-\sqrt{1-f_{uv}^2}~~~~{\Longrightarrow\atop{\rm
symmetric}}~~~~~-1 \ee

\section{Some useful series and integrals \label{dd}}

\be \sum_{n=0}^\infty L_n^\a(x) z^n =
\frac{e^{\frac{xz}{z-1}}}{(1-z)^{1+\a}} \ . \ee

\be \sum_{n=0}^\infty\frac{z^n}{\G(n+\a+1)}  L_n^\a(x) =
e^{z}(xz)^{-\frac{\a}{2}} J_{\a}(2 \sqrt{xz}) \ . \ee

\be \int_0^\infty dx e^{-\b x^2}x^{\n+1} L_n^\n(\a x^2) J_\n(xy) =
2^{-\n-1}\b^{-\n-n-1}(\b-\a)^n y^\n e^{-\frac{y^2}{4\b}} L^\n_n
\left [ \frac{\a y^2}{4\b(\a-\b)}\right ] \ . \label{ui1} \ee

\be \int_0^\infty dx e^{-x} x^\a L_n^\a(x) L_m^\a(x) = \d_{n,m}
\frac{\G(\a+n+1)}{n!} \ , \label{ui2} \ee

\be \int_0^\infty dx x J_m(sx)J_m(tx) = \frac{\d(s-t)}{s} \ , \ee

\be \int_0^\infty dx e^{- \s x} x^\a L_n^\a(\l x) L_m^\a(\m x) =
\frac{\G(m\!+\!n\!+\!\a\!+\!1)}{m!\ n!}
\frac{(\s\!-\!\l)^n(\s\!-\!\m)^m}{\s^{n+m+\a+1}}
F(\!-\!m,\!-\!n,\!-\!m\!-\!n\!-\!\a, \t) \ , \label{ui3} \ee where
\be \t = \frac{\s(\s-\l-\m)}{(\s-\l)(\s-\m)} \ . \label{ui4} \ee

\ba && \int_0^\infty dx x^{\n+1} e^{-\a x^2} L_m^{\n - \s}(\a
x^2)L_n^\s(\a x^2)
J_\n(xy) \nb \\
&=& (-1)^{m+n} (2 \a)^{-\n-1}y^\n e^{-\frac{y^2}{4 \a}} L^{\s
-m+n}_m \left ( \frac{y^2}{4\a} \right ) L^{\n - \s+m-n}_n \left (
\frac{y^2}{4\a} \right ) \ . \label{ui5} \ea

\ba && \int_0^\infty x^{\n+1} e^{-\b x^2}\left [
L_n^{\frac{\n}{2}}(\a x^2) \right ]^2 J_\n(xy) dx =
\frac{y^\n}{\pi n!} \G(n+1+\n/2) \frac{e^{-\frac{y^2}{4 \b}}}{(2 \b)^{\n+1}} \nb \\
&& \sum_{l=0}^n \frac{(-1)^l
\G(n-l+1/2)\G(l+1/2)}{\G(l+1+\n/2)(n-l)!} \left (
\frac{2\a-\b}{\b} \right )^{2l} L^\n_{2l} \left [ \frac{\a y^2}{2
\b(2\a-\b)}\right ] \ . \label{ui6} \ea

\be \sum_{n=0}^\infty n! \frac{L_n^\a(x)L_n^\a(y)z^n}{\G(n+\a+1)}
= \frac{(xyz)^{-\frac{\a}{2}}}{1-z} e^{-\frac{z(x+y)}{1-z}}
I_\a\left ( \frac{2 \sqrt{xyz}}{1-z} \right ) \ . \label{ui7} \ee

\be \sum_{j=0}^\infty
c_1^{j-l}L_l^{j-l}(b_1c_1)c_2^{l+n-j}L_j^{l+n-j}(b_2c_2) =
e^{-c_1b_2}(c_1+c_2)^nL_l^n\left [(b_1+b_2)(c_1+c_2)\right] \ .
\label{ui8} \ee

\ba \int_0^\infty dr r e^{-a \frac{r^2}{2}}&& \left [
\frac{r^2}{2} \right ]^{q-s-m+n} L_m^{q-m}\left (\frac{a r^2}{2}
\right )L_s^{n-s}\left ( \frac{b r^2}{2} \right
)L_{m+k-n}^{q+n-m-s}\left ( \frac{(a-b)r^2}{2} \right ) (-1)^{s+n-k}\nb \\
&=&
\frac{(k+q)!(k+m)!}{m!s!k!(k+m-n)!}\frac{(a-b)^{n-k-m}b^s}{a^{k+q+1}}
F\left(-k.-s,-q-k,\frac{a}{b}\right)\times\\
&&\times
F\left(-k,-m-k+n,-m-k,\frac{a}{a-b}\right) \ . \nb \label{ui9} \ea

\ba & & \int_{-\infty}^\infty du \, e^{-\frac{u^2}{2}+
\frac{iu(q_1-q_2)}{\sqrt{2}}}\left [u +
\frac{i(q_1+q_2)}{\sqrt{2}} \right ]^{n-m}L^{n-m}_m\left [ u^2 +
\frac{(q_1+q_2)^2}{2} \right ] \nb \\&=& \sqrt{\pi} i^{n+m}
\frac{e^{-\frac{(q_1-q_2)^2}{4}}}{2^{\frac{m+n-1}{2}}m!}
H_n(q_1)H_m(q_2) \ . \label{ui10} \ea

\be \int_{0}^\infty J_{\n}(ax) e^{ibx} dx = \frac{e^{i \n
\sin^{-1} \frac{b}{a}}}{\sqrt{a^2-b^2}} \ , \hspace{1cm} a > b \ .
\label{ui11} \ee

\be \int_{0}^\infty e^{-\a x^2} J_\n(\b x) dx =
\frac{1}{2}\sqrt{\frac{\pi}{\a}}e^{-\frac{\b^2}{8
\a}}I_{\n/2}\left (\frac{\b^2}{8 \a}\right ) \ , {\rm Re} (\a) >
0, \b > 0 , {\rm Re} (\n) > -1 \ .  \ee

\be \int_0^\infty ds \cos (bs) J_0(as) = \frac{1}{\sqrt{a^2-b^2}}
\ , \hspace{1cm} a > b \ . \ee

\be \int_0^{\infty} x~dx e^{-a~x^2}J_0(xy)={1\over
2a}~e^{-{y^2\over 4a}} \ee

\end{document}